# Standardizing Force Reconstruction in Dynamic Atomic Force Microscopy

Simon Laflamme[1], Bugrahan Guner[1], Omur E. Dagdeviren[1]*

[1]Department of Mechanical Engineering, École de technologie supérieure, University of Quebec, Canada, H3C 1K3

*Corresponding author's e-mail: omur.dagdeviren@etsmtl.ca

**Abstract**

Atomic force microscopy (AFM) enables high-resolution imaging and quantitative force measurement, which is critical for understanding nanoscale mechanical, chemical, and biological interactions. In dynamic AFM modes, however, interaction forces are not directly measured; they must be mathematically reconstructed from observables such as amplitude, phase, or frequency shift. Many reconstruction techniques have been proposed over the last two decades, but they rely on different assumptions and have been applied inconsistently, limiting reproducibility and cross-study comparison. Here, we systematically evaluate major force reconstruction methods in both frequency- and amplitude-modulation AFM, detailing their theoretical foundations, performance regimes, and sources of error. To support benchmarking and reproducibility, we introduce an open-source software package that unifies all widely used methods, enabling side-by-side comparisons across different formulations. This work represents a critical step toward achieving consistent and interpretable AFM force spectroscopy, thereby supporting the more reliable application of AFM in fields ranging from materials science to biophysics.

# 1. Introduction

Atomic force microscopy (AFM)[1], a central technique in scanning probe microscopy[2], enables nanoscale imaging and force measurements by probing the interaction between a sharp tip and a sample surface[3,4]. Over the past three decades, advances in tip functionalization[5-7] and data analysis[8-32] have extended AFM's capabilities to quantify interatomic and intermolecular forces with piconewton and picometer resolution[33-41]. However, in dynamic force microscopy, these interaction forces are not directly measured; instead, they are reconstructed from experimental observables using diverse mathematical models[8-32]. Despite numerous methodological developments, the field lacks a standardized framework for force reconstruction. Existing approaches are based on varying assumptions, offer differing levels of accuracy, and are often validated under inconsistent conditions. No systematic comparison or tool currently exists to evaluate their performance or reliability across experimental regimes.

In this analysis, we chronologically examine mathematical techniques developed for tip-sample force reconstruction from dynamic AFM measurements, beginning with early formulations from the early 2000s. We focus on the assumptions, limitations, and operational boundaries of each method, highlighting inconsistencies in their application. To address this gap, we present a software package that enables direct, side-by-side comparison of reconstruction methods using input data. Our aim is to foster reproducibility, establish clearer benchmarks, and move toward standardizing force reconstruction practices in dynamic AFM.

# 2. Results

## 2.1 Contact-Mode Atomic Force Microscopy-Based Force Quantification and Its Limitations

The first significant advance in measuring tip-sample interaction forces was the optical detection of cantilever deflection in contact-mode AFM[42,43]. This approach, rooted in Hooke's law, allows force calculation by multiplying the measured deflection by the known cantilever spring constant (Figure 1a). However, this method suffers from intrinsic limitations. As illustrated in Figure 1a, the finite tip-sample contact area, typically involving tens of atoms (shaded gray), leads to spatial averaging, obscuring interatomic-scale resolution. Moreover, mechanical instabilities such as jump-in and jump-out events (Figure 1b) hinder reproducibility, even though they may be applied to recover a fraction of local sample properties[44]. These arise because the cantilever spring constant is often lower than the effective gradient of the tip-sample interaction, leading to nonlinear dynamics during approach and retraction[10,40]. While using stiffer cantilevers may mitigate these effects, they reduce measurable deflection, limiting sensitivity. Consequently, contact-mode force spectroscopy fails to reliably quantify local tip-sample interactions with true atomic resolution.

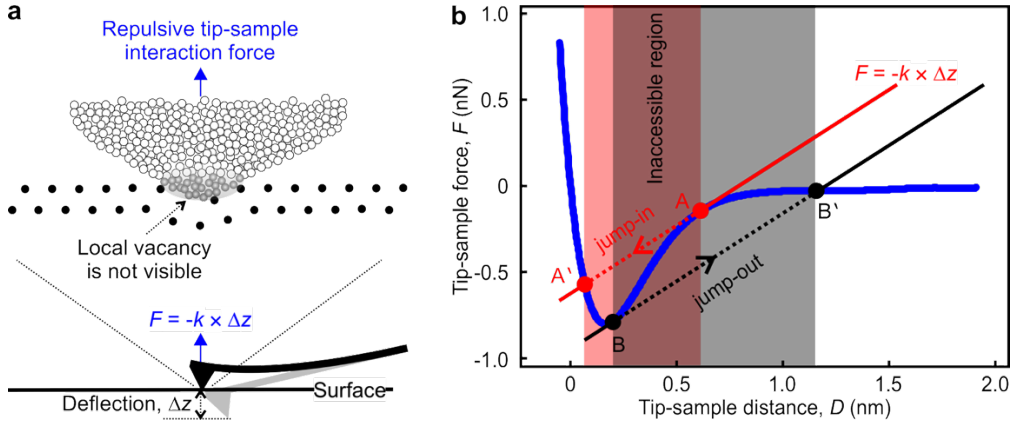

**Figure 1. Mechanical instabilities and limitations of contact mode atomic force microscopy (AFM) force measurements. (a)** In contact mode AFM, the cantilever tip remains in continuous contact with the sample surface, and the deflection of the cantilever is converted to force, $F$, using Hooke's law, $F = -k \times \Delta z$, where $k$ is the spring constant of the cantilever and $\Delta z$ is the cantilever's deflection. However, atomic-scale features such as local vacancies may remain undetected due to lateral averaging by the finite-size tip, as highlighted with the shaded region. **(b)** The tip–sample force curve (blue) shows typical instabilities: during approach, the tip abruptly "jumps-in" to contact (A → A′), and during retraction, it "jumps-out" (B → B′), resulting in a force–distance region that is inaccessible to direct measurement via contact mode AFM. These nonlinearities and discontinuities complicate accurate force quantification and highlight the limitations of contact mode for high-resolution force spectroscopy.

## 2.2 Dynamic Force Spectroscopy as an Alternative

To overcome contact-mode limitations, dynamic AFM modes[3,8] were developed to control and indirectly measure tip-sample forces through oscillatory motion (Figure 2). In amplitude modulation (AM-AFM), changes in oscillation amplitude ($A$) and phase ($\beta$) are monitored[3], whereas frequency modulation (FM-AFM) tracks resonance frequency shift ($\Delta f$)[8]. AM-AFM is more user-friendly, but conventional AM-based measurements suffer from low force sensitivity and long response times under vacuum[26]. FM-AFM, by contrast, offers higher resolution and mechanical stability at the cost of greater operational complexity[45]. In both operation modes, force spectroscopy is typically performed by recording $A$, $\beta$, or $\Delta f$ as a function of piezo displacement, which is eventually employed to recover tip-sample distance, $D$[8-32]. These experimental observables must then be converted into physical forces using mathematical reconstruction techniques, each with unique assumptions, strengths, and limitations.

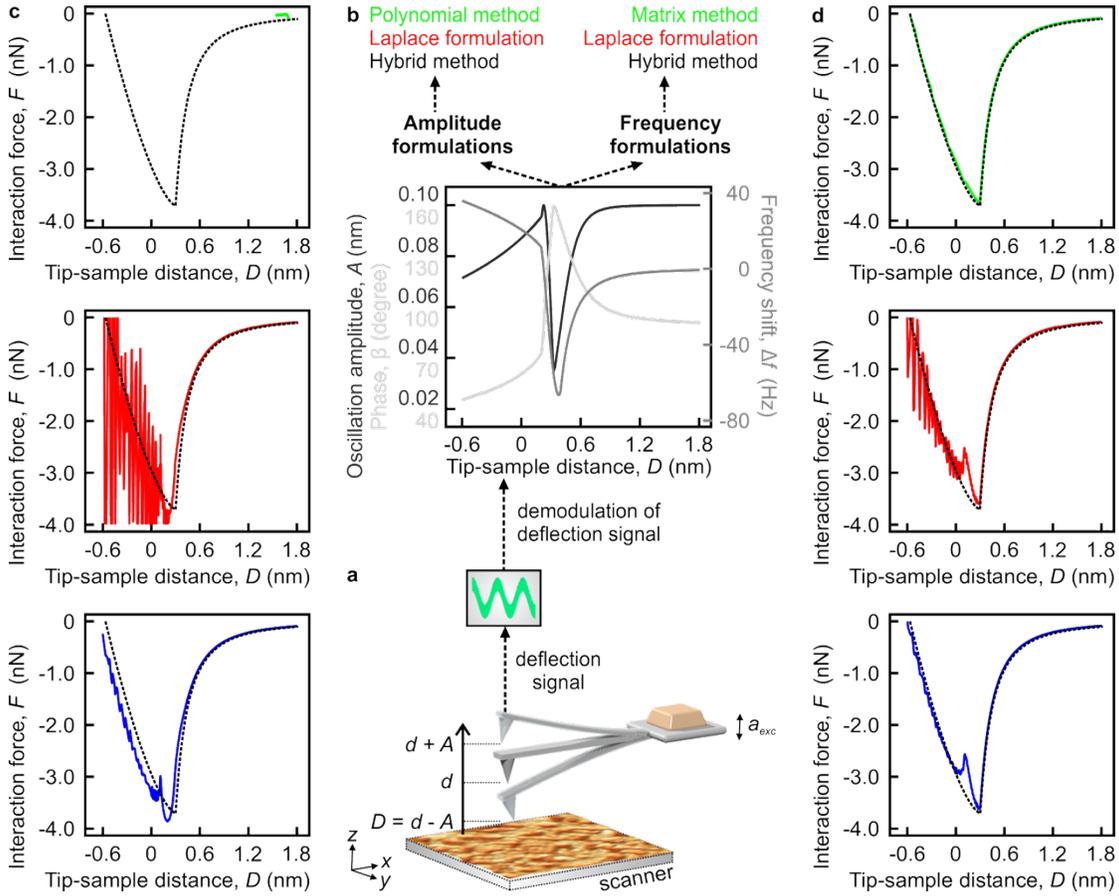

**Figure 2. Explanation of dynamic force spectroscopy methodologies. (a)** Schematic of dynamic atomic force microscopy (AFM), where the cantilever oscillates with amplitude $A$ near a sample surface, with an excitation signal $a_{exc}$. The tip–sample distance, $D$, is achieved by the average position $d$ and $A$. The deflection signal is demodulated to yield amplitude $A(D)$, phase $\beta(D)$, and frequency shift $\Delta f(D)$ observables, which are used for force reconstruction. **(b)** Force reconstruction methods are categorized by whether they rely on amplitude (left) or frequency (right) formulations. **(c)** Force reconstruction using amplitude-based methods: polynomial method (top, green)[23], Laplace formulation (middle, red)[24,25], and hybrid formulation (bottom, blue)[27]. **(d)** Frequency-based formulations: matrix method (top, green)[17], Laplace formulation (middle, red)[19], and hybrid formulation (bottom, blue)[27]. In all panels, the true model force is shown as a black dotted line. All reconstructions are computed under ideal, noise-free conditions to isolate intrinsic methodological differences[29]. The polynomial method only reproduces a small fraction of the tip-sample force. The Laplace formulation (red curves) exhibits pronounced oscillations and instability near the point of closest approach, especially in the amplitude-based implementation, indicating sensitivity to numerical ill-posedness even without noise. In contrast, the hybrid methods (blue curves) more accurately capture the true force profile with minimal deviation, benefiting from combined robustness and regularization. The matrix-based method (green curve) shows good agreement under ideal numerical conditions at intermediate distances but tends to deviate near contact or in regions with steep force gradients.

### 2.2.1 Frequency-Modulation-Based Dynamic Force Spectroscopy

Early FM-AFM studies used the assumption that $\Delta f$ is proportional to the force gradient, i.e., slowly varying with distance, valid under small-amplitude (small $A$) oscillations, which refers to $A$ relative to the decay length of the tip-sample interaction force[8,46,47]. This approximation simplifies analysis via Taylor expansion and perturbation theory but breaks down at small separations where higher-order terms become non-negligible[9,11,13,15]. While this method is intuitive and analytically tractable, it lacks generality and fails under non-ideal conditions.

Another early alternative to direct inversion was recursive modeling of the tip-sample force in a damped harmonic oscillator framework. By iteratively adjusting the model force to reproduce experimental $\Delta f(D)$ curves, this method aimed to approximate the real interaction[48]. However, its reliance on an accurate initial guess and high computational demand limited its practical adoption, especially in early applications lacking robust hardware.

General formulations based on classical mechanics[9,11,13,15] introduced integral equations linking $\Delta f$ to the interaction force (see Sections 1 and 2 of the Supplementary Information for classical mechanics background and detailed derivations). Distinct expressions were derived for small and large $A$ regimes. Small $A$ formulations, again, rely on Taylor expansion and yield $\Delta f \propto d^2U/dz^2$, suitable only near equilibrium[9,11,13,15]. In the large $A$ regime, $\Delta f$ reflects an averaged interaction across the oscillation cycle. The concept of normalized frequency shift emerged from this formulation, enabling comparison across experimental conditions and cantilever geometries. This formulation improved robustness but still required amplitude-specific adjustments.

The inverse Abel transform offers a theoretically grounded method for large $A$ data, assuming weak anharmonicity and slow variation of $\Delta f$[13,15]. Under these conditions, the tip-sample force can be directly recovered from the integral kernel of the frequency shift. Though elegant, the method is sensitive to noise and requires the force to be smooth and well-behaved over the integration domain. Experimental validation showed good agreement, though deviations occurred near sharp force transitions.

To bypass regime-specific formulations, discretization-based methods[17] were first developed (see Section 3 of the Supplementary Information for the detailed derivation). These link $\Delta f$ to the force gradient via a weighted kernel, independent of $A$. The method assumes stable oscillation amplitudes and smooth $\Delta f(z)$ data, often necessitating pre-processing such as curve fitting. While powerful, the method excludes higher harmonic effects and assumes first-order perturbation theory holds, limiting its generality. Finally, the discretization relied on tip-sample separation and a stable, i.e., constant, $A$ as a function of $z$ was assumed. With this approach, the force was assumed to be constant over several segments of the integrating range so that it could be taken out of the integral. It resulted in a matrix to be solved computationally, for which each term is standardized irrespective of the shape of the force.

The most widely adopted modern approach employs Laplace transforms[19-21,49] to recover force and potential energy from $\Delta f$ data for arbitrary $A$ (see Section 3 of the Supplementary Information for the detailed derivation). This formulation integrates contributions from small, medium, and large $A$ via a Padé approximant. While this approach unifies previous formulations and accommodates realistic oscillation amplitudes, it assumes constant $A$ across $z$ and depends on the accuracy of the inverse Laplace transformation. Medium $A$ reconstructions remain sensitive to the choice of approximant and decay length of the interaction force, introducing uncertainties. Nevertheless, the prudence of each of those terms in the final equation depends on the relative magnitude of the *A* compared to the decay of the tip sample interaction force and its distance dependence.

**2.2.2 Amplitude-Modulation-Based Dynamic Force Spectroscopy**

Conventional amplitude-modulation (AM)-based AFM spectroscopy suffers from mechanical instabilities and long settling times, particularly under vacuum conditions[3,26,50-52]. Despite these limitations, mathematical formulations for AM-based spectroscopy have evolved in parallel with those for frequency-modulation (FM)-based methods, including both small- and large-amplitude regimes (see Sections 4 and 5 of the Supplementary Information for the detailed derivation).

Laplace-transform-based formulations originally developed for FM-AFM have been adapted to AM-AFM[24,25]. For intermediate amplitudes, these rely on similar formulations and approximations but with a modified prefactor to reflect the different detection mechanism. The accuracy of these reconstructions depends strongly on the specific form of the tip-sample interaction force and the relative weighting of terms within the formulation.

A general formulation was introduced to describe both conservative and dissipative tip-sample interactions using Laplace transforms[22]. This approach requires the use of modified Bessel functions of the first kind, represented as a power series. Increasing the number of terms in the series improves accuracy, but the associated differential equations increase in order correspondingly, resulting in rapidly growing computational complexity.

An alternative polynomial-based approach approximates the force as a finite series of Chebyshev polynomials[23]. Each matrix element corresponds to an integral involving a Chebyshev polynomial of a specific order and a standard kernel. Solving the inverse matrix equation yields the polynomial coefficients, enabling analytical reconstruction of the force. This method avoids some of the simplifying assumptions made in earlier techniques and provides a continuous expression for the force rather than a discrete set of points. However, the matrix size scales quadratically with the number of polynomials, leading to significant increases in processing time for higher-resolution reconstructions.

**2.2.3 Limitations of Force Spectroscopy Techniques**

Deviations from general integral formulations were systematically observed as functions of oscillation amplitude, interaction strength, and tip–sample separation[27-31]. It was recognized that the general integral equation used to reconstruct tip–sample interaction forces constitutes an ill-posed inverse problem: small experimental inaccuracies, such as fluctuations in oscillation amplitude or frequency shift, can lead to disproportionately large errors in the reconstructed force. To quantify this instability, a diagnostic expression was introduced, incorporating the oscillation amplitude and the first and third derivatives of the tip–sample force curve[28]. This expression served to assess the validity of the reconstruction to experimental noise, thereby offering a predictive criterion for the stability of force recovery. Subsequently, a standardized protocol was proposed, enabling experimentalists to apply this ill-posedness test to their own data prior to force reconstruction[27,28]. Further analysis revealed that the small-amplitude term in the perturbative expansion contributes significantly to instability, especially in the short-range repulsive regime where force gradients are steep[27]. To mitigate this, a hybrid reconstruction approach[27] was developed wherein the small-amplitude term was replaced by a first-order approximation using Bessel functions. This modification significantly improved numerical stability for oscillation amplitudes on the order of tens of picometers and preserved reconstruction accuracy even within the repulsive regime of the interaction potential.

A summary of the mathematical formulations, along with their assumptions, advantages, limitations, and typical performance characteristics, is presented in Table 1. While dynamic AFM has become a cornerstone technique for probing tip–sample interaction forces, the lack of standardized analytical approaches has led to inconsistencies across studies and limited the reproducibility of quantitative force spectroscopy. To address this longstanding challenge, we developed an open-source software package that unifies all major dynamic force reconstruction methods for both FM and AM modes (see supplementary software). The platform integrates experimental parameters such as oscillation amplitude variation and signal fitting, and, crucially, allows for direct, quantitative comparison of all reconstruction algorithms under the same conditions (see supplementary video, explaining the use of the supplementary software).

**Table 1. Comparison of dynamic atomic force microscopy (AFM) tip-sample interaction force reconstruction formulations.** An overview of commonly used formulations in frequency-modulation (FM) and amplitude-modulation (AM) AFM, detailing underlying assumptions, advantages, limitations, and typical computation times. $A$ denotes oscillation amplitude, $Q$ is the quality factor, and $\beta$ is the phase shift. The table highlights the importance of standardized comparisons across models.

| Formulation and Modulation | Assumptions | Advantages | Drawbacks | Computation Time |
|---|---|---|---|---|
| Matrix (FM)[17] | Harmonic motion; constant force over integration steps, constant $A$. | Scalable; adjustable accuracy via step number | Sensitive to noise and ill-posedness; cannot account for amplitude variation. | < 1 minute |
| Laplace (FM)[19] | Harmonic motion; constant $A$ and $\beta$. | Less sensitive to ill-posedness for proper $A$. | Systematic error at mid-range $A$. | A few minutes |
| Polynomial (AM)[23] | Harmonic motion; $Q \geq 400$; large $A$. | Analytical solution; tunable accuracy via Chebyshev series. | Sensitive to ill-posedness; impractical for small $A$. | A few hours |
| Laplace (AM)[24,25] | Harmonic motion; slowly varying $A$ and $\beta$. | Less sensitive to ill-posedness for proper $A$. | Systematic error at mid-range $A$. | A few minutes |
| Hybrid (AM/FM)[27] | Harmonic motion; Bessel term for small $A$. | Improved stability and accuracy at small $A$; converges to Laplace formulation at larger $A$. | Systematic error at mid-range $A$. | A few minutes |

The software package represents a foundational step toward the standardization of dynamic AFM force spectroscopy, enabling reproducible and transparent benchmarking of reconstruction techniques. Without such a framework, reported force values can vary substantially depending on the chosen method and hidden assumptions, hindering cross-comparison between labs and slowing progress in the field. By providing the first publicly available tool for comprehensive, side-by-side assessment, our software lays the groundwork for a community-wide consensus on best practices, paving the way for more robust, reliable, and universally interpretable AFM force measurements. We expect this framework to become a reference in both method development and experimental validation in dynamic force microscopy.

## 3. Discussions

Despite the extensive development of mathematical formulations to reconstruct tip–sample interaction forces from dynamic AFM measurements, several fundamental limitations remain. For example, all current approaches are intrinsically sensitive to the oscillation amplitude regime. Small-amplitude terms often rely on linearized models, which become invalid at short-range interactions due to the increasing influence of higher-order force terms. Moreover, most reconstruction techniques assume specific analytical forms or decay characteristics of the interaction force, which may not hold in realistic or heterogeneous systems. As a result, accuracy degrades for force fields that deviate from idealized behavior. Furthermore, many formulations require either direct differentiation or curve fitting of noisy experimental data, which amplifies noise and introduces artifacts in the recovered force curves. Another common limitation is the constancy of oscillation amplitude during spectroscopy. However, in practice, the amplitude can vary with tip–sample separation due to feedback dynamics or nonlinear interactions, violating the core assumptions of several models and/or contributing to reconstruction errors. Several of the more recent and powerful formulations, such as those based on Chebyshev decompositions or Bessel series expansions, mitigate some of these issues, but they may introduce significant computational overhead. The accuracy of these methods increases with series or matrix order, but so does processing time and numerical instability, especially in low signal-to-noise scenarios.

A broader issue affecting all methodologies is the lack of standardization. No single technique offers universally robust performance across interaction regimes or environmental conditions. Moreover, the absence of a standardized framework to benchmark reconstruction accuracy hinders cross-comparison and method validation. Finally, while most models focus on conservative interactions, the accurate reconstruction of dissipative forces remains limited, particularly under ambient or liquid conditions where energy loss mechanisms are complex and spatially heterogeneous.

To address these limitations, future developments may prioritize the creation of hybrid reconstruction frameworks that adaptively combine small- and large-amplitude formulations depending on the local interaction profile. Machine learning models trained on synthetic datasets with known force fields may offer a route to noise-tolerant and model-agnostic force reconstructions. Further, there is an urgent need for standardized benchmark datasets and validation protocols to enable direct performance comparisons across mathematical approaches. Finally, incorporating multichannel measurements—such as simultaneous amplitude, phase, and frequency shift—may enhance the separation of conservative and dissipative forces, improving both spatial resolution and physical interpretability.

## 4. Conclusions

The accurate reconstruction of tip–sample interaction forces is essential for unlocking the full potential of dynamic atomic force microscopy (AFM) in nanoscience, materials, and molecular biology. Despite the diversity of existing reconstruction methods, their widespread adoption has been limited by inconsistent assumptions, ill-posedness, and the absence of standardized validation. In this study, we systematically dissected the mathematical foundations of widely used FM- and AM-based reconstruction techniques and identified key regimes where each method succeeds or fails. To address longstanding challenges in reproducibility and comparability, we developed an open-source software package that enables direct benchmarking of reconstruction methods using unified experimental inputs and diagnostics. This platform represents a foundational step toward standardizing dynamic AFM force spectroscopy and provides a common reference point for future methodological innovation. By enabling transparent, reproducible force reconstructions, our work paves the way for more robust and interpretable nanoscale measurements across disciplines.

**Methods**

*Mathematical Methods*

To reconstruct the tip-sample interaction force from experimental observables, we implemented different formulations for amplitude and frequency with a linear interpolation scheme for high-resolution integration. All analyses were performed within a custom MATLAB App Designer interface that communicates with a linked Simulink model (see supplementary software).

Raw amplitude, phase, and frequency shift data were first filtered to remove zero-valued and duplicate entries, which otherwise introduce numerical instability during interpolation and differentiation. The tip-sample distance was estimated as the difference between the piezo displacement and the measured amplitude. The amplitude, phase, and frequency shift values were interpolated using MATLAB's fitting function, providing a smooth and continuous representation of these parameters over the interaction range.

For both modulation techniques, the effective interaction stiffness, $K(z)$, was derived from experimental observables and the properties of the probe, where applicable, and was obtained by numerically integrating the corresponding potential energy gradient. All integrals were evaluated using MATLAB's integral function with absolute and relative tolerances set to $10^{-15}$, ensuring high numerical precision. To prevent divergence due to singularities near the turning point, the algorithm excludes duplicate or nearly identical values in the distance vector and confines integration to a user-defined interaction range.

The reconstructed force was then computed by finite differencing the potential energy values, with the total force comprising three contributions from distinct analytical forms derived from each method's theoretical foundation. The final force-distance curve was plotted and annotated with custom legends. Computation progress was monitored using real-time elapsed time estimation and gauge updating, with all results rendered dynamically in the App Designer interface. The utilization of the software was explained with a supplemental video. All source data for force calculations and the corresponding software interface are available as supplementary material.

**Acknowledgements**

This work was supported by Le Fonds de Recherche du Québec - Nature et Technologies, the Canada Economic Development Fund, and the Natural Sciences and Engineering Research Council of Canada. O.E.D. also gratefully acknowledges funds provided by the École de technologie supérieure, University of Quebec. We would like to thank Xavier Daxhelet for his help with classical mechanics and Amir Payam for his discussions.

**Author Contributions**

O.E.D. conceived the project. S.L. analyzed the data. O.E.D. and S.L. wrote the manuscript and prepared figures. All authors participated in the analysis and interpretation of the data and commented on the manuscript.

**Data and Code Availability**

Any relevant data are available from the authors upon reasonable request. The suite of MATLAB codes developed for this study can be accessed from the authors, and the report of its use should cite this work.

**Competing interests**

The authors declare no competing interests.

**References**

1. Binnig, G., Quate, C. F. & Gerber, C. Atomic Force Microscope. *Physical Review Letters* **56**, 930-933, (1986).
2. Wiesendanger, R. *Scanning Probe Microscopy and Spectroscopy: Methods and Applications*. Cambridge University Press, 1994.
3. Garcia, R. *Amplitude Modulation Atomic Force Microscopy*. Wiley-VCH, 2010.
4. Giessibl, F. J. Atomic Resolution of the Silicon (111)-(7x7) Surface by Atomic Force Microscopy. *Science* **267**, 68-71, (1995).
5. Gross, L., Mohn, F., Moll, N., Liljeroth, P. & Meyer, G. The Chemical Structure of a Molecule Resolved by Atomic Force Microscopy. *Science* **325**, 1110-1114, (2009).
6. Mönig, H. *et al.* Submolecular Imaging by Noncontact Atomic Force Microscopy with an Oxygen Atom Rigidly Connected to a Metallic Probe. *ACS Nano* **10**, 1201-1209, (2016).
7. Kawai, S. *et al.* Van der Waals interactions and the limits of isolated atom models at interfaces. *Nature Communications* **7**, 11559, (2016).
8. Albrecht, T. R., Grutter, P., Horne, D. & Rugar, D. Frequency modulation detection using high-Q cantilevers for enhanced force microscope sensitivity. *Journal of Applied Physics* **69**, 668–673 (1991).

9    Giessibl, F. J. Forces and frequency shifts in atomic-resolution dynamic-force microscopy. *Physical Review B* **56**, 16010-16015, (1997).

10   Giessibl, F. J., Bielefeldt, H., Hembacher, S. & Mannhart, J. Calculation of the optimal imaging parameters for frequency modulation atomic force microscopy. *Applied Surface Science* **140**, 352-357, (1999).

11   Hölscher, H., Schwarz, U. D. & Wiesendanger, R. Calculation of the frequency shift in dynamic force microscopy. *Applied Surface Science* **140**, 344-351, (1999).

12   Hölscher, H., Allers, W., Schwarz, U. D., Schwarz, A. & Wiesendanger, R. Determination of Tip-Sample Interaction Potentials by Dynamic Force Spectroscopy. *Physical Review Letters* **83**, 4780-4783, (1999).

13   Dürig, U. Relations between interaction force and frequency shift in large-amplitude dynamic force microscopy. *Applied Physics Letters* **75**, 433-435, (1999).

14   Giessibl, F. J. & Bielefeldt, H. Physical interpretation of frequency-modulation atomic force microscopy. *Physical Review B* **61**, 9968-9971, (2000).

15   Dürig, U. Interaction sensing in dynamic force microscopy. *New Journal of Physics* **2**, 5, (2000).

16   Hölscher, H., Schwarz, A., Allers, W., Schwarz, U. D. & Wiesendanger, R. Quantitative analysis of dynamic-force-spectroscopy data on graphite(0001) in the contact and noncontact regimes. *Physical Review B* **61**, 12678-12681, (2000).

17   Giessibl, F. J. A direct method to calculate tip–sample forces from frequency shifts in frequency-modulation atomic force microscopy. *Applied Physics Letters* **78**, 123-125, (2001).

18   Hölscher, H. *et al.* Measurement of conservative and dissipative tip-sample interaction forces with a dynamic force microscope using the frequency modulation technique. *Physical Review B* **64**, 075402, (2001).

19   Sader, J. E. & Jarvis, S. P. Accurate formulas for interaction force and energy in frequency modulation force spectroscopy. *Applied Physics Letters* **84**, 1801-1803, (2004).

20   Sader, J. E. & Jarvis, S. P. Interpretation of frequency modulation atomic force microscopy in terms of fractional calculus. *Physical Review B* **70**, 012303, (2004).

21   Sader, J.E. *et al.* Quantitative force measurements using frequency modulation atomic force microscopy—theoretical foundations. *Nanotechnology* **16**, 94, (2005).

22   Lee, M. & Jhe, W. General Theory of Amplitude-Modulation Atomic Force Microscopy. *Physical Review Letters* **97**, 036104, (2006).

23   Shuiqing, H. & Arvind, R. Inverting amplitude and phase to reconstruct tip–sample interaction forces in tapping mode atomic force microscopy. *Nanotechnology* **19**, 375704, (2008).

24   J. Katan, A., H. van Es, M. & H. Oosterkamp, T. Quantitative force versus distance measurements in amplitude modulation AFM: a novel force inversion technique. *Nanotechnology* **20**, 165703, (2009).

25   Payam, A., Daniel, M.-J. & Ricardo, G. Force reconstruction from tapping mode force microscopy experiments. *Nanotechnology* **26**, 185706, (2015).

26   Dagdeviren, O. E., Götzen, J., Hölscher, H., Altman, E. I. & Schwarz, U. D. Robust high-resolution imaging and quantitative force measurement with tuned-oscillator atomic force microscopy. *Nanotechnology* **27**, 065703, (2016).

27  Dagdeviren, O. E. Confronting interatomic force measurements. *Review of Scientific Instruments* **92**, 063703, (2021).

28  Sader, J. E., Hughes, B. D., Huber, F. & Giessibl, F. J. Interatomic force laws that evade dynamic measurement. *Nature Nanotechnology* **13**, 1088-1091, (2018).

29  Dagdeviren, O. E., Zhou, C., Altman, E. I. & Schwarz, U. D. Quantifying Tip-Sample Interactions in Vacuum Using Cantilever-Based Sensors: An Analysis. *Physical Review Applied* **9**, 044040, (2018).

30  Dagdeviren, O. E. & Schwarz, U. D. Accuracy of tip-sample interaction measurements using dynamic atomic force microscopy techniques: Dependence on oscillation amplitude, interaction strength, and tip-sample distance. *Review of Scientific Instruments* **90**, 033707, (2019).

31  Dagdeviren, O. E., Miyahara, Y., Mascaro, A., Enright, T. & Grütter, P. Amplitude Dependence of Resonance Frequency and its Consequences for Scanning Probe Microscopy. *Sensors* **19**, 4510, (2019).

32  Gisbert, V. G. & Garcia, R. Insights and guidelines to interpret forces and deformations at the nanoscale by using a tapping mode AFM simulator: dForce 2.0. *Soft Matter* **19**, 5857-5868, (2023).

33  Weymouth, A. J., Hofmann, T. & Giessibl, F. J. Quantifying Molecular Stiffness and Interaction with Lateral Force Microscopy. *Science* **343**, 1120, (2014).

34  Fukuma, T. & Garcia, R. Atomic- and Molecular-Resolution Mapping of Solid-Liquid Interfaces by 3D Atomic Force Microscopy. *ACS Nano* **12**, 11785-11797, (2018).

35  Martin-Jimenez, D., Chacon, E., Tarazona, P. & Garcia, R. Atomically resolved three-dimensional structures of electrolyte aqueous solutions near a solid surface. *Nature Communications* **7**, 12164, (2016).

36  Gross, L. *et al.* Atomic Force Microscopy for Molecular Structure Elucidation. *Angewandte Chemie International Edition* **57**, 3888-3908, (2018).

37  Gross, L. *et al.* Bond-Order Discrimination by Atomic Force Microscopy. *Science* **337**, 1326-1329, (2012).

38  Gross, L. Recent advances in submolecular resolution with scanning probe microscopy. *Nat Chem* **3**, 273-278, (2011).

39  Swart, I., Gross, L. & Liljeroth, P. Single-molecule chemistry and physics explored by low-temperature scanning probe microscopy. *Chemical Communications* **47**, 9011-9023, (2011).

40  Baykara, M. Z., Schwendemann, T. C., Altman, E. I. & Schwarz, U. D. Three-Dimensional Atomic Force Microscopy – Taking Surface Imaging to the Next Level. *Advanced Materials* **22**, 2838-2853, (2010).

41  Albers, B. J. *et al.* Three-dimensional imaging of short-range chemical forces with picometre resolution. *Nat Nano* **4**, 307-310, (2009).

42  Burnham, N. A., Colton, R. J. & Pollock, H. M. Interpretation of force curves in force microscopy. *Nanotechnology* **4**, 64, (1993).

43  Yongho, S. & Wonho, J. Atomic force microscopy and spectroscopy. *Reports on Progress in Physics* **71**, 016101, (2008).

44  Dokukin, M. E. & Sokolov, I. Quantitative Mapping of the Elastic Modulus of Soft Materials with HarmoniX and PeakForce QNM AFM Modes. *Langmuir* **28**, 16060-16071, (2012).

45 Giessibl, F. J. The qPlus sensor, a powerful core for the atomic force microscope. *Review of Scientific Instruments* **90**, 011101, (2019).

46 Martin, Y., Williams, C. C. & Wickramasinghe, H. K. Atomic force microscope–force mapping and profiling on a sub 100-Å scale. *Journal of Applied Physics* **61**, 4723-4729, (1987).

47 Erlandsson, R., McClelland, G. M., Mate, C. M. & Chiang, S. Atomic force microscopy using optical interferometry. *Journal of Vacuum Science & Technology A* **6**, 266-270, (1988).

48 Gotsmann, B., Anczykowski, B., Seidel, C. & Fuchs, H. Determination of tip–sample interaction forces from measured dynamic force spectroscopy curves. *Applied Surface Science* **140**, 314-319, (1999).

49 Sader, J. E. & Jarvis, S. P. Coupling of conservative and dissipative forces in frequency-modulation atomic force microscopy. *Physical Review B* **74**, 195424, (2006).

50 Rodríguez, T. R. & García, R. Theory of Q control in atomic force microscopy. *Applied Physics Letters* **82**, 4821-4823, (2003).

51 Hölscher, H., Ebeling, D. & Schwarz, U. D. Theory of Q-Controlled dynamic force microscopy in air. *Journal of Applied Physics* **99**, -, (2006).

52 Hölscher, H. & Schwarz, U. D. Theory of amplitude modulation atomic force microscopy with and without Q-Control. *International Journal of Non-Linear Mechanics* **42**, 608-625, (2007).

# Supplementary Information:
# Standardizing Force Reconstruction in Dynamic Atomic Force Microscopy

Simon Laflamme[1], Bugrahan Guner[1], Omur E. Dagdeviren[1]*

[1]Department of Mechanical Engineering, École de technologie supérieure, University of Quebec, Canada, H3C 1K3

*Corresponding author's e-mail: omur.dagdeviren@etsmtl.ca

Table of contents of Supplementary Materials

## Nomenclature

All variables in the supplementary material are listed alphabetically with their fundamental units in parentheses, where applicable. The underscore denotes a vector quantity; the dot above the variable represents its time derivative.

**Greek Symbols**

| | |
|---|---|
| $\beta$ | Phase, |
| $\delta$ | Variational operator, |
| $\varepsilon$ | Inverse Fourier transform of E, |
| $\zeta$ | Normalized point of closest approach between the tip and the sample, |
| $\kappa$ | Deflection of the cantilever's tip or position of its equivalent mass (meter), |
| $\mu$ | Independent variable of $\nu$, |
| $\nu$ | Inverse Abel transform of function A, |
| $\xi$ | Normalized tip-sample distance, |
| $\upsilon$ | Canonical coordinates transformation function, |
| $\Xi$ | Fourier transform of $\varepsilon$, |
| $\Pi$ | Generalized damping coefficient (kilogram.second$^{-1}$), |
| $\Sigma$ | Kernel in the convolution integral, |
| $\Phi_r$ | Generalized force associated to $q_r$, |
| $\Psi$ | Abel transform of a function $\nu$, |
| $\psi$ | Orbital of the tip's motion (meter), |
| $\psi_0$ | Tip-sample distance of the tip apex for an undeflected cantilever (meter), |
| $\Omega$ | Normalized frequency shift. |

**Other Variables**

| | |
|---|---|
| $a_d$ | Amplitude of the of the driving force (meter), |
| $a_n$ | Orbital function's coefficients (meter), |
| $A$ | Cantilever's tip oscillating amplitude under the influence of the tip-sample force (meter), |
| $A_0$ | Cantilever's tip oscillating amplitude far from the sample (meter), |
| $A_{\text{ratio}}$ | Normalized amplitude of the cantilever's motion, |
| $b$ | Damping coefficient (kilogram.second$^{-1}$), |

| | | |
|---|---|---|
| $C$ | Tip-sample force constant (Newton.meter$^n$), | |
| $C_i^{\text{eff}}$ | Spring constant of the perturbed cantilever due to the interaction force (Newton/meter), | |
| $C_n$ | Coefficients of Chebyshev polynomials of the first kind (Newton), | |
| $d$ | Tip-sample distance at the point where the cantilever is undeflected (meter), | |
| $D$ | Minimal distance between the tip and the sample (meter), | |
| $f$ | Frequency of the perturbed cantilever's tip motion (Hertz), | |
| $f_0$ | Frequency of the unperturbed harmonic oscillator, e.g., cantilever (Hertz), | |
| $f_d$ | Driving frequency (Hertz), | |
| $F_{\text{even}}$ | Even part of the tip-sample's interaction force (Newton), | |
| $F_{\text{int}}$ | Tip-sample interaction force (Newton), | |
| $F_{\text{int c}}$ | Tip-sample's conservative interaction force (Newton), | |
| $F_{\text{int nc}}$ | Tip-sample's non-conservative interaction force (Newton), | |
| $\underline{F_j}$ | Force vector acting on particle j (Newton), | |
| $F_{\text{odd}}$ | Odd part of the tip-sample's interaction force (Newton), | |
| $F_\downarrow$ | Tip-sample's interaction force when approaching the sample (Newton), | |
| $F_\uparrow$ | Tip-sample's interaction force when retracting from the sample (Newton), | |
| $G$ | Inverse Laplace transform of $F_{\text{int c}}$ (Newton.meter), | |
| $g$ | Gain factor (Newton) | |
| $H$ | Hamiltonian (joule), | |
| $H'$ | Hamiltonian following a canonical transformation (joule), | |
| $I_1$ | Modified Bessel function of the first kind of order one, | |
| $J$ | Action variable for the perturbed cantilever's tip motion (joule.second), | |
| $k$ | Spring constant (Newton.meter$^{-1}$), | |
| $\ell$ | Length scale of the tip-sample's interaction range (meter), | |
| $K$ | Kinetic energy (joule), | |
| $l$ | Independent variable of $\varepsilon$, | |
| $L$ | Lagrangian (joule), | |
| $L'$ | Lagrangian following a canonical transformation (joule), | |
| $M$ | Inverse Laplace transform of $B$ (Newton.second.meter$^{-1}$), | |
| $m$ | Cantilever's tip equivalent mass (kilogram), | |
| $m_j$ | Mass of particle j (kilogram), | |

| | |
|---|---|
| $P$ | Forward kernel function, |
| $p_r$ | Generalized momentum associated to $q_r$, |
| $P_r$ | Generalized momentum associated to $Q_r$, |
| $q_r$ | Generalized coordinate $r$ |
| $Q_0$ | Q factor without gain, |
| $Q_{eff}$ | Effective Q factor with gain, |
| $Q_r$ | Generalized coordinate $r$ following a transformation using generating function $\upsilon$, |
| $\underline{r_j}$ | Position vector of particle $j$ (meter), |
| $\delta \underline{r_j}$ | Virtual displacement of particle $j$ (meter), |
| $R(\hat{z})$ | Chebyshev polynomials of the first kind, |
| $s$ | Independent variable of E, |
| $t$ | Time (second), |
| $t_0$ | Time phase shifter (second), |
| $T_0$ | Period of oscillation of the cantilever's tip unperturbed motion (second), |
| $T$ | Period of oscillation of the perturbed cantilever's tip motion (second), |
| $U_{int}$ | Interaction potential energy of the cantilever's tip (joule), |
| $v_j$ | Velocity of particle j (meter.second$^{-1}$), |
| $V$ | Potential energy (joule), |
| $W$ | Work (joule), |
| $x$ | Independent variable of A, |
| $z$ | Position of the cantilever's tip (meter), |
| $z_0$ | Position of the force's step function discontinuity (meter). |

# Section 1 Review of classical mechanics

## 1.1 Hamilton's principle

In this part, an overview of classical mechanics is presented to explain the relation of the resonance frequency shift with tip-sample interaction force by employing mathematical methods based on Hamilton's principle.[1-4] $\delta$ is a variational operator while $d/dx$ is the differential one.[5,6]

A system of $n$ particles of mass $m_j$, located using the vector $r_j$ ($j = 1, \ldots, n$) has $m$ degrees of freedom ($m \leq 3n$) and thus, $q_r$ coordinates. If the system is formed by $n$ particles, the sum of all forces must be equal to zero to have a system in dynamic equilibrium. To this end, D'Alembert's principle states that:[1,5,6]

$$\sum_{j=1}^{n}\left(\underline{F}_j - m_j \underline{\ddot{r}}_j\right) \cdot \delta \underline{r}_j = 0 \qquad 1.1$$

The virtual work $\delta W$ is:

$$\delta W = \sum_{j=1}^{n} \underline{F}_j \cdot \delta \underline{r}_j \qquad 1.2$$

where $\delta \underline{r}_j$ is the virtual displacement of particle $j$.

Since we have:

$$\frac{d}{dt}\left(\underline{\dot{r}}_j \cdot \delta \underline{r}_j\right) = \underline{\dot{r}}_j \cdot \frac{d}{dt}\left[\delta \underline{r}_j\right] + \underline{\ddot{r}}_j \cdot \delta \underline{r}_j = \underline{\dot{r}}_j \cdot \delta \underline{\dot{r}}_j + \underline{\ddot{r}}_j \cdot \delta \underline{r}_j \qquad 1.3$$

$$\frac{d}{dt}\left(\underline{\dot{r}}_j \cdot \delta \underline{r}_j\right) = \frac{1}{2}\delta\left(\underline{\dot{r}}_j\right)^2 + \underline{\ddot{r}}_j \cdot \delta \underline{r}_j = \delta \frac{1}{2} v_j^2 + \underline{\ddot{r}}_j \cdot \delta \underline{r}_j \qquad 1.4$$

$$\underline{\ddot{r}}_j \cdot \delta \underline{r}_j = \frac{d}{dt}\left(\underline{\dot{r}}_j \cdot \delta \underline{r}_j\right) - \delta \frac{1}{2} v_j^2 \qquad 1.5$$

$$\delta W = \sum_{j=1}^{n} m_j \underline{\ddot{r}}_j \cdot \delta \underline{r}_j = \sum_{j=1}^{n} m_j \frac{d}{dt}\left[\underline{\dot{r}}_j \cdot \delta \underline{r}_j\right] - \delta \sum_{j=1}^{n}\left(\frac{1}{2} m_j v_j^2\right) \qquad 1.6$$

We get:

$$\sum_{j=1}^{n} m_j \frac{d}{dt}\left[\underline{\dot{r}}_j \cdot \delta \underline{r}_j\right] = \delta W + \delta K \qquad 1.7$$

Integrating over time from $t_0$ to $t_1$:

$$\int_{t_0}^{t_1}\sum_{j=1}^{n} m_j \frac{d}{dt}\left[\underline{\dot{r}}_j \cdot \delta \underline{r}_j\right] dt = \sum_{j=1}^{n} m_j \left[\underline{\dot{r}}_j \cdot \delta \underline{r}_j\right]_{t_0}^{t1} = \int_{t_0}^{t_1}(\delta W + \delta K) dt \qquad 1.8$$

According to the variational principles, the variations $\delta \underline{r_i}$ at the boundaries $t_0$ and $t_1$ are zero.[5,6] Thus, the integration over time is equal to:

$$\delta \int_{t_0}^{t_1} (W + K) dt = 0 \qquad 1.9$$

Hamilton's principle is obtained in the case of a force derived from a potential:

$$\delta \int_{t_0}^{t_1} (K - V) dt = \delta \int_{t_0}^{t_1} L\, dt = 0 \qquad 1.10$$

where $L = K\text{-}V$ is the Lagrangian.

## 1.2 Euler-Lagrange equation

The Lagrangian can be expressed as:[1,5,6]

$$L = L\left(q_r, \dot{q}_r, t\right) \qquad 1.11$$

The generalized momentum quantity associated to $q_r$ is defined as:

$$p_r = \frac{\partial L}{\partial \dot{q}_r} \qquad 1.12$$

In a conservative system, $V$ is independent of $\dot{q}_r$, and results in following equations:

$$p_r = \frac{\partial L}{\partial \dot{q}_r} = \frac{\partial K}{\partial \dot{q}_r} \qquad 1.13$$

$$\underline{r}_j = \underline{r}_j(q_1, \ldots, q_m) \qquad 1.14$$

$$\underline{\dot{r}}_j = \sum_{r=1}^{m} \frac{\partial \underline{r}_j}{\partial q_r} \dot{q}_r \qquad 1.15$$

$$K\left(q_1, \ldots, q_m, \dot{q}_1, \ldots, \dot{q}_m\right) = \frac{1}{2}\sum_{j=1}^{n} m_j \left(\underline{\dot{r}}_j\right)^2 = \frac{1}{2}\sum_{j=1}^{n} m_j \sum_{p=1}^{m}\sum_{s=1}^{m} \left\{ \frac{\partial \underline{r}_j}{\partial q_p} \frac{\partial \underline{r}_j}{\partial q_s} \dot{q}_p \dot{q}_s \right\} \qquad 1.16$$

$$K\left(q_1, \ldots, q_m, \dot{q}_1, \ldots, \dot{q}_m\right) = \frac{1}{2}\sum_{p=1}^{m}\sum_{s=1}^{m} \left\{ \sum_{j=1}^{n} m_j \frac{\partial \underline{r}_j}{\partial q_p} \frac{\partial \underline{r}_j}{\partial q_s} \right\} \dot{q}_p \dot{q}_s = \frac{1}{2}\sum_{p=1}^{m}\sum_{s=1}^{m} a_{ps} \dot{q}_p \dot{q}_s \qquad 1.17$$

where $a_{ps}$ is independent of velocities and consequently:

$$p_r = \frac{\partial K}{\partial \dot{q}_r} = \sum_{j=1}^{m} a_{jr} \dot{q}_j \qquad 1.18$$

for $r = 1, \ldots, m$, using matrices, we can find out the values of $\dot{q}_j$ according to $p_r$:

$$\begin{pmatrix} p_1 \\ \vdots \\ p_m \end{pmatrix} = \begin{pmatrix} a_{11} & \cdots & a_{m1} \\ \vdots & \ddots & \vdots \\ a_{1m} & \cdots & a_{mn} \end{pmatrix} \begin{pmatrix} \dot{q}_1 \\ \vdots \\ \dot{q}_m \end{pmatrix} \qquad 1.19$$

$$\begin{pmatrix} \dot{q}_1 \\ \vdots \\ \dot{q}_m \end{pmatrix} = \begin{pmatrix} a_{11} & \cdots & a_{m1} \\ \vdots & \ddots & \vdots \\ a_{1m} & \cdots & a_{mn} \end{pmatrix}^{-1} \begin{pmatrix} p_1 \\ \vdots \\ p_m \end{pmatrix} \qquad 1.20$$

Now, by using the variational operator on K:

$$\delta K\left(q_1,...,q_m,\dot{q}_1,...,\dot{q}_m\right) = \sum_{r=1}^{m}\left(\frac{\partial K}{\partial q_r}\delta q_r + \frac{\partial K}{\partial \dot{q}_r}\delta \dot{q}_r\right) \qquad 1.21$$

Employing Equation 1.2:

$$\delta W = \sum_{j=1}^{n} \underline{F}_j \cdot \delta \underline{r}_j$$

and

$$\delta \underline{r}_j = \sum_{r=1}^{m} \frac{\partial \underline{r}_j}{\partial q_r}\delta q_r \qquad 1.22$$

Combining Equations 1.2 and 1.22:

$$\delta W = \sum_{r=1}^{m}\left\{\sum_{j=1}^{n}\underline{F}_j \cdot \frac{\partial \underline{r}_j}{\partial q_r}\right\}\delta q_r = \sum_{r=1}^{m}\Phi_r \delta q_r \qquad 1.23$$

where

$$\Phi_r = \left\{\sum_{j=1}^{n}\underline{F}_j \cdot \frac{\partial \underline{r}_j}{\partial q_r}\right\} \qquad 1.24$$

$\Phi_r$ is the generalized force associated to $q_r$. Equations 1.21 and 1.24 are used with Hamilton's principle to obtain:

$$\int_{t_0}^{t_1}\delta L dt = 0 = \int_{t_0}^{t_1}\delta(K+W)dt = \int_{t_0}^{t_1}\sum_{r=1}^{m}\left\{\frac{\partial K}{\partial q_r}\delta q_r + \frac{\partial K}{\partial \dot{q}_r}\delta \dot{q}_r + \Phi_r \delta q_r\right\}dt \qquad 1.25$$

Integration by parts, and since the variations $\delta q_r$ are zero at the boundaries $t_0$ and $t_1$:

$$\int_{t_0}^{t_1}\frac{\partial K}{\partial \dot{q}_r}\delta \dot{q}_r dt = \left[\frac{\partial K}{\partial \dot{q}_r}\delta q_r\right]_{t_0}^{t_1} - \int_{t_0}^{t_1}\frac{d}{dt}\left[\frac{\partial K}{\partial \dot{q}_r}\right]\delta q_r dt = -\int_{t_0}^{t_1}\frac{d}{dt}\left[\frac{\partial K}{\partial \dot{q}_r}\right]\delta q_r dt \qquad 1.26$$

Finally, with Equations 1.25 and 1.26, it is obtained:

$$\int_{t_0}^{t_1}\sum_{r=1}^{m}\left\{\frac{\partial K}{\partial q_r} - \frac{d}{dt}\left[\frac{\partial K}{\partial \dot{q}_r}\right] + \Phi_r\right\}\delta q_r dt = 0 \qquad 1.27$$

Thus, the quantity in the curly bracket must be equal to zero.

$$\left\{ \frac{\partial K}{\partial q_r} - \frac{d}{dt}\left[\frac{\partial K}{\partial \dot{q}_r}\right] + \Phi_r \right\} = 0 \qquad 1.28$$

where $r = 1, \ldots, m$.

If forces are derived from a potential:

$$V = V(q_1, \ldots, q_m) \qquad 1.29$$

by employing Equation 1.23:

$$\delta W = -\delta V = -\sum_{r=1}^{m} \frac{\partial V}{\partial q_r} \delta q_r = \sum_{r=1}^{m} \Phi_r \delta q_r \qquad 1.30$$

$$\Phi_r = -\frac{\partial V}{\partial q_r} \qquad 1.31$$

Since $\partial V/\partial \dot{q}_r = 0$ for forces deriving from a potential and using Equation 1.31, with Equation 1.28, we have:

$$\left\{ \frac{d}{dt}\left[\frac{\partial L}{\partial \dot{q}_r}\right] - \frac{\partial L}{\partial q_r} \right\} = 0, r = 1, \ldots, m \qquad 1.32$$

Using Equation 1.12, finally, the following is obtained:

$$\dot{p}_r = \frac{\partial L}{\partial q_r} \qquad 1.33$$

## 1.3 Hamilton's equations

The Hamiltonian is defined as:

$$H\left(p_r, q_r, \dot{q}_r, t\right) = \sum_{r=1}^{m}\left[p_r \dot{q}_r\right] - L\left(q_r, \dot{q}_r, t\right) \qquad 1.34$$

Then,

$$dH = \sum_{r=1}^{m} p_r d\dot{q}_j + \sum_{r=1}^{m} \dot{q}_r dp_r - \sum_{r=1}^{m} \frac{\partial L}{\partial q_r} dq_r - \sum_{r=1}^{m} \frac{\partial L}{\partial \dot{q}_r} d\dot{q}_r - \frac{\partial L}{\partial t} dt \qquad 1.35$$

With Equations 1.12 and 1.33:

$$dH = \sum_{r=1}^{m} p_r d\dot{q}_j + \sum_{r=1}^{m} \dot{q}_r dp_r - \sum_{r=1}^{m} \dot{p}_r dq_r - \sum_{r=1}^{m} p_r d\dot{q}_r - \frac{\partial L}{\partial t} dt \qquad 1.36$$

$$dH = \sum_{r=1}^{m} \dot{q}_r dp_r - \sum_{r=1}^{m} \dot{p}_r dq_r - \frac{\partial L}{\partial t} dt \qquad 1.37$$

Since $\dot{q}_r$ can be expressed as function of the $p_r$ according to Equation 1.20:

$$H\left(p_r, q_r, \dot{q}_r, t\right) = H\left(p_r, q_r, t\right) \qquad 1.38$$

$$dH = \sum_{r=1}^{m} \frac{\partial H}{\partial p_r} dp_r + \sum_{r=1}^{m} \frac{\partial H}{\partial q_r} dq_r + \frac{\partial H}{\partial t} dt \qquad 1.39$$

From Equations 1.37 and 1.39, Hamilton's equations are obtained:

$$\dot{q}_r = \frac{\partial H\left(p_r, q_r, t\right)}{\partial p_r} \qquad 1.40$$

$$\dot{p}_r = -\frac{\partial H\left(p_r, q_r, t\right)}{\partial q_r} \qquad 1.41$$

$$\frac{\partial L}{\partial t} = -\frac{\partial H\left(p_r, q_r, t\right)}{\partial t} \qquad 1.42$$

According to Equation 1.40, it is concluded that if a coordinate $p_r$ does not appear in H, then $q_r$ is a constant.

Conversely, from Equation 1.41, if a coordinate $q_r$ is not included explicitly in H, then $p_r$ is a constant.

## 1.4 Canonical transformation equations

A change of coordinate system from $(p_r, q_r)$ to $(P_r, Q_r)$ is said to be canonical, and the following equations are obtained:[1,5,6]

$$\dot{Q}_r = \frac{\partial H'(P_r, Q_r, t)}{\partial P_r} \qquad 1.43$$

$$\dot{P}_r = -\frac{\partial H'(P_r, Q_r, t)}{\partial Q_r}, \qquad 1.44$$

$$\frac{\partial L}{\partial t} = -\frac{\partial H'(P_r, Q_r, t)}{\partial t} \qquad 1.45$$

This implies Hamilton's equations are still valid.

Since the two coordinate systems represents the same physical entity, Hamilton's principle must be respected for both. To this end, with Equation 1.10:

$$\delta \int_{t_0}^{t_1} (L - L') dt = 0 \qquad 1.46$$

We pose $L - L' = dv/dt$, where $v(q_1, \ldots, q_m, Q_1, \ldots, Q_m, t)$ is an arbitrary function. Starting with Equation 1.34:

$$H\left(p_r, q_r, \dot{q}_r, t\right) = \sum_{r=1}^{m} \left[p_r \dot{q}_r\right] - L\left(q_r, \dot{q}_r, t\right)$$

Equation 1.34 is applied to $H'$:

$$H'\left(P_r, Q_r, \dot{Q}_r, t\right) = \sum_{r=1}^{m} \left[P_r \dot{Q}_r\right] - L'\left(Q_r, \dot{Q}_r, t\right) \qquad 1.47$$

Subtracting Equation 1.34 and Equation 1.47 and using $L - L' = dv/dt$, the following equation is obtained:

$$\frac{dv}{dt} = \sum_{r=1}^{m} p_r \dot{q}_r - \sum_{r=1}^{m} P_r \dot{Q}_r + (H' - H) \qquad 1.48$$

Since $v$ is a function of $(q_1, \ldots, q_m, Q_1, \ldots, Q_m, t)$:

$$dv = \sum_{r=1}^{m} p_r dq_r - \sum_{r=1}^{m} P_r dQ_r + (H' - H) dt = \sum_{r=1}^{m} \frac{\partial v}{\partial q_r} dq_r + \sum_{r=1}^{m} \frac{\partial v}{\partial Q_r} dQ_r + \frac{\partial v}{\partial t} dt \qquad 1.49$$

Thus, the following equations are obtained:

$$P_r = -\frac{\partial v}{\partial Q_r} \qquad 1.50$$

$$p_r = \frac{\partial \upsilon}{\partial q_r} \qquad 1.51$$

$$H' - H = \frac{\partial \upsilon}{\partial t} \qquad 1.52$$

If a transformation such that $H' = 0$ is chosen, then from Equations 1.43 and all $P_r$ and $Q_r$ are constant over time in the new coordinate system. For this specific case, the Hamilton-Jacobi Equation is obtained:

$$H' - H = -H(p_r, q_r, t) = -H\left(\frac{\partial \upsilon}{\partial q_r}, q_r, t\right) = \frac{\partial \upsilon}{\partial t} \qquad 1.53$$

$$H\left(\frac{\partial \upsilon}{\partial q_r}, q_r, t\right) + \frac{\partial \upsilon}{\partial t} = 0 \qquad 1.54$$

## 1.5 Derivation of the harmonic oscillator's equation:

In this part, the Hamilton-Jacobi Equation is employed to find the harmonic oscillator's equation of motion.

$$H(\dot{\kappa},\kappa) = \frac{1}{2}m\dot{\kappa}^2 + \frac{1}{2}k\kappa^2 \qquad 1.55$$

$$p_\kappa = \frac{\partial K}{\partial \dot{\kappa}} = m\dot{\kappa} \qquad 1.56$$

$$H(p_\kappa,\kappa) = \frac{1}{2}\frac{p_\kappa^2}{m} + \frac{1}{2}k\kappa^2 \qquad 1.57$$

A canonical transformation function $\upsilon(\kappa,Q,t)$ allows to obtain:

$$Q = Q(\kappa, p_\kappa, t) \qquad 1.58$$

$$P = P(\kappa, p_\kappa, t) \qquad 1.59$$

Equation 1.54 can be written as:

$$H\left(\frac{\partial \upsilon}{\partial \kappa}, \kappa\right) + \frac{\partial \upsilon}{\partial t} = 0 \qquad 1.60$$

Equation 1.57 with Equation 1.51 yields:

$$H\left(\frac{\partial \upsilon}{\partial \kappa}, \kappa\right) = \frac{1}{2m}\left(\frac{\partial \upsilon}{\partial \kappa}\right)^2 + \frac{1}{2}k\kappa^2 \qquad 1.61$$

The type of solution should be in the following form to be able to separate variables:

$$\upsilon = \upsilon_1(\kappa) + \upsilon_2(t) \qquad 1.62$$

$$\frac{\partial \upsilon}{\partial \kappa} = \frac{d\upsilon_1}{d\kappa} \qquad 1.63$$

$$\frac{\partial \upsilon}{\partial t} = \frac{d\upsilon_2}{dt} \qquad 1.64$$

Using Equations 1.60 and 1.61:

$$\frac{1}{2m}\left(\frac{\partial \upsilon}{\partial \kappa}\right)^2 + \frac{1}{2}k\kappa^2 = -\frac{\partial \upsilon}{\partial t} \qquad 1.65$$

Both sides of the equation are equated to an arbitrary constant.

$$\frac{1}{2m}\left(\frac{d\upsilon_1}{d\kappa}\right)^2 + \frac{1}{2}k\kappa^2 = -\frac{d\upsilon_2}{dt} = Jf_0 \qquad 1.66$$

where $J$ is an arbitrary constant.

Since:

$$\frac{\partial \upsilon}{\partial t} = -Jf_0 \qquad 1.67$$

We can conclude from Equation 1.52 that $J.f_0$ has the units of energy. Thus, $J$ is an action variable with joule.second units.

Solving for $\upsilon_1(\kappa)$:

$$\upsilon_1(\kappa) = \int \sqrt{2m\left(Jf_0 - \frac{1}{2}k\kappa^2\right)}\,d\kappa \qquad 1.68$$

Then, solving for $\upsilon_2(t)$:

$$\upsilon_2(t) = -Jf_0 t \qquad 1.69$$

Combining Equations 1.68 and 1.69 with Equation 1.62:

$$\upsilon = \int \sqrt{2m\left(Jf_0 - \frac{1}{2}k\kappa^2\right)}\,d\kappa - Jf_0 t \qquad 1.70$$

The value of $Q$ is arbitrarily chosen in the new coordinate system as:

$$Q = Jf_0 \qquad 1.71$$

The value of $P$ is found as:

$$P = -\frac{\partial \upsilon}{\partial Q} = -m\int \frac{d\kappa}{\sqrt{2m\left(Jf_0 - \frac{1}{2}k\kappa^2\right)}} + t \qquad 1.72$$

In this new coordinate system, $P$ must be a constant with units in seconds.

We equal $P$ to $-\beta$ divided by $f_0$. Consequently, $\beta$ is unitless. It corresponds to the movement's initial phase.

$$P = -\frac{\beta}{f_0} = -m\int \frac{d\kappa}{\sqrt{2m\left(Jf_0 - \frac{1}{2}k\kappa^2\right)}} + t \qquad 1.73$$

$$\sqrt{\frac{m}{k}} \sin^{-1}\left(\kappa\sqrt{\frac{k}{2Jf_0}}\right) = t + \frac{\beta}{f_0} \qquad 1.74$$

$$\kappa(t) = \sqrt{\frac{2Jf_0}{k}} \sin\left(\sqrt{\frac{k}{m}}\left(t + \frac{\beta}{f_0}\right)\right) = \sqrt{\frac{2Jf_0}{k}} \sin(2\pi(f_0 t + \beta)) \qquad 1.75$$

From the Hamilton equations in the new coordinate system (i.e., Equations 1.43-1.45) and with Equations 1.73 and 1.71, the following is obtained:

$$\dot{Q} = f_0 \frac{dJ}{dt} = \frac{\partial H'}{\partial P} = -f_0 \frac{\partial H'}{\partial \beta} \qquad 1.76$$

$$\dot{P} = -\frac{1}{f_0} \frac{d\beta}{dt} = -\frac{\partial H'}{\partial Q} = -\frac{1}{f_0} \frac{\partial H'}{\partial J} \qquad 1.77$$

## 1.6 Introduction of a perturbation

Suppose that the Hamiltonian $H$ in the original coordinate system is perturbed by a quantity $\Delta H$.

$$H' = H + \Delta H + \frac{\partial \upsilon}{\partial t} \qquad 1.78$$

If we use the same transformation function $\upsilon$, we employ Equation 1.60:

$$H + \frac{\partial \upsilon}{\partial t} = 0$$

Then, from Equation 1.78:

$$H' = \Delta H \qquad 1.79$$

Now, since $H' \neq 0$, $J$, and are no longer constant but are varying slowly over time. For this reason, by using Equations 1.76 and 1.77:

$$\frac{dJ}{dt} = -\frac{\partial H'}{\partial \beta} = -\frac{\partial \Delta H}{\partial \beta} \qquad 1.80$$

$$\frac{d\beta}{dt} = \frac{\partial H'}{\partial J} = \frac{\partial \Delta H}{\partial J} \qquad 1.81$$

# Section 2 Mathematical approaches for the calculation of Δ*f* and tip-sample interaction force based on classical mechanics for frequency-modulation atomic force microscopy

Attempts to understand the nature of tip-sample interaction forces date back to the late '90s.[2-4] First mathematical proposals to reveal tip-sample interaction force from the measured resonance frequency shift data are based on classical mechanics.[2-4] In this section, we provide an overview of formulations based on classical mechanics in their chronological order.

## 2.1 Calculation of the Δ*f* using perturbation theory

Initial attempts to understand the tip-sample interaction forces in dynamic microscopy were via the calculation of the Δ*f* for a pre-determined force field. To this end, the Δ*f* was derived by using perturbation theory in 1997 by Giessibl.[2] In this formulation, *D* represents the minimum tip-sample distance, so $z(t) = \kappa(t) + D + A$. From Equation 1.75, the following equality is obtained:

$$\kappa(t) = \sqrt{\frac{2Jf_0}{k}} \sin(2\pi(f_0 t + \beta)) \qquad 2.1$$

$$p = m\frac{d\kappa}{dt} = \sqrt{\frac{kJ}{2\pi^2 f_0}} \cos(2\pi(f_0 t + \beta)) \qquad 2.2$$

Since, $\kappa = A\cos(2\pi f_0 t)$:

$$A = \sqrt{\frac{2Jf_0}{k}} \qquad 2.3$$

If the following perturbation is analyzed:

$$\Delta H = U_{int}(z) = \frac{-C}{(n-1)z^{n-1}} \quad \text{for } n>1 \qquad 2.4$$

$$f = f_0 + \Delta f \qquad 2.5$$

$$ft = f_0 t + \Delta ft = f_0 t + \beta \qquad 2.6$$

Employing Equation 1.81:

$$\Delta f = \left\langle \frac{d\beta}{dt} \right\rangle = \left\langle \frac{\partial \Delta H}{\partial J} \right\rangle = \left\langle \frac{\partial U_{int}}{\partial J} \right\rangle = \left\langle \frac{\partial U_{int}}{\partial z}\frac{\partial z}{\partial J} \right\rangle = -\frac{f_0}{kA^2}\langle F_{int}\kappa \rangle \qquad 2.7$$

With $\kappa(t) = A\cos(2\pi f_0 t)$:

$$\Delta f(D,k,A,f_0,n) = \frac{f_0}{kA^2}\frac{1}{T_0}\int_0^{T_0} \frac{CA\cos(2\pi f_0 t)}{\{D+A[\cos(2\pi f_0 t)+1]\}^n} dt \qquad 2.8$$

S18

Using $x = 2\pi f_0 t$:

$$\Delta f(D,k,A,f_0,n) = \frac{f_0 C}{kA 2\pi D^n} \int_0^{2\pi} \frac{\cos(x)}{\left\{1 + \frac{A}{D}[\cos(x)+1]\right\}^n} dx \qquad 2.9$$

Equation 2.9 is solved for both large and small values of $A$ compared to $D$.

If $A \ll D$, then:

$$\left\{1 + \frac{A}{D}[\cos(x)+1]\right\}^{-n} \approx 1 - n\frac{A}{D}[\cos(x)+1] \qquad 2.10$$

$$\Delta f(D,k,A,f_0,n) \approx \frac{-f_0 C n}{2kD^{n+1}} \qquad 2.11$$

If $A \gg D$, $\cos(x)$ can be approximated by its value at $\pi$. Hence, the force acting on the tip will only be significant at close distances:

Therefore, for $A \gg D$, we can approximate:

$$1 + \frac{A}{D}[\cos(x)+1] = 1 + (x-\pi)\left(-\frac{A}{D}\sin\pi\right) + \frac{(x-\pi)^2}{2}\left(-\frac{A}{D}\cos\pi\right) + \ldots$$
$$\approx 1 + (x-\pi)^2 \frac{A}{2D} \qquad 2.12$$

With the substitution $x' = x - \pi$, we get an approximation for $\Delta f$ which remains valid for $A \gg D$:

$$\Delta f(D,k,A,f_0,n) \approx -\frac{f_0 C}{2\pi kAD^n} \int_{-\pi}^{\pi} \frac{dx'}{\left(1 + (x')^2 \frac{A}{2D}\right)^n} \qquad 2.13$$

Posing $y^2 = (A/2D)x'^2$, the following equations are obtained:

$$\Delta f(D,k,A,f_0,n) \approx -\frac{f_0 C}{\sqrt{2}\pi kA^{3/2}D^{n-1/2}} \int_{-\pi\sqrt{\frac{A}{2D}}}^{\pi\sqrt{\frac{A}{2D}}} \frac{dy}{(1+y^2)^n}$$
$$\approx -\frac{f_0 C}{\sqrt{2}\pi kA^{3/2}D^{n-1/2}} \int_{-\infty}^{\infty} \frac{dy}{(1+y^2)^n} \qquad 2.14$$

$$\Delta f(D,k,A,f_0,n) \approx -\frac{f_0 C}{\sqrt{2}\pi kA^{3/2}D^{n-1/2}} I_1(n) \qquad 2.15$$

If the interaction force is the following:

S19

$$F_{int}(q) = -\sum_{n=1}^{\infty} \frac{C_n}{z^n} \qquad 2.16$$

Finally, the following equations are obtained:

$$\Delta f(D,k,A,f_0,n) = \frac{f_0}{kA2\pi D^n} \sum_{n=1}^{\infty} \int_0^{2\pi} C_n \frac{\cos(x)}{\left\{1+\frac{A}{D}[\cos(x)+1]\right\}^n} dx \qquad 2.17$$

Equation 2.17 can be reduced to a normalized $\Delta f$ (i.e., Equation 2.18, which relates experimental parameters to tip-sample interaction forces).

$$\Omega(D) \approx -\frac{1}{\sqrt{2\pi}} \sum_{n=1}^{\infty} \frac{C_n I_1(n)}{D^{n-\frac{1}{2}}} \qquad 2.18$$

## 2.2 Analytical calculation of the $\Delta f$ for tip-sample interaction forces with the approximation of inverse power laws

Equation 2.17 is valid if the oscillation of the cantilever is harmonic under the tip-sample interaction forces. Towards this end, Hölscher et al.[3] solved Equation 2.17 analytically for tip-sample interaction forces with the approximation of inverse power laws. In this formulation, $D$ is the minimum tip-sample distance, so that $D \approx d-A$ for $A << D$. Also, $x = 2\pi f_0 t$ and is unitless. The following is the tip-sample interaction force:

$$F_{int}(z) = -Cz^{-1} \qquad 2.19$$

Equation 2.17 becomes:

$$\Delta f(d,k,A,f_0,n=1) = \frac{f_0 C_1}{2\pi kA} \int_0^{2\pi} \frac{\cos(x)}{d + A\cos(x)} dx \qquad 2.20$$

To solve this equation, we need to use the tangent half-angle substitution (the so-called Weierstrass substitution):[7]

$$t = \tan\left(\frac{x}{2}\right), \cos(x) = \frac{1-t^2}{1+t^2}, \sin(x) = \frac{2t}{1+t^2}, dx = \frac{2}{1+t^2} dt, \qquad 2.21$$

$$\int_0^{2\pi} \frac{\cos(x)}{d + A[\cos(x)]} dx = 2\int_0^{\pi} \frac{\cos(x)}{d + A[\cos(x)]} dx \qquad 2.22$$

$$2\int \frac{\cos(x)}{d + A[\cos(x)]} dx = \frac{4}{d} \int \frac{(1-t^2)}{(1+t^2)^2 \left(1 + a\left(\frac{1-t^2}{1+t^2}\right)\right)} dt \qquad 2.23$$

where $a = A/d$.

$$\frac{4}{d} \int \frac{(1-t^2)}{(1+t^2)^2 \left(1 + a\left(\frac{1-t^2}{1+t^2}\right)\right)} dt = \frac{4}{d} \int \frac{(1-t^2)}{(1+t^2)(1+a+t^2(1-a))} dt \qquad 2.24$$

$$= \frac{4}{d} \int \frac{(1-t^2)}{(1+t^2)(1-a)\left(\frac{1+a}{1-a}+t^2\right)} dt$$

$$= \frac{4}{d} \int \frac{(1-t^2)}{(1+t^2)(1-a)\left[\left[\left(\frac{1+a}{1-a}\right)^{\frac{1}{2}}\right]^2 + t^2\right]} dt \qquad 2.25$$

We use a particular partial fraction decomposition:

$$= \frac{4}{d(1-a)} \int \left(\frac{1-a}{a}\right)\left(\frac{1}{1+t^2}\right) - \frac{1}{a\left[\left[\left(\frac{1+a}{1-a}\right)^{\frac{1}{2}}\right]^2 + t^2\right]} dt \qquad 2.26$$

$$2 \int_0^{\pi} \frac{\cos(x)}{d + A[\cos(x)]} dx = \frac{4}{d(1-a)}$$

$$\left[\left|\left(\frac{1-a}{a}\right)\tan^{-1}\left[\tan\left(\frac{x}{2}\right)\right]\right|_0^{\pi} - \left|\frac{1}{a}\left(\frac{1-a}{1+a}\right)^{\frac{1}{2}}\tan^{-1}\left[\left(\frac{1-a}{1+a}\right)^{\frac{1}{2}}\tan\left[\frac{x}{2}\right]\right]\right|_0^{\pi}\right] \qquad 2.27$$

$$= \frac{4}{d(1-a)}\left[\left(\frac{1-a}{a}\right)\frac{\pi}{2} - \frac{1}{a}\left(\frac{1-a}{1+a}\right)^{\frac{1}{2}}\left[\tan^{-1}(\infty) - \tan^{-1}(0)\right]\right] \qquad 2.28$$

$$= \frac{2\pi}{d(1-a)}\left[\left(\frac{1-a}{a}\right) - \frac{1}{a}\left(\frac{1-a}{1+a}\right)^{\frac{1}{2}}\right] = \frac{2\pi}{A}\left(1 - \frac{d}{(d^2-A^2)^{\frac{1}{2}}}\right) \qquad 2.29$$

$$\Delta f(d,k,A,f_0,n=1) = \frac{f_0 C_1}{kA^2}\left(1 - \frac{d}{(d^2-A^2)^{\frac{1}{2}}}\right) \qquad 2.30$$

This analytical solution can be extended to $A \gg D$.[3] The accuracy of the solution for the $A \gg D$ case increases with amplitude and approximation terms, as explained elsewhere.[3]

## 2.3 Relation between the Δf and the tip-sample interaction force using the least-action principle

### 2.3.1 The inverse Abel transforms

The Abel transform is fundamental to retrieve the force from the resonance frequency shift data.[8] For its derivation, we first define the Fourier transform, its inverse and one derivative property:[8]

$$\Xi(s) = \int_{-\infty}^{\infty} \varepsilon(l) e^{-i2\pi l s} dl \qquad 2.31$$

$$\varepsilon(l) = \int_{-\infty}^{\infty} \Xi(s) e^{i2\pi l s} ds \qquad 2.32$$

$$\frac{d}{dl}\varepsilon(l) = \int_{-\infty}^{\infty} i2\pi s\, \Xi(s) e^{i2\pi l s} ds = F^{-1}(i2\pi s\, \Xi(s)) \qquad 2.33$$

The Abel transform $\Psi(x)$ of a function $v(\mu)$ is defined as:[8]

$$\Psi(x) = \int_{x}^{\infty} 2\mu (\mu^2 - x^2)^{-1/2} v(\mu) d\mu \qquad 2.34$$

To further pursue the demonstration, it is convenient to do the following change of variables: $o = x^2$, $\rho = \mu^2$, $d\rho = 2\mu d\mu$, $\Psi(x) = \alpha(x^2)$, $v(\mu) = N(\mu^2)$.

$$\alpha(o) = \int_{-\infty}^{\infty} \Sigma(o - \rho) N(\rho) d\rho = (\Sigma * N)(o) \qquad 2.35$$

Where the Kernel is:

$$\Sigma(o) = \begin{cases} (-o)^{-1/2} & , o < 0 \\ 0 & , o \geq 0 \end{cases} \qquad 2.36$$

Applying the convolution theorem to Equation 2.35:

$$F\{\alpha(o)\} = F\{\Sigma(o)\} F\{N(\rho)\} \qquad 2.37$$

$$F\{\Sigma(o)\} = \int_{-\infty}^{\infty} \Sigma(o) e^{-i2\pi o s} do = \int_{-\infty}^{0} \frac{1}{(-o)^{1/2}} e^{-i2\pi o s} do = \int_{0}^{\infty} \frac{1}{(o)^{1/2}} e^{i2\pi o s} do \qquad 2.38$$

With the variable change $\tau^2 = o$, $do = 2\tau d\tau$

$$\int_{0}^{\infty} \frac{1}{\tau} e^{i2\pi \tau^2 s} 2\tau d\tau = 2\int_{0}^{\infty} e^{i2\pi \tau^2 s} d\tau = 2\left(\sqrt{\frac{i\pi}{8\pi s}}\right) = \frac{1}{\sqrt{-i2s}} \qquad 2.39$$

$$F\{\alpha(o)\} = F\{\Sigma(o)\}F\{N(\rho)\} = \frac{F\{N(\rho)\}}{\sqrt{-i2s}} = \frac{\pi\sqrt{-i2s}}{\pi(-i2s)}F\{N(\rho)\} \qquad 2.40$$

$$F\{N(\rho)\} = F\{\alpha(o)\}\frac{\pi(-i2s)}{\pi\sqrt{-i2s}} = -\frac{1}{\pi}F\{\Sigma(\rho)\}F\left\{\frac{d\alpha(o)}{do}\right\}$$
$$= -\frac{1}{\pi}F\left((\Sigma * \frac{d\alpha}{do})(\rho)\right) \qquad 2.41$$

$$N(\rho) = -\frac{1}{\pi}\int_{-\infty}^{\infty}\Sigma(\rho-o)\frac{d\alpha(o)}{do}do = -\frac{1}{\pi}\int_{\rho}^{\infty}\frac{1}{(o-\rho)^{1/2}}\frac{d\alpha(o)}{do}do \qquad 2.42$$

Again with: $o = x^2$, $do = 2xdx$, $\rho = \mu^2$, $d\rho = 2\mu d\mu$, $\Psi(x) = \alpha(x^2)$, $\nu(\mu) = N(\mu^2)$:

$$N(\mu^2) = -\frac{1}{\pi}\int_{\mu}^{\infty}\frac{1}{(x^2-\mu^2)^{1/2}}\frac{d\alpha(x^2)}{d(x^2)}2x\,dx \qquad 2.43$$

By applying:

$$\frac{d\Psi(x)}{dx} = \frac{d\alpha(x^2)}{d(x)} = \frac{d\alpha(x^2)}{d(x^2)}\frac{d(x^2)}{dx} = \frac{d\alpha(x^2)}{d(x^2)}2x \qquad 2.44$$

We finally get:

$$\nu(\mu) = -\frac{1}{\pi}\int_{\mu}^{\infty}\frac{1}{(x^2-\mu^2)^{1/2}}\frac{d\Psi(x)}{dx}dx \qquad 2.45$$

The inverse Abel transforms of the function $\Psi(x)$ can be calculated by employing Equation 2.45.

### 2.3.2 Relation between $\Delta f$ and tip-sample interaction force employing the least-action principle

The oscillation of the cantilever under the tip-sample interaction can be expressed in terms of a Fourier series. This enables to inclusion of anharmonic correction terms systematically.[4] Dürig provided an alternative calculation of the $\Delta f$ using classical mechanics, i.e., the least action principle.[4]

To this end, the derivation starting with Hamilton's principle, i.e., Equation 1.10:

$$\delta\int_{t_0}^{t_1}Ldt = 0 \qquad 2.46$$

In this particular case, the Lagrangian is:[4]

$$L(\psi,\dot{\psi}) = T - V = \frac{1}{2}m\left(\dot{\psi}(t)\right)^2 - \frac{1}{2}k\left(\psi(t)\right)^2 - U_{int}\left(\psi_0 + \psi(t)\right) \qquad 2.47$$

$$\delta L = \frac{\partial L}{\partial \psi}\delta\psi + \frac{\partial L}{\partial \dot{\psi}}\delta\dot{\psi} \qquad 2.48$$

In Equations 2.47 and 2.48, $\psi(t)$ is the orbital of the tip's motion, $-k(\psi(t))$ is the harmonic spring force, with $k$ a positive value. $\psi_0$ is the tip-sample's distance at the point where the cantilever is undeflected. The integration by parts is possible, since variations $\delta\psi$ at the boundaries $t = 0$ and $t = T$ are zero:[5]

$$\int_0^T \frac{\partial L}{\partial \dot{\psi}}\delta\dot{\psi}\,dt = \left[\frac{\partial L}{\partial \dot{\psi}}\delta\psi\right]_0^T - \int_0^T \frac{d}{dt}\left(\frac{\partial L}{\partial \dot{\psi}}\right)\delta\psi\,dt = -\int_0^T \frac{d}{dt}\left(\frac{\partial L}{\partial \dot{\psi}}\right)\delta\psi\,dt \qquad 2.49$$

$$\delta\int_0^T \left(\frac{-L}{m}\right)dt = -\frac{1}{m}\int_0^T \left[\frac{\partial L}{\partial \psi}\delta\psi - \frac{d}{dt}\left(\frac{\partial L}{\partial \dot{\psi}}\right)\delta\psi\right]dt \qquad 2.50$$

Since $\omega_0 = (k/m)^{1/2}$;

$$\frac{\partial L(\psi,\dot{\psi})}{\partial \psi}\delta\psi = -k\psi(t)\delta\psi + F_{int}\left(\psi_0 + \psi(t)\right)\delta\psi \qquad 2.51$$

$$\frac{d}{dt}\left(\frac{\partial L(\psi,\dot{\psi})}{\partial \dot{\psi}}\right) = m\ddot{\psi}(t) \qquad 2.52$$

The following is obtained:

$$\delta\int_0^T \left(\frac{-L}{m}\right)dt = \int_0^T \left[\ddot{\psi}(t) + \omega_0^2\psi(t) - \frac{\omega_0^2}{k}F_{int}\left(\psi_0 + \psi(t)\right)\right]\delta\psi\,dt = 0 \qquad 2.53$$

Introducing a periodic ansatz for the orbital function $\psi(t)$ and substituting Equation 2.54 into Equation 2.53, Equation 2.55 is obtained.

$$\psi(t) = \sum_{n=1}^{\infty} a_n \cos(n\omega t) \qquad 2.54$$

$$\delta \int_0^T \left(\frac{-L}{m}\right) dt = \omega_0^2 \int_0^T \sum_{m=1}^{\infty} a_m \cos(m\omega t) \sum_{n=1}^{\infty} \delta a_n \cos(n\omega t) dt$$

$$+ \int_0^T \sum_{m=1}^{\infty} \left(-m^2 \omega^2\right) a_m \cos(m\omega t) \sum_{n=1}^{\infty} \delta a_n \cos(n\omega t) dt \qquad 2.55$$

$$- \frac{\omega_0^2}{k} \int_0^T F_{int}\left(\psi_0 + \sum_{m=1}^{\infty} a_m \cos(m\omega t)\right) \sum_{n=1}^{\infty} \delta a_n \cos(n\omega t) dt$$

$$\delta \int_0^T \left(\frac{-L}{m}\right) dt = \omega_0^2 \sum_{n=1}^{\infty} a_n \left(\frac{\pi}{\omega}\right) \delta a_n + \sum_{n=1}^{\infty} a_n \left(\frac{\pi}{\omega}\right)\left(-n^2 \omega^2\right) \delta a_n$$

$$- \frac{\omega_0^2}{k} \sum_{n=1}^{\infty} \delta a_n \int_0^T F_{int}\left(\psi_0 + \sum_{m=1}^{\infty} a_m \cos(m\omega t)\right) \cos(n\omega t) dt \qquad 2.56$$

Employing the nomenclature[4], the following is defined, where $a_n$ stands for the oscillation amplitude for each expansion term:

$$S(a_1, a_2, ..., a_n, t) = \frac{-L}{m} \qquad 2.57$$

$$\delta S = \frac{\partial S}{\partial a_1} \delta a_1 + \frac{\partial S}{\partial a_2} \delta a_2 + ... + \frac{\partial S}{\partial a_n} \delta a_n + \frac{\partial S}{\partial t} \delta t \qquad 2.58$$

Since in our case the Lagrangian $L$ doesn't depend on $t$, terms with time derivatives in Equation 2.58 are equal to zero.

$$\delta S = \frac{\partial S}{\partial a_1} \delta a_1 + \frac{\partial S}{\partial a_2} \delta a_2 + ... + \frac{\partial S}{\partial a_n} \delta a_n \qquad 2.59$$

Moreover, because $\delta S$ must be equal to 0, notwithstanding the value of each $\delta a_n$:

$$\frac{\partial S}{\partial a_1} = \frac{\partial S}{\partial a_2} = ... = \frac{\partial S}{\partial a_n} = 0 \qquad 2.60$$

If we consider only the terms with $n < 2$:

$$\int_0^T (\delta S)dt = \int_0^T \left(\frac{\partial S}{\partial a_1}\delta a_1\right)dt$$

$$= \left(\frac{\pi}{\omega}a_1(\omega_0^2 - \omega^2) - \frac{\omega_0^2}{k}\int_0^T F_{int}(\psi_0 + a_1\cos(\omega t))\cos(\omega t)dt\right)\delta a_1 \qquad 2.61$$

$$= 0$$

This equation must be valid for any $\delta a_1$. That is only possible if:

$$\frac{\pi}{\omega}a_1(\omega_0^2 - \omega^2) - \frac{\omega_0^2}{k}\int_0^T F_{int}(\psi_0 + a_1\cos(\omega t))\cos(\omega t)dt = 0 \qquad 2.62$$

$$k = \frac{\omega_0^2 \omega}{\pi a_1(\omega_0^2 - \omega^2)}\int_0^T F_{int}(\psi_0 + a_1\cos(\omega t))\cos(\omega t)dt \qquad 2.63$$

With:

$$\frac{\omega}{\omega_0} = \frac{(k + C_i^{eff})^{1/2}}{k^{1/2}} \qquad 2.64$$

$$C_i^{eff} = k\frac{(\omega^2 - \omega_0^2)}{\omega_0^2} \qquad 2.65$$

By employing Equations 2.63 and 2.65:

$$C_i^{eff} = -\frac{\omega}{\pi a_1}\int_0^T F_{int}(\psi_0 + a_1\cos(\omega t))\cos(\omega t)dt \qquad 2.66$$

If $\omega \approx \omega_0$ is assumed:

$$C_i^{eff} = k\frac{(\omega + \omega_0)(\omega - \omega_0)}{\omega_0^2} \approx \frac{2k\Delta\omega}{\omega} \qquad 2.67$$

$$\Delta f \approx -\frac{f_0}{ka_1^2}\frac{1}{T}\int_0^T F_{int}(\psi_0 + a_1\cos(\omega t))a_1\cos(\omega t)dt \qquad 2.68$$

Equation 2.68 correlates the measured $\Delta f$ to the effective spring constant of the cantilever under the tip-sample interaction force. The result presented in Equation 2.68 can be elucidated by introducing the following substitution:

$$u = \cos(\omega t) \qquad 2.69$$

$$du = -\omega \sin(\omega t) dt = -\omega \sqrt{1-u^2} dt \qquad (2.70)$$

$$\psi_0 = D + a_1 \qquad (2.71)$$

and considering that the integrand is even:

$$C_i^{eff} = -\frac{2\omega}{\pi a_1} \int_0^{T/2} F_{int}(\psi_0 + a_1 \cos(\omega t)) \cos(\omega t) dt \qquad (2.72)$$

$$C_i^{eff} = -\frac{2}{\pi a_1} \int_{-1}^{1} F_{int}(D + a_1(1+u)) \frac{u}{\sqrt{1-u^2}} du \qquad (2.73)$$

If we have a linear interaction force $F_{int}(z) = -kz$ with $k$ a positive value:

$$C_i^{eff} = \frac{2\omega}{\pi a_1} \int_0^{T/2} (k\psi_0 + ka_1 \cos(\omega t)) \cos(\omega t) dt = \frac{2\omega}{\pi a_1}\left(\frac{a_1 kT}{4}\right) = k \qquad (2.74)$$

It is important to note that Equation 2.68 is the same as Equation 2.7, derived using a different approach.

### 2.3.3 Force reconstruction employing the inverse Abel transform

The force acting on the tip will be dominated at close distances, which occurs when $t = T/2$ or $u = -1$, where $u$ is a convenient change of variable for demonstration purposes.[2,4]

With the transformations $\xi = z/\ell$ and $\zeta = D/\ell$:

$$z = D + a_1(1+u) = \xi \ell \approx D \qquad (2.75)$$

$$u = \frac{\xi \ell - D - a_1}{a_1} \qquad (2.76)$$

$$du = \frac{d\xi \ell}{a_1} \qquad (2.77)$$

$$\frac{u}{\sqrt{1-u^2}} = \frac{\xi \ell - D - a_1}{\left(a_1^2 - (\xi \ell - D - a_1)^2\right)^{1/2}} \qquad (2.78)$$

$$ua_1 = \xi \ell - D - a_1 \approx -a_1 \qquad (2.79)$$

$$-(\xi \ell - D - a_1)^2 = -a_1^2 \left(1 - \frac{(\xi \ell - D)}{a_1}\right)^2 \approx -a_1^2 \left(1 - \frac{2(\xi \ell - D)}{a_1}\right) \qquad (2.80)$$

$$\frac{u}{\sqrt{1-u^2}} = \frac{\xi\ell - D - a_1}{\left(a_1^2 - (\xi\ell - D - a_1)^2\right)^{1/2}} \approx \frac{-a_1}{\left(2a_1(\xi\ell - D)\right)^{1/2}} = -\sqrt{\frac{a_1}{2\ell}}\frac{1}{(\xi-\zeta)^{1/2}} \qquad 2.81$$

$$u = -1 \Rightarrow \xi = \frac{D}{\ell} = \zeta$$
$$u = 1 \Rightarrow \xi = \frac{(2a_1 + D)}{\ell} \to \infty \qquad 2.82$$

$$C_i^{\text{eff}} = -\frac{2}{\pi a_1}\int_{-1}^{1} F_{\text{int}}(D + a_1(1+u))\frac{u}{\sqrt{1-u^2}}du \approx \frac{\sqrt{2\ell}}{\pi a_1^{3/2}}\int_{\zeta}^{\infty}\frac{F_{\text{int}}(\xi\ell)}{\sqrt{\xi-\zeta}}d\xi \qquad 2.83$$

Equation 2.83 is the lowest-order harmonic approximation of the orbital function.[4] Moreover, Equation 2.83 is generalized, as there is no restriction on the interaction or oscillation amplitude (see above for an example with a linear mass-spring system).[4]

Equation 2.83 can be further improved by introducing anharmonic corrections. If Equation 2.54 is retained, Equation 2.84 is valid for the $n^{\text{th}}$ term:

$$\frac{\partial S}{\partial a_n} = 0 = \frac{\pi}{\omega}a_n(\omega_0^2 - n^2\omega^2) - \frac{\omega_0^2}{k}\int_0^T F_{\text{int}}(\psi_0 + a_1\cos(\omega t))\cos(n\omega t)dt \qquad 2.84$$

Transferring the first term to the left side and employing Equation 2.47, Equation 2.85 is obtained:

$$\frac{a_n(\omega_0^2 - n^2\omega^2)}{a_1(\omega_0^2 - \omega^2)} = \frac{\int_0^T F_{\text{int}}(\psi_0 + a_1\cos(\omega t))\cos(n\omega t)dt}{\int_0^T F_{\text{int}}(\psi_0 + a_1\cos(\omega t))\cos(\omega t)dt} \qquad 2.85$$

Developing at $t = T/2$:

$$u = \cos(\omega t) \approx -1 + \frac{\omega^2(t - T/2)^2}{2} \qquad 2.86$$

$$\cos(n\omega t) \approx -1 + \frac{n^2\omega^2(t - T/2)^2}{2} \qquad 2.87$$

$$\cos(n\omega t) \approx -1 + n^2 + n^2 u \qquad 2.88$$

Since $a_1 \gg D$, i.e., the first harmonic term of the oscillation amplitude $A$:

$$\frac{z}{a_1} = \frac{\xi\ell}{a_1} = \frac{D}{a_1} + (1+u) \approx 1+u \qquad 2.89$$

Higher anharmonic terms can be calculated as follows:

$$dt = \frac{-1}{\omega\sqrt{1-u^2}}du \qquad 2.90$$

$$\int_0^T F_{int}\left(\psi_0 + a_1\cos(\omega t)\right)\cos(n\omega t)dt$$
$$\approx \frac{2}{\omega}\int_{-1}^{1} F_{int}\left(D + a_1(1+u)\right)\left(1 - n^2(1+u)\right)\frac{u}{\sqrt{1-u^2}}du \qquad 2.91$$

$$\frac{2}{\omega}\int_{-1}^{1} F_{int}\left(D + a_1(1+u)\right)\left(1 - n^2(1+u)\right)\frac{u}{\sqrt{1-u^2}}du$$
$$= \frac{-2}{\omega}\sqrt{\frac{\ell}{2a_1}}\int_{\zeta}^{\infty}\frac{F_{int}(\xi\ell)\left(1 - n^2\frac{\xi\ell}{a_1}\right)}{\sqrt{\xi-\zeta}}d\xi \qquad 2.92$$

$$\frac{a_n\left(\omega_0^2 - n^2\omega^2\right)}{a_1\left(\omega_0^2 - \omega^2\right)} = \frac{\int_{\zeta}^{\infty}\frac{F_{int}(\xi\ell)\left(1 - n^2\frac{\xi\ell}{a_1}\right)}{\sqrt{\xi-\zeta}}d\xi}{\int_{\zeta}^{\infty}\frac{F_{int}(\xi\ell)}{\sqrt{\xi-\zeta}}d\xi} \qquad 2.93$$

By employing Equation 2.83:

$$a_1^{3/2} = \frac{\sqrt{2\ell}}{\pi C_i^{eff}}\int_{\zeta}^{\infty}\frac{F_{int}(\xi\ell)}{\sqrt{\xi-\zeta}}d\xi = -\frac{\sqrt{2\ell}\omega_0^2}{\pi k\left(\omega_0^2 - \omega^2\right)}\int_{\zeta}^{\infty}\frac{F_{int}(\xi\ell)}{\sqrt{\xi-\zeta}}d\xi \qquad 2.94$$

The integral term in the denominator of Equation 2.93 can be replaced:

$$\frac{a_n\left(\omega_0^2 - n^2\omega^2\right)}{a_1} = -\frac{\sqrt{2\ell}\omega_0^2}{a_1^{3/2}\pi k}\int_{\zeta}^{\infty}\frac{F_{int}(\xi\ell)\left(1 - n^2\frac{\xi\ell}{a_1}\right)}{\sqrt{\xi-\zeta}}d\xi \qquad 2.95$$

In case that $n^2\ell/a_1 \ll 1$ and $\omega \approx \omega_0$:

$$\frac{a_n\left(\omega_0^2 - n^2\omega_0^2\right)}{a_1} = -\frac{\sqrt{2\ell}\omega_0^2}{a_1^{3/2}\pi k}\int_{\zeta}^{\infty}\frac{F_{int}(\xi\ell)}{\sqrt{\xi-\zeta}}d\xi = -\frac{C_i^{eff}\omega_0^2}{k} \qquad 2.96$$

$$a_n = \frac{C_i^{eff} a_1}{k(n^2 - 1)} \qquad 2.97$$

By employing Equation 2.96, it is possible to correlate the tip-sample interaction force and energy to the effective spring constant:

$$C_i^{eff}(\zeta\ell) = \frac{\sqrt{2\ell}}{\pi a_1^{3/2}} \int_\zeta^\infty \frac{F_{int}(\xi\ell)}{\sqrt{\xi - \zeta}} d\xi = \frac{\sqrt{2}}{\pi a_1^{3/2}} \int_{\ell\zeta}^\infty \frac{F_{int}(\xi\ell)}{\sqrt{\ell\xi - \ell\zeta}} d\ell\xi \qquad 2.98$$

$$C_i^{eff}(\zeta\ell) \frac{a_1^{3/2}}{\sqrt{2}} = -\frac{1}{\pi} \int_{\ell\zeta}^\infty \frac{\frac{dU_{int}(\xi\ell)}{d\xi\ell}}{\sqrt{\ell\xi - \ell\zeta}} d\ell\xi \qquad 2.99$$

Most importantly, by employing the inverse Abel transform, the measured resonance frequency shift can be employed to reconstruct tip-sample interaction energy (starting from Equation 2.45 and proceeding backward to Equation 2.37):

$$\frac{a_1^{3/2}}{\sqrt{2}} F\{C_i^{eff}(\zeta\ell)\} = -\frac{1}{\pi} F\left((P * \frac{dU_{int}}{d\xi\ell})(\ell\zeta)\right) = -\frac{1}{\pi} F\{P(\zeta\ell)\} F\left\{\frac{dU_{int}(\xi\ell)}{d\xi\ell}\right\} \qquad 2.100$$

$$F\{U_{int}(\xi\ell)\} = \frac{a_1^{3/2}}{\sqrt{2}} F\{P(\xi\ell)\} F\{C_i^{eff}(\zeta\ell)\} \qquad 2.101$$

$$U_{int}(\xi\ell) = \frac{a_1^{3/2}}{\sqrt{2}} (P * C_i^{eff})(\xi\ell)$$

$$= \frac{a_1^{3/2}}{\sqrt{2}} \int_{-\infty}^\infty P(\xi\ell - \zeta\ell) C_i^{eff}(\zeta\ell) d\zeta\ell \qquad 2.102$$

$$= \frac{a_1^{3/2}}{\sqrt{2}} \int_{\xi\ell}^\infty \frac{C_i^{eff}(\zeta\ell)}{(\zeta\ell - \xi\ell)^{1/2}} d\zeta\ell$$

$$U_{int}(\xi\ell) = \frac{a_1^{3/2} \ell^{1/2}}{\sqrt{2}} \int_\xi^\infty \frac{C_i^{eff}(\zeta\ell)}{(\zeta - \xi)^{1/2}} d\zeta \qquad 2.103$$

By using the Abel transform, the relation between the measured resonance frequency shift and the tip-sample interaction force can be found as follows (starting from Equation 2.35):

$$C_i^{eff}(\zeta\ell) = \frac{\sqrt{2}}{\pi a_1^{3/2}} \int_{\ell\zeta}^{\infty} \frac{F_{int}(\xi\ell)}{\sqrt{\ell\xi - \ell\zeta}} d\ell\xi$$

$$= \frac{\sqrt{2}}{\pi a_1^{3/2}} \int_{-\infty}^{\infty} P(\zeta\ell - \xi\ell) F_{int}(\xi\ell) d\xi\ell \qquad 2.104$$

$$= \frac{\sqrt{2}}{\pi a_1^{3/2}} (P * F_{int})(\zeta\ell)$$

$$F\{C_i^{eff}(\zeta\ell)\} = \frac{\sqrt{2}}{\pi a_1^{3/2}} F\{P(\zeta\ell)\} F\{F_{int}(\xi\ell)\} \qquad 2.105$$

$$F\{F_{int}(\xi\ell)\} = -\frac{a_1^{3/2}}{\sqrt{2}} F\{P(\xi\ell)\} F\left\{\frac{dC_i^{eff}(\zeta\ell)}{d\zeta\ell}\right\}$$

$$= -\frac{a_1^{3/2}}{\sqrt{2}} F\left((P * \frac{dC_i^{eff}(\zeta\ell)}{d\zeta\ell})(\xi\ell)\right) \qquad 2.106$$

$$F_{int}(\xi\ell) = -\frac{a_1^{3/2}}{\sqrt{2}} \int_{-\infty}^{\infty} P(\xi\ell - \zeta\ell) \frac{dC_i^{eff}(\zeta\ell)}{d\zeta\ell} d\zeta\ell$$

$$= -\frac{a_1^{3/2}}{\sqrt{2}} \int_{\xi\ell}^{\infty} \frac{1}{\sqrt{\zeta\ell - \xi\ell}} \frac{dC_i^{eff}(\zeta\ell)}{d\zeta\ell} d\zeta\ell \qquad 2.107$$

Eventually, Dürig's final equation, which can be used to reconstruct tip-sample interaction force from the resonance frequency data for large oscillation amplitudes ($a_1 \gg D$) is reached:[4]

$$F_{int}(\xi\ell) = -\frac{a_1^{3/2}}{\sqrt{2\ell}} \int_{\xi}^{\infty} \frac{\frac{dC_i^{eff}(\zeta\ell)}{d\zeta}}{\sqrt{\zeta - \xi}} d\zeta \qquad 2.108$$

## Section 3 Amplitude independent formulations for frequency modulation atomic force microscopy

### 3.1 Direct calculation of tip-sample interaction force from the frequency shift data with discretization

Although earlier attempts were made to reconstruct tip-sample interaction force from the experimental data, they are best suited for large oscillation amplitudes.[9,10] Further methods based on perturbation theory to reconstruct the tip-sample interaction force for small oscillation amplitudes require demodulation of higher harmonics.[11] The discretization of the tip-sample interaction force measurement eliminated the requirement of higher harmonic demodulation.[12] The discretization method is based on the weight averaging of the force gradient.[12] As presented in Section 4, the $\Delta f$ can be calculated by Equation 4.25.

$$\Delta f \approx -\frac{2f_0^2}{CA} \int_0^{T/2} F_{int}(D + A + A\cos(2\pi ft))\cos(2\pi ft)dt \qquad 3.1$$

If the harmonic motion of the cantilever is approximated with the following equation:

$$z(t \gg 0) = D + A + A\cos(2\pi ft) = D + p \qquad 3.2$$

$$dp = -2\pi fA\sin(2\pi ft)dt = -2\pi f\sqrt{A^2 - (p-A)^2}dt \qquad 3.3$$

Substituting Equations 3.2 and 3.3 into Equation (4.25), we obtain:

$$\Delta f(D) = \frac{f_0}{\pi CA^2}\int_{2A}^{0}\frac{F_{int}(D+p)(p-A)}{\sqrt{A^2-(p-A)^2}}dp \qquad 3.4$$

The $\Delta f$ measurement is not continuous but discrete experimentally. Hence, for each value of $F_{int}$, we have discrete experimental points of $\Delta f$ and $z$ for a total of $N$ values. Consequently, Equation 3.4 can be discretized and implemented in a matrix form.[12] For the discretization, $\Delta$ is the distance between each measured value on the $z$-axis, i.e., upon approach to the sample surface. We define two integration parameters: $\iota = \text{round}(A/\Delta)$ and the integration step $\phi = 2A / (2\iota+1)$. It is assumed that the $F_{int}$ remains constant over $\phi$. Two examples, where we have the particular case $2A = N\Delta$, can be seen in Figure S1 below. Here, the number of integration steps will be $N$. We will see further down that this will result in a $N \times N$ matrix.

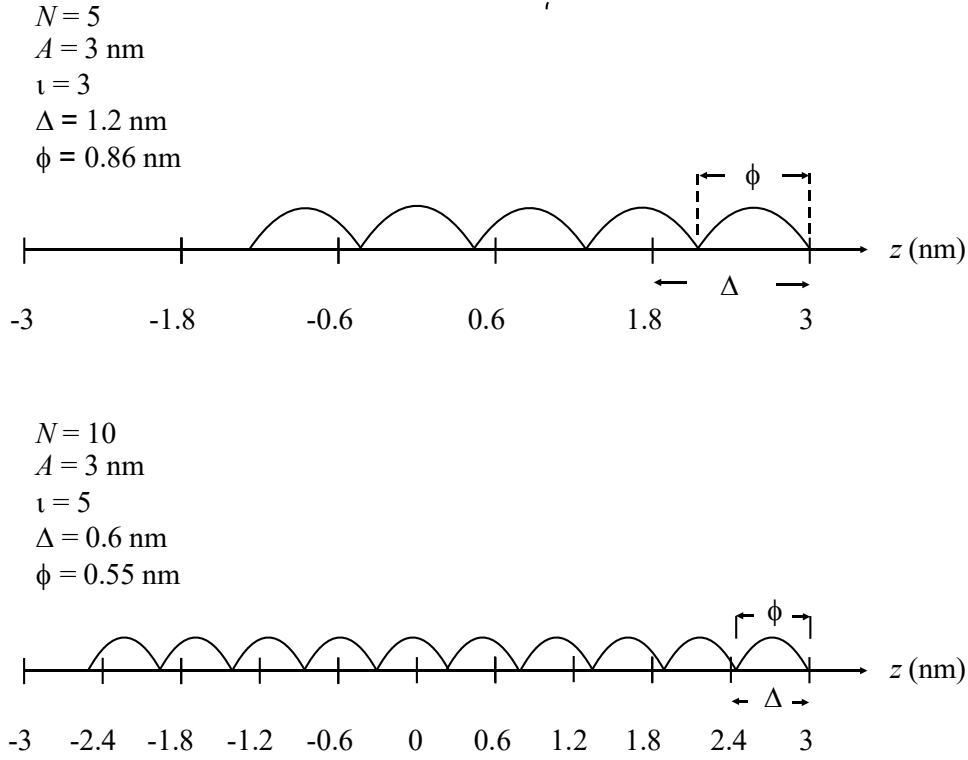

**Figure S1. Discretization scheme for numerical force reconstruction for $2A = N\Delta$.** Illustration of the discretization of the frequency shift ($\Delta f$) measurement along the z-axis for numerical integration of the interaction force $F_{int}$. Two cases are shown with different numbers of data points ($N$): **(a)**, $N = 5$, corresponding to an amplitude $A = 3$ nm, discretization step $\Delta = 1.2$ nm, and integration step $\phi = 0.86$ nm; **(b)**, $N = 10$ with $\Delta = 0.6$ nm and $\phi = 0.55$ nm. The integration interval spans $2A$, centered around the oscillation midpoint. The figure illustrates how finer discretization (higher $N$) improves resolution in the numerical evaluation of the force from $\Delta f$.

As $N \rightarrow \infty$ we have $\phi \rightarrow \Delta$.

By posing $\tau = (p-A)/A$ in Equation 3.4, we get:

$$\Delta f(D) = \frac{f_0}{\pi C A^2} \int_{2A}^{0} \frac{F_{int}(D+p)(p-A)}{\sqrt{A^2 - (p-A)^2}} dp = \frac{f_0}{\pi C A} \int_{-1}^{1} \frac{F_{int}(D+A-\tau A)\tau}{\sqrt{1-\tau^2}} d\tau \qquad 3.5$$

If several values of $D$ are used, each will be separated by $\Delta$. It is assumed that $F_{int}$ is constant throughout integration. We have $N$ normalized integration steps with a value equal to $\phi / A = 2 / (2\iota+1)$. $D_N$ is the first distance with $\Delta f \neq 0$, i.e., for all $D > D_N$, $\Delta f = 0$.

$$\Delta f(D_1) \approx \frac{f_0}{\pi C A} \left[ F_{int}(D_1) \int_{1-2/(2\iota+1)}^{1} \frac{\tau}{\sqrt{1-\tau^2}} d\tau \right] \qquad 3.6$$

$$\Delta f(D_2) \approx \frac{f_0}{\pi CA}\left[F_{int}(D_1)\int_{1-4/(2\iota+1)}^{1-2/(2\iota+1)}\frac{\tau}{\sqrt{1-\tau^2}}d\tau + F_{int}(D_2)\int_{1-2/(2\iota+1)}^{1}\frac{\tau}{\sqrt{1-\tau^2}}d\tau\right] \qquad 3.7$$

...

$$\begin{aligned}\Delta f(D_{N-1}) \approx \frac{f_0}{\pi CA}&\left[F_{int}(D_1)\int_{1-2(N-1)/(2\iota+1)}^{1-2(N-2)/(2\iota+1)}\frac{\tau}{\sqrt{1-\tau^2}}d\tau + ...\right.\\ &+F_{int}(D_{N-2})\int_{1-4/(2\iota+1)}^{1-2/(2\iota+1)}\frac{\tau}{\sqrt{1-\tau^2}}d\tau\\ &\left.+F_{int}(D_{N-1})\int_{1-2/(2\iota+1)}^{1}\frac{\tau}{\sqrt{1-\tau^2}}d\tau\right]\end{aligned} \qquad 3.8$$

$$\begin{aligned}\Delta f(D_N) \approx \frac{f_0}{\pi CA}&\left[F_{int}(D_1)\int_{1-2N/(2\iota+1)}^{1-2(N-1)/(2\iota+1)}\frac{\tau}{\sqrt{1-\tau^2}}d\tau +\right.\\ &F_{int}(D_2)\int_{1-2(N-1)/(2\iota+1)}^{1-2(N-2)/(2\iota+1)}\frac{\tau}{\sqrt{1-\tau^2}}d\tau + ...\\ &+F_{int}(D_{N-1})\int_{1-4/(2\iota+1)}^{1-2/(2\iota+1)}\frac{\tau}{\sqrt{1-\tau^2}}d\tau\\ &\left.+F_{int}(D_N)\int_{1-2/(2\iota+1)}^{1}\frac{\tau}{\sqrt{1-\tau^2}}d\tau\right]\end{aligned} \qquad 3.9$$

As an example, if we have twelve $\Delta f$ and $D$ values, we obtain:

$$\Delta f(D_1) \approx \frac{f_0}{\pi CA}\left[F_{int}(D_1)\int_{1-2/(2\iota+1)}^{1}\frac{\tau}{\sqrt{1-\tau^2}}d\tau\right] \qquad 3.10$$

$$\Delta f(D_2) \approx \frac{f_0}{\pi CA}\left[F_{int}(D_1)\int_{1-4/(2\iota+1)}^{1-2/(2\iota+1)}\frac{\tau}{\sqrt{1-\tau^2}}d\tau + F_{int}(D_2)\int_{1-2/(2\iota+1)}^{1}\frac{\tau}{\sqrt{1-\tau^2}}d\tau\right] \qquad 3.11$$

...

$$\Delta f(D_{11}) \approx \frac{f_0}{\pi CA}\left[ F_{int}(D_1) \int_{1-22/(2\iota+1)}^{1-20/(2\iota+1)} \frac{\tau}{\sqrt{1-\tau^2}} d\tau + ...\right.$$

$$+ F_{int}(D_{10}) \int_{1-4/(2\iota+1)}^{1-2/(2\iota+1)} \frac{\tau}{\sqrt{1-\tau^2}} d\tau$$

$$\left. + F_{int}(D_{11}) \int_{1-2/(2\iota+1)}^{1} \frac{\tau}{\sqrt{1-\tau^2}} d\tau \right] \qquad 3.12$$

$$\Delta f(D_{12}) \approx \frac{f_0}{\pi CA}\left[ F_{int}(D_1) \int_{1-24/(2\iota+1)}^{1-22/(2\iota+1)} \frac{\tau}{\sqrt{1-\tau^2}} d\tau + \right.$$

$$F_{int}(D_2) \int_{1-22/(2\iota+1)}^{1-20/(2\iota+1)} \frac{\tau}{\sqrt{1-\tau^2}} d\tau + ...$$

$$+ F_{int}(D_{11}) \int_{1-4/(2\iota+1)}^{1-2/(2\iota+1)} \frac{\tau}{\sqrt{1-\tau^2}} d\tau \qquad 3.13$$

$$\left. + F_{int}(D_{12}) \int_{1-2/(2\iota+1)}^{1} \frac{\tau}{\sqrt{1-\tau^2}} d\tau \right]$$

Finally, with $i$ the rank of $\Delta f$, $j$ the rank of the component in the associated equation:

$$w_{ij} = \begin{cases} \dfrac{f_0}{\pi CA} \displaystyle\int_{1-[2(i-j+1)/(2\iota+1)]}^{1-[2(i-j)/(2\iota+1)]} \dfrac{\tau}{\sqrt{1-\tau^2}} d\tau & 0 \leq i-j \leq 2\iota \\ 0 & \text{else} \end{cases}$$

It is now possible to write the equations in a triangular matrix form with the matrix dimensions being $N \times N$:

$$\begin{pmatrix} \Delta f(D_1) \\ \Delta f(D_2) \\ \Delta f(D_3) \\ \Delta f(D_4) \\ \Delta f(D_5) \\ \Delta f(D_6) \\ \Delta f(D_7) \\ \Delta f(D_8) \\ \Delta f(D_9) \\ \Delta f(D_{10}) \\ \Delta f(D_{11}) \\ \Delta f(D_{12}) \end{pmatrix} = \begin{bmatrix} w_{11} & 0 & 0 & 0 & 0 & 0 & 0 & 0 & 0 & 0 & 0 & 0 \\ w_{21} & w_{22} & 0 & 0 & 0 & 0 & 0 & 0 & 0 & 0 & 0 & 0 \\ w_{31} & w_{32} & w_{33} & 0 & 0 & 0 & 0 & 0 & 0 & 0 & 0 & 0 \\ w_{41} & w_{42} & w_{43} & w_{44} & 0 & 0 & 0 & 0 & 0 & 0 & 0 & 0 \\ w_{51} & w_{41} & w_{42} & w_{43} & w_{55} & 0 & 0 & 0 & 0 & 0 & 0 & 0 \\ w_{61} & w_{62} & w_{63} & w_{64} & w_{65} & w_{66} & 0 & 0 & 0 & 0 & 0 & 0 \\ w_{71} & w_{72} & w_{73} & w_{74} & w_{75} & w_{76} & w_{77} & 0 & 0 & 0 & 0 & 0 \\ w_{81} & w_{82} & w_{83} & w_{84} & w_{85} & w_{86} & w_{87} & w_{88} & 0 & 0 & 0 & 0 \\ w_{91} & w_{92} & w_{93} & w_{94} & w_{95} & w_{96} & w_{97} & w_{98} & w_{99} & 0 & 0 & 0 \\ w_{10\,1} & w_{10\,2} & w_{10\,3} & w_{10\,4} & w_{10\,5} & w_{10\,6} & w_{10\,7} & w_{10\,8} & w_{10\,9} & w_{10\,10} & 0 & 0 \\ w_{11\,1} & w_{11\,2} & w_{11\,3} & w_{11\,4} & w_{11\,5} & w_{11\,6} & w_{11\,7} & w_{11\,8} & w_{11\,9} & w_{11\,10} & w_{11\,11} & 0 \\ w_{12\,1} & w_{12\,2} & w_{12\,3} & w_{12\,4} & w_{12\,5} & w_{12\,6} & w_{12\,7} & w_{12\,8} & w_{12\,9} & w_{12\,10} & w_{12\,11} & w_{12\,12} \end{bmatrix} \begin{pmatrix} F_{int}(D_1) \\ F_{int}(D_2) \\ F_{int}(D_3) \\ F_{int}(D_4) \\ F_{int}(D_5) \\ F_{int}(D_6) \\ F_{int}(D_7) \\ F_{int}(D_8) \\ F_{int}(D_9) \\ F_{int}(D_{10}) \\ F_{int}(D_{11}) \\ F_{int}(D_{12}) \end{pmatrix} \quad 3.14$$

With a matrix inversion, $F_{int}$ can be retrieved:

$$\begin{pmatrix} F_{int}(D_1) \\ F_{int}(D_2) \\ F_{int}(D_3) \\ F_{int}(D_4) \\ F_{int}(D_5) \\ F_{int}(D_6) \\ F_{int}(D_7) \\ F_{int}(D_8) \\ F_{int}(D_9) \\ F_{int}(D_{10}) \\ F_{int}(D_{11}) \\ F_{int}(D_{12}) \end{pmatrix} = \begin{bmatrix} w_{11} & 0 & 0 & 0 & 0 & 0 & 0 & 0 & 0 & 0 & 0 & 0 \\ w_{21} & w_{22} & 0 & 0 & 0 & 0 & 0 & 0 & 0 & 0 & 0 & 0 \\ w_{31} & w_{32} & w_{33} & 0 & 0 & 0 & 0 & 0 & 0 & 0 & 0 & 0 \\ w_{41} & w_{42} & w_{43} & w_{44} & 0 & 0 & 0 & 0 & 0 & 0 & 0 & 0 \\ w_{51} & w_{41} & w_{42} & w_{43} & w_{55} & 0 & 0 & 0 & 0 & 0 & 0 & 0 \\ w_{61} & w_{62} & w_{63} & w_{64} & w_{65} & w_{66} & 0 & 0 & 0 & 0 & 0 & 0 \\ w_{71} & w_{72} & w_{73} & w_{74} & w_{75} & w_{76} & w_{77} & 0 & 0 & 0 & 0 & 0 \\ w_{81} & w_{82} & w_{83} & w_{84} & w_{85} & w_{86} & w_{87} & w_{88} & 0 & 0 & 0 & 0 \\ w_{91} & w_{92} & w_{93} & w_{94} & w_{95} & w_{96} & w_{97} & w_{98} & w_{99} & 0 & 0 & 0 \\ w_{10\,1} & w_{10\,2} & w_{10\,3} & w_{10\,4} & w_{10\,5} & w_{10\,6} & w_{10\,7} & w_{10\,8} & w_{10\,9} & w_{10\,10} & 0 & 0 \\ w_{11\,1} & w_{11\,2} & w_{11\,3} & w_{11\,4} & w_{11\,5} & w_{11\,6} & w_{11\,7} & w_{11\,8} & w_{11\,9} & w_{11\,10} & w_{11\,11} & 0 \\ w_{12\,1} & w_{12\,2} & w_{12\,3} & w_{12\,4} & w_{12\,5} & w_{12\,6} & w_{12\,7} & w_{12\,8} & w_{12\,9} & w_{12\,10} & w_{12\,11} & w_{12\,12} \end{bmatrix}^{-1} \begin{pmatrix} \Delta f(D_1) \\ \Delta f(D_2) \\ \Delta f(D_3) \\ \Delta f(D_4) \\ \Delta f(D_5) \\ \Delta f(D_6) \\ \Delta f(D_7) \\ \Delta f(D_8) \\ \Delta f(D_9) \\ \Delta f(D_{10}) \\ \Delta f(D_{11}) \\ \Delta f(D_{12}) \end{pmatrix} \quad 3.15$$

We now consider the more general cases where $2A$ is smaller than the range of experimental values. Again, $\Delta$ is the step between each value. This leads to several integral sweeps instead of just one. Suppose that we have twelve $\Delta f$ and $D$ values. $D_2 = D_1 + \Delta$, $D_3 = D_2 + \Delta$, ..., $D_{12} = D_{11} + \Delta$. We show two sweeps in red and black on the figure below. In this general case, the number of normalized integration step is equal to floor($2\iota$) +1.

$N = 12$
$A = 3$ nm
$\iota = 3$
$\Delta = 1$ nm
$\phi = 0.86$ nm

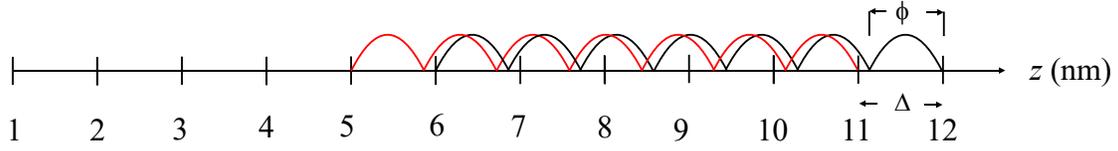

**Figure S2. Discretization scheme for numerical force reconstruction if $2A < N\Delta$.** Illustration of the discretization of $\Delta f$ measurement along the z-axis for numerical integration of the interaction force $F_{int}$. $N = 12$, corresponding to an amplitude $A = 3$ nm, discretization step $\Delta = 1$ nm, and integration step $\phi = 0.86$ nm.

As an example:

$$\Delta f(D_1) \approx \frac{f_0}{\pi C A} \left[ F_{int}(D_1) \int_{1-2/(2\iota+1)}^{1} \frac{\tau}{\sqrt{1-\tau^2}} d\tau \right] \qquad 3.16$$

$$\Delta f(D_2) \approx \frac{f_0}{\pi C A} \left[ F_{int}(D_1) \int_{1-4/(2\iota+1)}^{1-2/(2\iota+1)} \frac{\tau}{\sqrt{1-\tau^2}} d\tau + F_{int}(D_2) \int_{1-2/(2\iota+1)}^{1} \frac{\tau}{\sqrt{1-\tau^2}} d\tau \right] \qquad 3.17$$

...

$$\Delta f(D_6) \approx \frac{f_0}{\pi C A} \left[ F_{int}(D_1) \int_{1-12/(2\iota+1)}^{1-10/(2\iota+1)} \frac{\tau}{\sqrt{1-\tau^2}} d\tau + \ldots \right.$$

$$+ F_{int}(D_5) \int_{1-4/(2\iota+1)}^{1-2/(2\iota+1)} \frac{\tau}{\sqrt{1-\tau^2}} d\tau \qquad 3.18$$

$$\left. + F_{int}(D_6) \int_{1-2/(2\iota+1)}^{1} \frac{\tau}{\sqrt{1-\tau^2}} d\tau \right]$$

$$\Delta f(D_7) \approx \frac{f_0}{\pi C A}\left[ F_{int}(D_1) \int_{1-14/(2\iota+1)}^{1-12/(2\iota+1)} \frac{\tau}{\sqrt{1-\tau^2}} d\tau + \right.$$

$$F_{int}(D_2) \int_{1-12/(2\iota+1)}^{1-10/(2\iota+1)} \frac{\tau}{\sqrt{1-\tau^2}} d\tau + \ldots$$

$$+ F_{int}(D_6) \int_{1-4/(2\iota+1)}^{1-2/(2\iota+1)} \frac{\tau}{\sqrt{1-\tau^2}} d\tau$$

$$\left. + F_{int}(D_7) \int_{1-2/(2\iota+1)}^{1} \frac{\tau}{\sqrt{1-\tau^2}} d\tau \right] \quad\quad 3.19$$

Importantly, we see here that unlike when $N\Delta=2A$, the matrix is not triangular and the elements repeat themselves starting from the eighth line.

$$\begin{pmatrix} \Delta f(D_1) \\ \Delta f(D_2) \\ \Delta f(D_3) \\ \Delta f(D_4) \\ \Delta f(D_5) \\ \Delta f(D_6) \\ \Delta f(D_7) \\ \Delta f(D_8) \\ \Delta f(D_9) \\ \Delta f(D_{10}) \\ \Delta f(D_{11}) \\ \Delta f(D_{12}) \end{pmatrix} = \begin{bmatrix} w_{11} & 0 & 0 & 0 & 0 & 0 & 0 & 0 & 0 & 0 & 0 & 0 \\ w_{21} & w_{22} & 0 & 0 & 0 & 0 & 0 & 0 & 0 & 0 & 0 & 0 \\ w_{31} & w_{32} & w_{33} & 0 & 0 & 0 & 0 & 0 & 0 & 0 & 0 & 0 \\ w_{41} & w_{42} & w_{43} & w_{44} & 0 & 0 & 0 & 0 & 0 & 0 & 0 & 0 \\ w_{51} & w_{52} & w_{53} & w_{54} & w_{55} & 0 & 0 & 0 & 0 & 0 & 0 & 0 \\ w_{61} & w_{62} & w_{63} & w_{64} & w_{65} & w_{66} & 0 & 0 & 0 & 0 & 0 & 0 \\ w_{71} & w_{72} & w_{73} & w_{74} & w_{75} & w_{76} & w_{77} & 0 & 0 & 0 & 0 & 0 \\ 0 & w_{71} & w_{72} & w_{73} & w_{74} & w_{75} & w_{76} & w_{77} & 0 & 0 & 0 & 0 \\ 0 & 0 & w_{71} & w_{72} & w_{73} & w_{74} & w_{75} & w_{76} & w_{77} & 0 & 0 & 0 \\ 0 & 0 & 0 & w_{71} & w_{72} & w_{73} & w_{74} & w_{75} & w_{76} & w_{77} & 0 & 0 \\ 0 & 0 & 0 & 0 & w_{71} & w_{72} & w_{73} & w_{74} & w_{75} & w_{76} & w_{77} & 0 \\ 0 & 0 & 0 & 0 & 0 & w_{71} & w_{72} & w_{73} & w_{74} & w_{75} & w_{76} & w_{77} \end{bmatrix} \begin{pmatrix} F_{int}(D_1) \\ F_{int}(D_2) \\ F_{int}(D_3) \\ F_{int}(D_4) \\ F_{int}(D_5) \\ F_{int}(D_6) \\ F_{int}(D_7) \\ F_{int}(D_8) \\ F_{int}(D_9) \\ F_{int}(D_{10}) \\ F_{int}(D_{11}) \\ F_{int}(D_{12}) \end{pmatrix} \quad 3.20$$

With a matrix inversion, $F_{int}$ can be retrieved:

$$\begin{pmatrix} F_{int}(D_1) \\ F_{int}(D_2) \\ F_{int}(D_3) \\ F_{int}(D_4) \\ F_{int}(D_5) \\ F_{int}(D_6) \\ F_{int}(D_7) \\ F_{int}(D_8) \\ F_{int}(D_9) \\ F_{int}(D_{10}) \\ F_{int}(D_{11}) \\ F_{int}(D_{12}) \end{pmatrix} = \begin{bmatrix} w_{11} & 0 & 0 & 0 & 0 & 0 & 0 & 0 & 0 & 0 & 0 & 0 \\ w_{21} & w_{22} & 0 & 0 & 0 & 0 & 0 & 0 & 0 & 0 & 0 & 0 \\ w_{31} & w_{32} & w_{33} & 0 & 0 & 0 & 0 & 0 & 0 & 0 & 0 & 0 \\ w_{41} & w_{42} & w_{43} & w_{44} & 0 & 0 & 0 & 0 & 0 & 0 & 0 & 0 \\ w_{51} & w_{52} & w_{53} & w_{54} & w_{55} & 0 & 0 & 0 & 0 & 0 & 0 & 0 \\ w_{61} & w_{62} & w_{63} & w_{64} & w_{65} & w_{66} & 0 & 0 & 0 & 0 & 0 & 0 \\ w_{71} & w_{72} & w_{73} & w_{74} & w_{75} & w_{76} & w_{77} & 0 & 0 & 0 & 0 & 0 \\ 0 & w_{71} & w_{72} & w_{73} & w_{74} & w_{75} & w_{76} & w_{77} & 0 & 0 & 0 & 0 \\ 0 & 0 & w_{71} & w_{72} & w_{73} & w_{74} & w_{75} & w_{76} & w_{77} & 0 & 0 & 0 \\ 0 & 0 & 0 & w_{71} & w_{72} & w_{73} & w_{74} & w_{75} & w_{76} & w_{77} & 0 & 0 \\ 0 & 0 & 0 & 0 & w_{71} & w_{72} & w_{73} & w_{74} & w_{75} & w_{76} & w_{77} & 0 \\ 0 & 0 & 0 & 0 & 0 & w_{71} & w_{72} & w_{73} & w_{74} & w_{75} & w_{76} & w_{77} \end{bmatrix}^{-1} \begin{pmatrix} \Delta f(D_1) \\ \Delta f(D_2) \\ \Delta f(D_3) \\ \Delta f(D_4) \\ \Delta f(D_5) \\ \Delta f(D_6) \\ \Delta f(D_7) \\ \Delta f(D_8) \\ \Delta f(D_9) \\ \Delta f(D_{10}) \\ \Delta f(D_{11}) \\ \Delta f(D_{12}) \end{pmatrix} \quad 3.21$$

Equation 3.15 and 3.21 assumes that the oscillation of the cantilever has weak perturbations and low noise is preserved for $\Delta f$ measurement for a constant peak-to-peak oscillation amplitude.[12] It was proposed that the measured $\Delta f$ can be smoothed to reduce the noise.[12]

## 3.2 Formulation based on inverse Laplace transformation

Equation 4.25 can be used to calculate the frequency shift for a FM-driven AFM under a tip-sample interaction and a small frequency shift:

$$\Delta f \approx -\frac{f_0^2}{kA} \int_0^{1/f_0} F_{intc}(d + A\cos(2\pi ft))\cos(2\pi ft)dt$$

$$= -\frac{2f_0^2}{kA} \int_0^{T/2} F_{intc}(d + A\cos(2\pi ft))\cos(2\pi ft)dt$$

By posing $u = \cos(\omega t)$ and assuming $f_0 \approx f$

$$\int_{-1}^{1} F_{intc}(D + A + Au)\frac{u}{\sqrt{1-u^2}}du \approx -\pi kA \frac{\Delta f}{f_0} \qquad 3.22$$

Laplace transform is applied to pursue:

$$F_{intc}(D) = \int_0^{\infty} G(\lambda)e^{-\lambda D}d\lambda \qquad 3.23$$

Substituting Equation 3.23 in Equation 3.22:

$$\int_{-1}^{1}\int_0^{\infty} G(\lambda)e^{-(D+A+Au)\lambda}d\lambda \frac{u}{\sqrt{1-u^2}}du = \int_0^{\infty}\int_{-1}^{1} G(\lambda)e^{-(D+A+Au)\lambda}\frac{u}{\sqrt{1-u^2}}dud\lambda$$

$$= \int_0^{\infty}\int_{-1}^{1} e^{-(A+Au)\lambda} \frac{u}{\sqrt{1-u^2}} du\, G(\lambda)e^{-D\lambda}d\lambda \qquad 3.24$$

$$= -\pi kA \frac{\Delta f}{f_0}$$

$$\int_0^{\infty} T(\lambda A)G(\lambda)e^{-D\lambda}d\lambda = -\pi kA \frac{\Delta f}{f_0} \qquad 3.25$$

Where:

$$T(\lambda A) = \int_{-1}^{1} e^{-(A+Au)\lambda} \frac{u}{\sqrt{1-u^2}} du \qquad 3.26$$

The modified Bessel function of the first kind of order one is:

$$I_1(x) = \frac{1}{\pi}\int_0^{\pi} e^{x\cos\theta}\cos(\theta)d\theta = \frac{1}{\pi}\int_{-1}^{1} \frac{e^{xu}u}{\sqrt{1-u^2}}du \qquad 3.27$$

So,

$$T(\lambda A) = e^{-A\lambda}\pi I_1(-A\lambda) = -e^{-A\lambda}\pi I_1(A\lambda) \qquad 3.28$$

Then, using the asymptotic limits of Equation 3.28 and a Padé approximant we get:[13]

$$T(\lambda A) \approx -\pi\frac{\lambda A}{2}\left(1 + \alpha\sqrt{\lambda A} + \sqrt{\frac{\pi}{2}}(\lambda A)^{3/2}\right)^{-1} \qquad 3.29$$

We choose $\alpha = 1/8$ as the coefficient for the second term. Introducing Equation 3.29 into Equation 3.25:

$$\int_0^\infty \left[\frac{\lambda A}{2}\left(1 + \frac{1}{8}\sqrt{\lambda A} + \sqrt{\frac{\pi}{2}}(\lambda A)^{3/2}\right)^{-1}\right]G(\lambda)e^{-D\lambda}d\lambda = kA\frac{\Delta f}{f_0} \qquad 3.30$$

The integral of the left term is the Laplace transform, and if written explicitly:

$$L\left\{\left[\frac{\lambda}{2}\left(1 + \frac{1}{8}\sqrt{\lambda A} + \sqrt{\frac{\pi}{2}}(\lambda A)^{3/2}\right)^{-1}\right]G(\lambda)\right\} = k\frac{\Delta f}{f_0} \qquad 3.31$$

Defining:

$$\Omega = \frac{\Delta f}{f_0} \qquad 3.32$$

$\Omega$ is dependent on $z$ via $\Delta f$. We take the inverse Laplace transform of Equation 3.31:

$$G(\lambda) = \frac{2k}{\lambda}\left(1 + \frac{1}{8}\sqrt{\lambda A} + \sqrt{\frac{\pi}{2}}(\lambda A)^{3/2}\right)L^{-1}\{\Omega\} \qquad 3.33$$

Employing Equation 3.23, we get:

$$F_{int c}(D) = L\left\{\frac{2k}{\lambda}\left(1 + \frac{1}{8}\sqrt{\lambda A} + \sqrt{\frac{\pi}{2}}(\lambda A)^{3/2}\right)L^{-1}\{\Omega\}\right\} \qquad 3.34$$

$$\frac{F_{int c}(D)}{2k} = L\left\{\lambda^{-1}L^{-1}\{\Omega\}\right\} + \frac{1}{8}L\left\{\lambda^{-1/2}L^{-1}\{\sqrt{A}\Omega\}\right\} + \sqrt{\frac{\pi}{2}}L\left\{\lambda^{1/2}L^{-1}\{A^{3/2}\Omega\}\right\} \qquad 3.35$$

We can transform this equation using the following results of fractional calculus theory: [14,15]

$$L\left\{\lambda^{-1}L^{-1}\{\Omega\}\right\} = I_-^1\left\{L\left\{L^{-1}\{\Omega\}\right\}\right\} = I_-^1\{\Omega\} \qquad 3.36$$

$$L\left\{\lambda^{-1/2}L^{-1}\{\Omega\}\right\} = I_-^{1/2}\left\{L\left\{L^{-1}\{\Omega\}\right\}\right\} = I_-^{1/2}\{\Omega\} \qquad 3.37$$

$$L\left\{\lambda^{1/2}L^{-1}\{\Omega\}\right\} = D_-^{1/2}\left\{L\left\{L^{-1}\{\Omega\}\right\}\right\} = D_-^{1/2}\{\Omega\} \qquad 3.38$$

$$I_-^{\alpha}\{\Omega(D)\} = \frac{1}{\Gamma(\alpha)} \int_D^{\infty} \frac{\Omega(z)dz}{(z-D)^{1-\alpha}} \qquad 3.39$$

$$D_-^{\alpha}\{\Omega(D)\} = \frac{(-1)^n}{\Gamma(n-\alpha)} \frac{d^n}{dD^n} \int_D^{\infty} \frac{\Omega(z)dz}{(z-D)^{\alpha-n+1}} \qquad 3.40$$

$\Gamma(\alpha)$ is the gamma function, and $\alpha > 0$ is any real positive number. $n = [\alpha]+1$, where $[\alpha]$ is the integer component of $\alpha$. Equation 3.39 where $\alpha = 1$ yields:

$$I_-^1\{\Omega(D)\} = \int_D^{\infty} \Omega(z)dz \qquad 3.41$$

Equation 3.39 where $\alpha = 1/2$ yields:

$$I_-^{1/2}\{\Omega(D)\} = \frac{1}{\sqrt{\pi}} \int_D^{\infty} \frac{\Omega(z)}{\sqrt{z-D}} dz \qquad 3.42$$

Finally, Equation 3.40 where $\alpha = 1/2$ yields:

$$D_-^{1/2}\{\Omega(D)\} = -\frac{1}{\sqrt{\pi}} \frac{d}{dD} \int_D^{\infty} \frac{\Omega(z)dz}{\sqrt{z-D}} \qquad 3.43$$

Inserting Equations 3.41, 3.42, and 3.43 into Equation 3.35, we get:

$$F_{intc}(D) = 2k \left[ \int_D^{\infty} \Omega(z)dz + \frac{1}{8\sqrt{\pi}} \int_D^{\infty} \frac{\sqrt{A}\Omega(z)}{\sqrt{z-D}} dz - \frac{1}{\sqrt{2}} \frac{d}{dD} \int_D^{\infty} \frac{A^{3/2}\Omega(z)dz}{\sqrt{z-D}} \right] \qquad 3.44$$

Where:

$$\Omega(z) = \frac{\Delta f}{f_0} \qquad 3.45$$

We can further develop the last term of Equation 3.44:

$$\int_D^{\infty} \frac{\Omega(z)dz}{\sqrt{z-D}} = 2\left[\Omega(z)\sqrt{z-D}\right]_D^{\infty} - 2\int_D^{\infty} \sqrt{z-D} \frac{d\Omega(z)}{dz} dz \qquad 3.46$$

As $z$ tends toward $\infty$, if $\Omega(z)$ tends faster toward 0 than $z^{1/2}$ toward $\infty$:

$$\int_D^{\infty} \frac{\Omega(z)dz}{\sqrt{z-D}} = -2\int_D^{\infty} \sqrt{z-D} \frac{d\Omega(z)}{dz} dz \qquad 3.47$$

With Leibniz integral rule:

$$\frac{d}{dD}\int_D^\infty \frac{\Omega(z)dz}{\sqrt{z-D}} = -2\frac{d}{dD}\left[\int_D^\infty \sqrt{z-D}\frac{d\Omega(z)}{dz}dz\right] = \int_D^\infty \frac{1}{\sqrt{z-D}}\frac{d\Omega(z)}{dz}dz \qquad 3.48$$

Inserting Equation 3.48 into Equation 3.44:

$$F_{intc}(D) = 2k\left[\int_D^\infty \Omega(z)dz + \frac{1}{8\sqrt{\pi}}\int_D^\infty \frac{\sqrt{A}\Omega(z)}{\sqrt{z-D}}dz - \int_D^\infty \frac{A^{3/2}}{\sqrt{2(z-D)}}\frac{d\Omega(z)}{dz}dz\right] \qquad 3.49$$

This is the most commonly used formulation to convert experimental measurements to tip-sample interaction forces. The basic assumption of this methodology is the well-posedness of the tip-sample interaction force with a constant oscillation amplitude regardless of the tip-sample separation. This Laplace transform-based formulation was later extended to non-conservative forces.[16] For non-conservative interaction forces, a driving force and a small frequency shift, we use Equation 5.29, using the variable $\Pi$ instead of $\Lambda$ for clarity:

$$\int_{-1}^1 \Pi(D+A+Au)\sqrt{1-u^2}du = \frac{k}{4Q_0 f_0} + \frac{kA_0}{4Q_0 A f_d}\sin(\beta) = \frac{\pi k}{2Q_0 \omega_0} + \frac{ka_d}{4Af_d}\sin(\beta)$$

Assuming that the damping coefficient is $b = \omega_0 m / Q_0$, the driving force amplitude is $F_0 = a_d k$ and $\beta = \pi/2$:

$$\int_{-1}^1 \Pi(D+A+Au)\sqrt{1-u^2}du = \frac{\pi k}{2Q_0 \omega_0} + \frac{ka_d}{4Af_d} = \frac{b\pi}{2} + \frac{TF_0}{4A} \qquad 3.50$$

$$\frac{2}{b\pi}\int_{-1}^1 \Pi(D+A+Au)\sqrt{1-u^2}du = 1 + \frac{F_0}{\omega_d bA} \qquad 3.51$$

$$\frac{2}{b\pi}\int_{-1}^1 \Delta\Pi(D+A+Au)\sqrt{1-u^2}du = \frac{1}{bA}\frac{(\omega_d \Delta F_0 - F_0 \Delta\omega)}{\omega_d^2} \qquad 3.52$$

If we insert Equation 4.44 with $\beta = \pi/2$ into Equation 4.45 with $g=0$:

$$mA(\omega_d)^2 \sin(\omega_d t) - \frac{(2\pi)^2 f_0 m A f_d}{Q_0}\cos(\omega_d t) - kA\sin(\omega_d t)$$
$$= F_{int}(z(t), \dot{z}(t)) + a_d k \cos(\omega_d t) \qquad 3.53$$

$F_{int}$ is very small compared to the driving force and thus equating cosine terms in Equation 3.53:

$$-\frac{\omega_0 m A \omega_d}{Q_0} = a_d k \qquad 3.54$$

$$-bA\omega_d = a_d k = F_0 \qquad 3.55$$

Then with Equation 3.52:

$$\frac{2}{b\pi}\int_{-1}^{1}\Delta\Pi(D+A+Au)\sqrt{1-u^2}\,du = \frac{1}{bA}\frac{(\omega_d\Delta F_0 - F_0\Delta\omega)}{\omega_d^2} = -\frac{\Delta F_0}{F_0} + \frac{\Delta\omega}{\omega_d} \qquad 3.56$$

$$\int_{-1}^{1}\Delta\Pi(D+A+Au)\sqrt{1-u^2}\,du = \frac{b\pi}{2}\left(-\frac{\Delta F_0}{F_0} + \frac{\Delta\omega}{\omega_d}\right) \qquad 3.57$$

Posing:

$$\int_{-1}^{1}\Delta\Pi(D+A+Au)\sqrt{1-u^2}\,du = \frac{b\pi}{2}\left(\frac{\Delta\omega}{\omega_d} - \frac{\Delta F_0}{F_0}\right) = \pi k\varepsilon \qquad 3.58$$

Where:

$$\varepsilon = \frac{b}{2k}\left(\frac{\Delta\omega}{\omega_d} - \frac{\Delta F_0}{F_0}\right) \qquad 3.59$$

Employing the function $B(x)$:

$$B(x) = 2\int_{x}^{\infty}\Delta\Gamma(x)\,dx \qquad 3.60$$

$$\frac{dB(x)}{dx} = 2\Delta\Gamma(x) \qquad 3.61$$

Inserting Equation 3.61 into the left term of Equation 3.58:

$$\int_{-1}^{1}\Delta\Gamma(D+A+Au)\sqrt{1-u^2}\,du = \frac{1}{2}\int_{-1}^{1}\frac{dB(D+A+Au)}{d(D+A+Au)}\sqrt{1-u^2}\,du \qquad 3.62$$

Further developing:

$$\frac{dB(D+A+Au)}{d(D+A+Au)} = \frac{dB(D+A+Au)}{du}\frac{du}{d(D+A+Au)} = \frac{1}{A}\frac{dB(D+A+Au)}{du} \qquad 3.63$$

$$\int_{-1}^{1}\Delta\Gamma(D+A+Au)\sqrt{1-u^2}\,du = \frac{1}{2A}\int_{-1}^{1}\frac{dB(D+A+Au)}{du}\sqrt{1-u^2}\,du \qquad 3.64$$

Integrating by parts:

$$\int_{-1}^{1}\Delta\Gamma(D+A+Au)\sqrt{1-u^2}\,du = \frac{1}{2A}\left[\left[B(D+A+Au)\sqrt{1-u^2}\right]_{-1}^{1} + \int_{-1}^{1}\frac{B(D+A+Au)u}{\sqrt{1-u^2}}\,du\right] \qquad 3.65$$

$$\int_{-1}^{1} \Delta\Gamma(D+A+Au)\sqrt{1-u^2}\,du = \frac{1}{2A}\int_{-1}^{1}\frac{B(D+A+Au)u}{\sqrt{1-u^2}}\,du \qquad 3.66$$

Using the Laplace transform of $M(\lambda)$:

$$B(D) = \int_0^\infty M(\lambda)e^{-\lambda D}\,d\lambda \qquad 3.67$$

So, the right term of Equation 3.66 becomes:

$$\begin{aligned}\frac{1}{2A}\int_{-1}^{1}\frac{B(D+A+Au)u}{\sqrt{1-u^2}}\,du &= \frac{1}{2A}\int_{-1}^{1}\int_{0}^{\infty}M(\lambda)e^{-\lambda D-\lambda A-\lambda Au}\,d\lambda\,\frac{u}{\sqrt{1-u^2}}\,du \\ &= \frac{1}{2A}\int_{0}^{\infty}\int_{-1}^{1}e^{-(A+Au)\lambda}\frac{u}{\sqrt{1-u^2}}\,du\,M(\lambda)e^{-\lambda D}\,d\lambda\end{aligned} \qquad 3.68$$

Using Equation 3.26, Equation (3.58) becomes:

$$\int_0^\infty T(\lambda A)M(\lambda)e^{-D\lambda}\,d\lambda = 2A\pi k\varepsilon \qquad 3.69$$

Again, using the asymptotic limits of $T(\lambda A)$ and a Padé approximant:

$$T(\lambda A) \approx -\pi\frac{\lambda A}{2}\left(1+\alpha\sqrt{\lambda A}+\sqrt{\frac{\pi}{2}}(\lambda A)^{3/2}\right)^{-1} \qquad 3.70$$

We choose $\alpha = 1/8$ as the coefficient for the second term. Inserting Equation 3.70 into Equation 3.69:

$$\int_0^\infty \left[\frac{\lambda}{2}\left(1+\frac{1}{8}\sqrt{\lambda A}+\sqrt{\frac{\pi}{2}}(\lambda A)^{3/2}\right)^{-1}\right]M(\lambda)e^{-D\lambda}\,d\lambda = -2k\varepsilon = \Theta b \qquad 3.71$$

Following the same development as for the conservative forces (i.e. starting from Equation 3.30):

$$B(D) = -2bL\left\{\frac{1}{\lambda}\left(1+\frac{1}{8}\sqrt{\lambda A}+\sqrt{\frac{\pi}{2}}(\lambda A)^{3/2}\right)L^{-1}\{\Theta\}\right\} \qquad 3.72$$

$$B(D) = -2b\left[I_-^1\{\Theta\}+\frac{1}{8}I_-^{1/2}\{\sqrt{A}\Theta\}+\sqrt{\frac{\pi}{2}}D_-^{1/2}\{A^{3/2}\Theta\}\right] \qquad 3.73$$

$$B(D) = -2b\left[\int_D^\infty \Theta(z)\,dz + \frac{1}{8\sqrt{\pi}}\int_D^\infty \frac{\sqrt{A}\Theta(z)}{\sqrt{z-D}}\,dz - \frac{1}{\sqrt{2}}\frac{d}{dD}\int_D^\infty \frac{A^{3/2}\Theta(z)\,dz}{\sqrt{z-D}}\right] \qquad 3.74$$

$$\Delta\Gamma(D) = b\frac{\partial}{\partial D}\left[\int_D^\infty \Theta(z)dz + \frac{1}{8\sqrt{\pi}}\int_D^\infty \frac{\sqrt{A}\Theta(z)}{\sqrt{z-D}}dz - \frac{1}{\sqrt{2}}\frac{d}{dD}\int_D^\infty \frac{A^{3/2}\Theta(z)dz}{\sqrt{z-D}}\right] \quad 3.75$$

$$\Delta\Gamma(D) = -b\frac{\partial}{\partial D}\int_D^\infty Y(z)dz - \frac{b}{8}\frac{\partial}{\partial D}\int_D^\infty \frac{\sqrt{A}\Theta(z)}{\sqrt{\pi(z-D)}}dz + b\frac{\partial^2}{\partial D^2}\int_D^\infty \frac{A^{3/2}\Theta(z)dz}{\sqrt{2(z-D)}} \quad 3.76$$

Using the same development as the one starting with Equation 5.29:

$$\Delta\Pi(D) = -b\frac{\partial}{\partial D}\int_D^\infty \Theta(z)dz - \frac{b}{8}\frac{\partial}{\partial D}\int_D^\infty \frac{\sqrt{A}\Theta(z)}{\sqrt{\pi(z-D)}}dz + b\frac{\partial^2}{\partial D^2}\int_D^\infty \frac{A^{3/2}\Theta(z)dz}{\sqrt{2(z-D)}} \quad 3.77$$

Where:

$$\Theta = \left(\frac{\Delta F_0}{F_0} - \frac{\Delta\omega}{\omega_d}\right) \quad 3.78$$

With Leibniz integral rule (see Equation 3.48):

$$\frac{d}{dD}\int_D^\infty \frac{\Theta(z)dz}{\sqrt{z-D}} = \int_D^\infty \frac{1}{\sqrt{z-D}}\frac{d\Theta(z)}{dz}dz \quad 3.79$$

$$\Delta\Pi(D) = -b\frac{\partial}{\partial D}\int_D^\infty \Theta(z)dz - \frac{b}{8}\frac{\partial}{\partial D}\int_D^\infty \frac{\sqrt{A}\Theta(z)}{\sqrt{\pi(z-D)}}dz$$

$$+ b\frac{\partial}{\partial D}\int_D^\infty \frac{A^{3/2}}{\sqrt{2(z-D)}}\frac{d\Theta(z)}{dz}dz \quad 3.80$$

**Section 4 Representation of dynamic atomic force microscopy with differential equations**

**4.1 Case with no tip-sample interaction force and no $Q$-control**

The general differential equation for AFM measurements is:[17,18]

$$m\ddot{z}(t) + \frac{2\pi f_0 m}{Q_0}\dot{z}(t) + k(z(t) - d) + gkz(t - t_0)$$
$$= F_{int}(z(t), \dot{z}(t)) + a_d k \cos(2\pi f_d t) \qquad 4.1$$

If $F_{int} = 0$ and $g = 0$ we have the standard damped harmonic oscillator's equation:

$$m\ddot{z}(t) + \frac{2\pi f_0 m}{Q_0}\dot{z}(t) + k(z(t) - d) = a_d k \cos(2\pi f_d t) \qquad 4.2$$

The homogeneous solution is:[19]

$$z_h(t) = Ce^{-\alpha t}\cos(\omega_h t - \delta) \qquad 4.3$$

It becomes negligible for large $t$ values.

Proposing the following particular solution:

$$z_p(t) = d + K_1 \cos(\omega_d t) + K_2 \sin(\omega_d t) \qquad 4.4$$

Inserting Equation 4.4 in Equation 4.2:

$$-m\omega_d^2 K_1 \cos(\omega_d t) - m\omega_d^2 K_2 \sin(\omega_d t) - \omega_d \frac{2\pi f_0 m}{Q_0}K_1 \sin(\omega_d t) + \omega_d \frac{2\pi f_0 m}{Q_0}K_2 \cos(\omega_d t)$$
$$+ K_1 k \cos(\omega_d t) + kK_2 \sin(\omega_d t) = a_d k \cos(\omega_d t) \qquad 4.5$$

Thus:

$$-m\omega_d^2 K_1 + \omega_d \frac{2\pi f_0 m}{Q_0}K_2 + K_1 k = a_d k$$
$$-m\omega_d^2 K_2 - \omega_d \frac{2\pi f_0 m}{Q_0}K_1 + K_2 k = 0 \qquad 4.6$$

With $\omega_0^2 = k/m$

$$K_1 = \frac{ma_d k(\omega_0^2 - \omega_d^2)}{m^2(\omega_0^2 - \omega_d^2)^2 + \omega_d^2\left(\frac{2\pi f_0 m}{Q_0}\right)^2} \qquad 4.7$$

$$K_2 = \frac{\omega_d a_d k \left(\dfrac{2\pi f_0 m}{Q_0}\right)}{m^2(\omega_0^2 - \omega_d^2)^2 + \omega_d^2 \left(\dfrac{2\pi f_0 m}{Q_0}\right)^2} \qquad 4.8$$

$$z(t) = z_p(t) = d + K_1 \cos(\omega_d t) + K_2 \sin(\omega_d t)$$
$$= d + \gamma \cos(\beta)\cos(\omega_d t) + \gamma \sin(\beta)\sin(\omega_d t) \qquad 4.9$$

$$K_1 = \gamma \cos(\beta), K_2 = \gamma \sin(\beta) \qquad 4.10$$

The familiar solution is achieved:

$$z(t) = d + \gamma \cos(\omega_d t - \beta) \qquad 4.11$$

$$\gamma = \sqrt{K_1^2 + K_2^2} = \frac{a_d k}{\sqrt{m^2\left(\omega_0^2 - \omega_d^2\right)^2 + \left(\dfrac{2\pi f_0 m}{Q_0}\right)^2 \omega_d^2}} = \frac{a_d}{\sqrt{\left(1 - \dfrac{f_d^2}{f_0^2}\right)^2 + \left(\dfrac{f_d}{f_0}\right)^2 \dfrac{1}{Q_0}}} \qquad 4.12$$

$$\beta = \tan^{-1}\left(\frac{K_2}{K_1}\right) = \frac{\omega_0}{Q_0}\frac{\omega_d}{\left(\omega_0^2 - \omega_d^2\right)} = \frac{1}{Q_0}\frac{f_d/f_0}{\left(1 - \dfrac{f_d^2}{f_0^2}\right)} \qquad 4.13$$

## 4.2 Case with Q-control but without an external excitation

Without external excitation but with gain and tip-sample interaction, we obtain the differential equation describing the FM-AFM. With constant amplitude mode:

$$m\ddot{z}(t) + \frac{2\pi f_0 m}{Q_0}\dot{z}(t) + k(z(t)-d) = F_{int}(z(t),\dot{z}(t)) + gkz(t-t_0) \qquad 4.14$$

We make the ansatz:

$$z(t \gg 0) = d + A\cos(2\pi ft) \qquad 4.15$$

Then:

$$-mA(2\pi f)^2 \cos(2\pi ft) - \frac{(2\pi)^2 f_0 mAf}{Q_0}\sin(2\pi ft) + kA\cos(2\pi ft)$$
$$= F_{int}(z(t),\dot{z}(t)) + gkA\cos(2\pi f(t-t_0)) + gkd \qquad 4.16$$

Multiplying both sides by $\cos(2\pi ft)$ and integrating over a full period $T$ we obtain:

$$-mA(2\pi f)^2 \frac{T}{2} + kA\frac{T}{2} = \int_0^T F_{int}(z(t),\dot{z}(t))\cos(2\pi ft)dt + \frac{gkAT}{2}\cos(2\pi ft_0) \qquad 4.17$$

Since $\omega_0^2 = k/m$ and $f = 1/T$:

$$g\cos(2\pi ft_0) = \frac{f_0^2 - f^2}{f_0^2} - \frac{2f}{kA}\int_0^T F_{int}(z(t),\dot{z}(t))\cos(2\pi ft)dt \qquad 4.18$$

Multiplying both sides by $\sin(2\pi ft)$ and integrating over a full period $T$ we obtain:

$$g\sin(2\pi ft_0) = \frac{-f}{f_0 Q_0} - \frac{2f}{kA}\int_0^T F_{int}(z(t),\dot{z}(t))\sin(2\pi ft)dt \qquad 4.19$$

Far from the sample $F_{int} \approx 0$:

$$g\cos(2\pi ft_0) = \frac{f_0^2 - f^2}{f_0^2} \qquad 4.20$$

$$g\sin(2\pi ft_0) = \frac{-f}{f_0 Q_0} \qquad 4.21$$

If we assume that the frequency shift is small and $t_0 = \pi/2$ then $\sin(2\pi ft_0) \approx \pm 1$, $\cos(2\pi ft_0) \approx 0$ and $f \approx f_0$.

$$\frac{f_0^2 - f^2}{f_0^2} = \frac{(f_0-f)(f_0+f)}{f_0^2} \approx -\frac{2f_0 \Delta f}{f_0^2} \qquad 4.22$$

$$-\frac{2f_0}{kA}\int_0^T F_{int}(z(t),\dot{z}(t))\cos(2\pi ft)dt \approx \frac{2f_0\Delta f}{f_0^2} \qquad 4.23$$

$$g = -\frac{1}{Q_0} - \frac{2f_0}{kA}\int_0^{1/f_0} F_{int}(z(t),\dot{z}(t))\sin(2\pi ft)dt \qquad 4.24$$

$$\Delta f \approx -\frac{f_0^2}{kA}\int_0^{1/f_0} F_{int}(z(t),\dot{z}(t))\cos(2\pi ft)dt \qquad 4.25$$

## 4.3 Case with external excitation and far from the sample

In this case the tip-sample interaction force is negligible [20]:

$$m\ddot{z}(t) + \frac{2\pi f_0 m}{Q_0}\dot{z}(t) + k(z(t) - d) + gkz(t - t_0) = a_d k \cos(2\pi f_d t) \qquad 4.26$$

In this case, ansatz is

$$z(t \gg 0) = d + A\cos(\omega_d t + \beta) = d + A\cos(\omega_d t)\cos(\beta) - A\sin(\omega_d t)\sin(\beta) \qquad 4.27$$

$$z(t \gg 0) = d + \eta\cos(\omega_d t) + \mu\sin(\omega_d t) \qquad 4.28$$

Inserting Equation 4.28 in Equation 4.26:

$$-\eta\omega_d^2 \cos(\omega_d t) - \mu\omega_d^2 \sin(\omega_d t) - \frac{\omega_0 \omega_d}{Q_0}\eta\sin(\omega_d t) + \frac{\omega_0 \omega_d}{Q_0}\mu\cos(\omega_d t)$$

$$+\omega_0^2 \eta\cos(\omega_d t) + \omega_0^2 \mu\sin(\omega_d t) + g\frac{k}{m}d + g\frac{k}{m}\eta\cos(\omega_d(t - t_0)) \qquad 4.29$$

$$+g\frac{k}{m}\mu\sin(\omega_d(t - t_0)) = \frac{a_d k}{m}\cos(\omega_d t)$$

Multiplying by $\cos(\omega_d t)$ and integrating over a period T:

$$-\eta\omega_d^2 + \frac{\omega_0 \omega_d}{Q_0}\mu + \omega_0^2 \eta + g\frac{k}{m}\eta\cos(\omega_d t_0) - g\frac{k}{m}\mu\sin(\omega_d t_0) = \frac{a_d k}{m} \qquad 4.30$$

Multiplying by $\sin(\omega_d t)$ and integrating over a period T:

$$-\mu\omega_d^2 - \frac{\omega_0 \omega_d}{Q_0}\eta + \omega_0^2 \mu + g\frac{c_z}{m}\eta\sin(\omega_d t_0) + g\frac{k}{m}\mu\cos(\omega_d t_0) = 0 \qquad 4.31$$

We find η using Equation 4.30 and Equation 4.31

$$\eta = \frac{\frac{a_d k}{m} - \mu\left(\frac{\omega_0 \omega_d}{Q_0} - g\frac{k}{m}\sin(\omega_d t_0)\right)}{\omega_0^2 - \omega_d^2 + g\frac{k}{m}\cos(\omega_d t_0)} \qquad 4.32$$

Inserting η in Equation 4.31:

$$\mu\omega_d^4 - \mu\omega_0^2\omega_d^2 - \mu\omega_d^2\frac{k}{m}g\cos(\omega_d t_0) - \frac{\omega_0\omega_d a_d k}{mQ_0} + \frac{\mu\omega_0^2\omega_d^2}{Q_0^2}$$

$$-\mu\omega_0\omega_d\frac{k}{Q_0 m}g\sin(\omega_d t_0) - \mu\omega_0^2\omega_d^2 + \mu\omega_0^4 + \mu\omega_0^2\frac{k}{m}g\cos(\omega_d t_0)$$

$$+\frac{a_d k^2}{m^2}g\sin(\omega_d t_0) - \mu\omega_0\omega_d\frac{k}{Q_0 m}g\sin(\omega_d t_0) + \mu\frac{k^2}{m^2}g^2\sin^2(\omega_d t_0)$$

$$-\mu\omega_d^2\frac{k}{m}g\cos(\omega_d t_0) + \mu\frac{k}{m}\omega_0^2 g\cos(\omega_d t_0) + \mu\frac{k^2}{m^2}g^2\cos^2(\omega_d t_0) = 0$$

4.33

$$\mu\left(\omega_0^2 - \omega_d^2\right)^2 + 2\mu g\frac{k}{m}\left(\omega_0^2 - \omega_d^2\right)\cos(\omega_d t_0) + \mu\frac{k^2}{m^2}g^2\cos^2(\omega_d t_0)$$

$$+\mu\frac{k^2}{m^2}g^2\sin^2(\omega_d t_0) + \frac{\mu\omega_0^2\omega_d^2}{Q_0^2} - \frac{2g\mu\omega_0\omega_d k}{Q_0 m}\sin(\omega_d t_0)$$

$$=\frac{\omega_0\omega_d a_d k}{mQ_0} - \frac{a_d k^2}{m^2}g\sin(\omega_d t_0)$$

4.34

$$\mu = \frac{\dfrac{\omega_0\omega_d a_d k}{mQ_0} - \dfrac{a_d k^2}{m^2}g\sin(\omega_d t_0)}{\left[\left(\omega_0^2 - \omega_d^2\right) + \dfrac{k}{m}g\cos(\omega_d t_0)\right]^2 + \left[\dfrac{\omega_0\omega_d}{Q_0} - \dfrac{k}{m}g\sin(\omega_d t_0)\right]^2}$$

4.35

$$\eta = \frac{\dfrac{a_d k}{m}\left(\omega_0^2 - \omega_d^2 + \dfrac{k}{m}g\cos(\omega_d t_0)\right)^2}{\left[\omega_0^2 - \omega_d^2 + \dfrac{k}{m}g\cos(\omega_d t_0)\right]\left[\left[\left(\omega_0^2 - \omega_d^2\right) + \dfrac{k}{m}g\cos(\omega_d t_0)\right]^2 + \left[\dfrac{\omega_0\omega_d}{Q_0} - \dfrac{k}{m}g\sin(\omega_d t_0)\right]^2\right]}$$

4.36

$$A = \sqrt{\eta^2 + \mu^2}$$

4.37

$$A^2 = \frac{a_d^2 k^2}{m^2} \frac{\left[\left[\omega_0^2 - \omega_d^2 + \frac{k}{m}g\cos(\omega_d t_0)\right]^4 + \left[\frac{\omega_0 \omega_d}{Q_0} - \frac{k}{m}g\sin(\omega_d t_0)\right]^2 \left[(\omega_0^2 - \omega_d^2) + \frac{k}{m}g\cos(\omega_d t_0)\right]^2\right]}{\left[\omega_0^2 - \omega_d^2 + \frac{k}{m}g\cos(\omega_d t_0)\right]^2 \left[\left[(\omega_0^2 - \omega_d^2) + \frac{k}{m}g\cos(\omega_d t_0)\right]^2 + \left[\frac{\omega_0 \omega_d}{Q_0} - \frac{k}{m}g\sin(\omega_d t_0)\right]^2\right]^2} \quad 4.38$$

$$A = \frac{a_d}{\sqrt{\left[\left(1 - \frac{f_d^2}{f_0^2}\right) + g\cos(\omega_d t_0)\right]^2 + \left[\frac{f_d}{f_0 Q_0} - g\sin(\omega_d t_0)\right]^2}} \quad 4.39$$

$$\tan(\beta) = \frac{\mu}{\eta} = \frac{\frac{f_d}{Q_0 f_0} - g\sin(\omega_d t_0)}{1 - \frac{f_d^2}{f_0^2} + g\cos(\omega_d t_0)} \quad 4.40$$

If we define:

$$Q_{eff}(g, t_0) = \frac{1}{1/Q_0 - g\sin(\omega_d t_0)} \quad 4.41$$

The free oscillation amplitude at resonance and $t_0 = f_0/4$ is:

$$A_0 = a_d Q_{eff} \quad 4.42$$

S53

## 4.4 General form of differential equation and the derivation of large amplitude term for amplitude modulated atomic force microscopy

Finally, if we add the tip-sample interaction force we must solve the full differential equation:[17,21]

$$m\ddot{z}(t) + \frac{2\pi f_0 m}{Q_0}\dot{z}(t) + k(z(t)-d) + gkz(t-t_0)$$
$$= F_{int}(z(t),\dot{z}(t)) + a_d k \cos(2\pi f_d t) \quad\quad 4.43$$

Again ansatz is:

$$z(t \gg 0) = d + A\cos(\omega_d t + \beta) \quad\quad 4.44$$

Then inserting Equation 4.44 into Equation 4.43:

$$-mA(\omega_d)^2 \cos(\omega_d t + \beta) - \frac{(2\pi)^2 f_0 m A f_d}{Q_0}\sin(\omega_d t + \beta) + kA\cos(\omega_d t + \beta)$$
$$+ gkA\cos(\omega_d(t-t_0)+\beta) + gkd = F_{int}(z(t),\dot{z}(t)) + a_d k \cos(\omega_d t) \quad\quad 4.45$$

Multiplying (4.45) by $\cos(\omega_d t + \beta)$ and integrating over a period T we find:

$$-mA(\omega_d)^2 \frac{T}{2} + kA\frac{T}{2} + gkA\frac{T}{2}\cos(\omega_d t_0)$$
$$= \int_0^{1/f_d} F_{int}(z(t),\dot{z}(t))\cos(\omega_d t + \beta)dt + a_d k \frac{T}{2}\cos(\beta) \quad\quad 4.46$$

$$\frac{f_0^2 - f_d^2}{f_0^2} + g\cos(\omega_d t_0) = \frac{2f_d}{Ak}\int_0^{1/f_d} F_{int}(z(t),\dot{z}(t))\cos(\omega_d t + \beta)dt + \frac{a_d}{A}\cos(\beta) \quad\quad 4.47$$

Multiplying Equation 4.45 by $\sin(\omega_d t + \beta)$ and integrating over a period T we find:

$$-\frac{f_d}{Q_0 f_0} + g\sin(\omega_d t_0) = \frac{2f_d}{Ak}\int_0^{1/f_d} F_{int}(z(t),\dot{z}(t))\sin(\omega_d t + \beta)dt + \frac{a_d}{A}\sin(\beta) \quad\quad 4.48$$

In all cases, we can decompose $F_{int}$ into an even and an odd part.[16] The even part is such that the force on retraction is equal to the one on approach. For the odd part, the force on retraction is equal in magnitude but of opposite in sign to the one on approach.

$$F_{even} = \frac{F_\downarrow + F_\uparrow}{2} \quad\quad 4.49$$

$$F_{odd} = \frac{F_\downarrow - F_\uparrow}{2} \quad\quad 4.50$$

Since the position vector of the tip changes sign when approaching versus retracting, the work done by $F_{even}$ over a full cycle will be zero. We thus associate $F_{even}$ to the conservative component of the force per cycle. $F_{odd}$ is associated to the non-conservative component of the force per cycle.

Further developing, we obtain by inserting Equation 4.49 and Equation 4.50 into Equation 4.47 and Equation 4.48:

$$\frac{f_0^2 - f_d^2}{f_0^2} + g\cos(\omega_d t_0) = \frac{2f_d}{Ak} \int_0^{1/2f_d} (F_\downarrow + F_\uparrow)\cos(\omega_d t + \beta)dt + \frac{a_d}{A}\cos(\beta) \qquad 4.51$$

$$-\frac{f_d}{Q_0 f_0} + g\sin(\omega_d t_0) = \frac{2f_d}{Ak} \int_0^{1/2f_d} (F_\downarrow - F_\uparrow)\sin(\omega_d t + \beta)dt + \frac{a_d}{A}\sin(\beta) \qquad 4.52$$

Using ansatz (i.e. Equation 4.44) with:

$$d = D + A \qquad 4.53$$

We find:

$$\cos(\omega_d t + \beta) = \frac{z - D - A}{A} \qquad 4.54$$

$$\sin(\omega_d t + \beta) = \frac{\sqrt{A^2 - (z - D - A)^2}}{A} \qquad 4.55$$

$$dt = \frac{-dz}{A\omega_d \sin(\omega_d t + \beta)} = \frac{-dz}{\omega_d \sqrt{A^2 - (z - D - A)^2}} \qquad 4.56$$

If we define

$$I_+(d, A) = \frac{1}{\pi A^2 k} \int_{d-A}^{d+A} (F_\downarrow + F_\uparrow) \frac{(z - D - A)dz}{\sqrt{A^2 - (z - D - A)^2}} \qquad 4.57$$

and,

$$I_-(d, A) = \frac{f_d}{A^2 \pi k} \int_{d-A}^{d+A} (F_\downarrow - F_\uparrow) dz \qquad 4.58$$

Inserting $I_+(d, A)$ and $I_-(d, A)$ into Equation 4.51 and Equation 4.52:

$$\frac{f_0^2 - f_d^2}{f_0^2} + g\cos(\omega_d t_0) = I_+(d, A) + \frac{a_d}{A}\cos(\beta) \qquad 4.59$$

$$-\frac{f_d}{Q_0 f_0} + g\sin(\omega_d t_0) = I_-(d, A) + \frac{a_d}{A}\sin(\beta) \qquad 4.60$$

Finally, by solving Equation 4.59 and Equation 4.60 we can find $A$ and $\beta$:

$$A = \frac{a_d}{\sqrt{\left[1-\frac{f_d^2}{f_0^2}+g\cos(\omega_d t_0)-I_+(d,A)\right]^2 + \left[\frac{f_d}{Q_0 f_0}-g\sin(\omega_d t_0)+I_-(d,A)\right]^2}} \quad 4.61$$

Assuming a negative phase shift $\beta$ :

$$\tan(\beta) = \frac{\frac{f_d}{Q_0 f_0} - g\sin(\omega_d t_0) + I_-(d,A)}{1-\frac{f_d^2}{f_0^2}+g\cos(\omega_d t_0)-I_+(d,A)} \quad 4.62$$

In the particular case of strictly conservative forces (i.e., $I_-(d,A) = 0$) and $t_0 = 1/4f_d$, the equation for $A$ can be simplified.

$$A = \frac{a_d}{\sqrt{\left[1-\frac{f_d^2}{f_0^2}-I_+(d,A)\right]^2 + \left[\frac{f_d}{Q_0 f_0} - \frac{1}{Q_0} + \frac{1}{Q_{eff}}\right]^2}} \quad 4.63$$

$$\frac{a_d^2}{A^2} = 1 - \frac{f_d^2}{f_0^2} - I_+(d,A) - \frac{f_d^2}{f_0^2} + \left(\frac{f_d^2}{f_0^2}\right)^2 + \frac{f_d^2}{f_0^2} I_+(d,A) - I_+(d,A)$$

$$+ I_+(d,A)\frac{f_d^2}{f_0^2} + I_+^2(d,A) + \frac{1}{Q_0^2}\frac{f_d^2}{f_0^2} + \frac{f_d}{Q_{eff}Q_0 f_0} - \frac{1}{Q_0^2}\frac{f_d}{f_0} + \frac{f_d}{Q_{eff}Q_0 f_0} \quad 4.64$$

$$+ \frac{1}{Q_{eff}^2} - \frac{1}{Q_{eff}Q_0} - \frac{1}{Q_0^2}\frac{f_d}{f_0} - \frac{1}{Q_{eff}Q_0} + \frac{1}{Q_0^2}$$

$$\left(\frac{f_d^2}{f_0^2}\right)^2 + \frac{f_d^2}{f_0^2}(-2 + 2I_+(d,A) + \frac{1}{Q_0^2}) + \frac{f_d}{f_0}(\frac{2}{Q_{eff}Q_0} - \frac{2}{Q_0^2})$$

$$+ 1 - 2I_+(d,A) + I_+^2(d,A) + (\frac{1}{Q_{eff}} - \frac{1}{Q_0})^2 - \frac{a_d^2}{A^2} = 0 \quad 4.65$$

If $g$ is small then $Q_{eff} \approx Q_0$

$$\left(\frac{f_d^2}{f_0^2}\right)^2 + \frac{f_d^2}{f_0^2}(-2 + 2I_+(d,A) + \frac{1}{Q_{eff}^2}) + 1 - 2I_+(d,A) + I_+^2(d,A) - \frac{a_d^2}{A^2} = 0 \quad 4.66$$

$$\frac{f_d^2}{f_0^2} = 1 - I_+(d,A) - \frac{1}{2Q_{eff}^2} \pm \sqrt{-\frac{1}{Q_{eff}^2} + \frac{I_+(d,A)}{Q_{eff}^2} + \frac{1}{4Q_{eff}^4} + \frac{a_d^2}{A^2}} \quad 4.67$$

S56

If $f_d = f_0$ and $t_0 = 1/4f_d$, then from Equation 4.63 and Equation 4.42:

$$a_d^2 = A^2 \left( [I_+(d,A)]^2 + \left[\frac{1}{Q_0} - g\right]^2 \right) = A^2 \left( [I_+(d,A)]^2 + \frac{1}{Q_{eff}^2} \right) \quad 4.68$$

$$A_0 = A\sqrt{[Q_{eff} I_+(d,A)]^2 + 1} \quad 4.69$$

If the amplitude A is much larger than the interaction force, we can expand at $z \approx D$:

$$z - D - A \approx -A \quad 4.70$$

$$-(z-D-A)^2 = -A^2\left(1 - \frac{(z-D)}{A}\right)^2 \approx -A^2\left(1 - \frac{2(z-D)}{A}\right) \quad 4.71$$

$$\frac{z-D-A}{\sqrt{A^2 - (z-D-A)^2}} \approx -\sqrt{\frac{A}{2(z-D)}} \quad 4.72$$

Inserting Equation 4.72 into equation 4.57

$$I_+(d,A) \approx -\frac{1}{\sqrt{2\pi} A^{3/2} k} \int_D^{D+2A} (F_\downarrow + F_\uparrow) \frac{dz}{\sqrt{z-D}} = -\frac{\sqrt{2}}{\pi A^{3/2} k} \int_D^{D+2A} F_{int}(z) \frac{dz}{\sqrt{z-D}} \quad 4.73$$

Assuming $g = 0$ and inserting the above value of $I_+(d,A)$ in Equation 4.59, we obtain:

$$\frac{A^{3/2} k}{\sqrt{2}} \left[ \frac{f_0^2 - f_d^2}{f_0^2} - \frac{a_d}{A} \cos(\beta) \right] = \sigma(D) = \frac{1}{\pi} \int_D^{D+2A} F_{int}(z) \frac{dz}{\sqrt{z-D}} \quad 4.74$$

Using the inverse Abel transform[8]:

$$\frac{1}{\pi} \int_D^{D+2A} F_{int}(z) \frac{dz}{\sqrt{z-D}} \approx \frac{1}{\pi} \int_D^\infty F_{int}(z) \frac{dz}{\sqrt{z-D}} = -\frac{1}{\pi} \int_D^\infty \frac{dU_{int}(z)}{dz} \frac{dz}{\sqrt{z-D}} = \sigma(D) \quad 4.75$$

$$F\{\sigma(D)\} = -\frac{1}{\pi} F\left[ (K * \frac{dU_{int}}{dz})(D) \right] = F\{K(D)\} F\left\{ \frac{dU_{int}(z)}{dz} \right\} \quad 4.76$$

$$F\{U_{int}(z)\} = F\{K(z)\} F\{\sigma(D)\} \quad 4.77$$

$$U_{int}(z) = (K * \sigma)(z) = \int_{-\infty}^\infty K(z-D)\sigma(D)dD = \int_z^\infty \frac{\sigma(D)}{(D-z)^{1/2}} dD \quad 4.78$$

$$F_{int}(z) = -\frac{\partial}{\partial z} \int_z^\infty \frac{\sigma(D)}{(D-z)^{1/2}} dD \quad 4.79$$

S57

$$F_{int}(D) = -\frac{\partial}{\partial D}\int_D^\infty \frac{\sigma(z)}{(z-D)^{1/2}}dz \approx -\frac{\partial}{\partial D}\int_D^{D+2A} \frac{\sigma(z)}{(z-D)^{1/2}}dz \qquad 4.80$$

## Section 5 Formulations for amplitude modulation atomic force microscopy

### 5.1 Formulation based on the inverse Laplace transform

These formulations were originally developed for FM-AFM[13,14,16]. Moreover, similar formulations were later proposed with assumptions employed for FM-AFM[22]. Now, let's generalize and solve without any assumption on the magnitude of the amplitude[22,23].

We go back to the Equation 4.47:

$$\frac{f_0^2 - f_d^2}{f_0^2} + g\cos(\omega_d t_0) = \frac{2f_d}{Ak}\int_0^{1/f_d} F_{int}(z(t),\dot{z}(t))\cos(\omega_d t + \beta)dt + \frac{a_d}{A}\cos(\beta)$$

With $g = 0$, conservative forces and using $A_0 = a_d Q_0$:

$$(\pi Ak)\frac{\omega_0^2 - \omega_d^2}{\omega_0^2} = \omega_d \int_0^{1/f_d} F_{intc}(d + A\cos(\omega_d t + \beta))\cos(\omega_d t + \beta)dt$$

$$+ (\pi k)\frac{A_0}{Q_0}\cos(\beta) \qquad 5.1$$

We pose $\iota = \omega_d t + \beta$, $d\iota = \omega_d dt$ and assuming that the integrand is even

$$(\pi Ak)\frac{\omega_0^2 - \omega_d^2}{\omega_0^2} = 2\int_0^\pi F_{intc}(d + A\cos\iota)\cos(\iota)d\iota + (\pi k)\frac{A_0}{Q_0}\cos(\beta) \qquad 5.2$$

Let's introduce the variable change $u=\cos\iota$, $du=-d\iota(1-u^2)^{1/2}$

$$\int_{-1}^1 F_{intc}(D+A+Au)\frac{u}{\sqrt{1-u^2}}du = \left(\frac{\pi Ak}{2}\right)\left(\frac{\omega_0^2 - \omega_d^2}{\omega_0^2}\right) - \frac{A_0\pi k}{2Q_0}\cos(\beta) \qquad 5.3$$

To further pursue, we use the Laplace transform:

$$F_{intc}(D) = \int_0^\infty G(\lambda)e^{-\lambda D}d\lambda \qquad 5.4$$

Inserting equation (5.4) in (5.3):

$$\int_{-1}^1\int_0^\infty G(\lambda)e^{-(D+A+Au)\lambda}d\lambda\frac{u}{\sqrt{1-u^2}}du = \int_0^\infty\int_{-1}^1 G(\lambda)e^{-(D+A+Au)\lambda}\frac{u}{\sqrt{1-u^2}}dud\lambda$$

$$= \int_0^\infty\int_{-1}^1 e^{-(A+Au)\lambda}\frac{u}{\sqrt{1-u^2}}du\, G(\lambda)e^{-D\lambda}d\lambda \qquad 5.5$$

$$= \left(\frac{\pi Ak}{2}\right)\left(\frac{\omega_0^2 - \omega_d^2}{\omega_0^2}\right) - \frac{A_0\pi k}{2Q_0}\cos(\beta)$$

$$\int_0^\infty T(\lambda A)G(\lambda)e^{-D\lambda}d\lambda = \left(\frac{\pi Ak}{2}\right)\left(\frac{\omega_0^2-\omega_d^2}{\omega_0^2}\right) - \frac{A_0\pi k}{2Q_0}\cos(\beta) \qquad 5.6$$

Where:

$$T(\lambda A) = \int_{-1}^{1} e^{-(A+Au)\lambda}\frac{u}{\sqrt{1-u^2}}du \qquad 5.7$$

The modified Bessel function of the first kind of order one is:

$$I_1(x) = \frac{1}{\pi}\int_0^\pi e^{x\cos\theta}\cos(\theta)d\theta = \frac{1}{\pi}\int_{-1}^{1}\frac{e^{xu}u}{\sqrt{1-u^2}}du \qquad 5.8$$

So

$$T(\lambda A) = e^{-A\lambda}\pi I_1(-A\lambda) = -e^{-A\lambda}\pi I_1(A\lambda) \qquad 5.9$$

Then, using the asymptotic limits of Equation 5.9 and a Padé approximant we get:[13]

$$T(\lambda A) \approx -\pi\frac{\lambda A}{2}\left(1+\alpha\sqrt{\lambda A}+\sqrt{\frac{\pi}{2}}(\lambda A)^{3/2}\right)^{-1} \qquad 5.10$$

We choose $\alpha=1/8$ as the coefficient for the second term. Introducing Equation 5.10 into Equation 5.6:

$$\int_0^\infty\left[\frac{\lambda A}{2}\left(1+\frac{1}{8}\sqrt{\lambda A}+\sqrt{\frac{\pi}{2}}(\lambda A)^{3/2}\right)^{-1}\right]G(\lambda)e^{-D\lambda}d\lambda$$
$$= \frac{A_0 k}{2Q_0}\cos(\beta) - \left(\frac{Ak}{2}\right)\left(\frac{\omega_0^2-\omega_d^2}{\omega_0^2}\right) \qquad 5.11$$

The integral of the left term is the Laplace transform. Let's write it explicitly:

$$L\left\{\left[\frac{\lambda}{2}\left(1+\frac{1}{8}\sqrt{\lambda A}+\sqrt{\frac{\pi}{2}}(\lambda A)^{3/2}\right)^{-1}\right]G(\lambda)\right\} = \frac{A_0 k}{2AQ_0}\cos(\beta) - \left(\frac{k}{2}\right)\frac{\omega_0^2-\omega_d^2}{\omega_0^2} \qquad 5.12$$

Defining:

$$X = \frac{A_0}{2AQ_0}\cos(\beta) - \frac{\omega_0^2-\omega_d^2}{2\omega_0^2} \qquad 5.13$$

$X$ is dependent on $z$ via $A$ and $\phi$. We take the inverse Laplace transform of Equation 5.12:

$$G(\lambda) = \frac{2k}{\lambda}\left(1+\frac{1}{8}\sqrt{\lambda A}+\sqrt{\frac{\pi}{2}}(\lambda A)^{3/2}\right)L^{-1}\{X\} \qquad 5.14$$

Using Equation 5.4, we get:

$$F_{int\,c}(D) = L\left\{\frac{2k}{\lambda}\left(1 + \frac{1}{8}\sqrt{\lambda A} + \sqrt{\frac{\pi}{2}}(\lambda A)^{3/2}\right)L^{-1}\{X\}\right\} \qquad 5.15$$

$$\frac{F_{int\,c}(D)}{2k} = L\left\{\lambda^{-1}L^{-1}\{X\}\right\} + \frac{1}{8}L\left\{\lambda^{-1/2}L^{-1}\{\sqrt{A}X\}\right\} + \sqrt{\frac{\pi}{2}}L\left\{\lambda^{1/2}L^{-1}\{A^{3/2}X\}\right\} \qquad 5.16$$

We can transform this equation using the following results of fractional calculus theory:[14,15]

$$L\left\{\lambda^{-1}L^{-1}\{X\}\right\} = I_-^1\left\{L\{L^{-1}\{X\}\}\right\} = I_-^1\{X\} \qquad 5.17$$

$$L\left\{\lambda^{-1/2}L^{-1}\{X\}\right\} = I_-^{1/2}\left\{L\{L^{-1}\{X\}\}\right\} = I_-^{1/2}\{X\} \qquad 5.18$$

$$L\left\{\lambda^{1/2}L^{-1}\{X\}\right\} = D_-^{1/2}\left\{L\{L^{-1}\{X\}\}\right\} = D_-^{1/2}\{X\} \qquad 5.19$$

$$I_-^\alpha\{X(D)\} = \frac{1}{\Gamma(\alpha)}\int_D^\infty \frac{X(z)dz}{(z-D)^{1-\alpha}} \qquad 5.20$$

$$D_-^\alpha\{X(D)\} = \frac{(-1)^n}{\Gamma(n-\alpha)}\frac{d^n}{dD^n}\int_D^\infty \frac{X(z)dz}{(z-D)^{\alpha-n+1}} \qquad 5.21$$

$\Gamma(\alpha)$ is the gamma function and $\alpha > 0$ is any real positive number. $n = [\alpha] + 1$, where $[\alpha]$ is the integer component of $\alpha$. Equation (5.20) where $\alpha = 1$ yields:

$$I_-^1\{X(D)\} = \int_D^\infty X(z)dz \qquad 5.22$$

Equation 5.20 where $\alpha = 1/2$ yields:

$$I_-^{1/2}\{X(D)\} = \frac{1}{\sqrt{\pi}}\int_D^\infty \frac{X(z)}{\sqrt{z-D}}dz \qquad 5.23$$

Finally, Equation 5.21 where $\alpha = 1/2$ yields:

$$D_-^{1/2}\{X(D)\} = -\frac{1}{\sqrt{\pi}}\frac{d}{dD}\int_D^\infty \frac{X(z)dz}{\sqrt{z-D}} \qquad 5.24$$

Inserting Equations 5.22, 5.23 and 5.24 into Equation 5.16, we get:

$$F_{int\,c}(D) = 2k\left[\int_D^\infty X(z)dz + \frac{1}{8\sqrt{\pi}}\int_D^\infty \frac{\sqrt{A}X(z)}{\sqrt{z-D}}dz - \frac{1}{\sqrt{2}}\frac{d}{dD}\int_D^\infty \frac{A^{3/2}X(z)dz}{\sqrt{z-D}}\right] \qquad 5.25$$

The same formulation can be extended to non-conservative forces[22]. Let's get back to Equation 4.48:

$$-\frac{f_d}{Q_0 f_0} + g\sin(\omega_d t_0) = \frac{2f_d}{Ak}\int_0^{1/f_d} F_{int}(z(t),\dot{z}(t))\sin(\omega_d t + \beta)dt + \frac{a_d}{A}\sin(\beta)$$

With $g=0$, $A_0=a_d Q_0$:

$$\int_0^{1/f_d} F_{int}(z(t),\dot{z}(t))\sin(\omega_d t + \beta)dt = -\frac{Ak}{2Q_0 f_0} - \frac{kA_0}{Q_0 2f_d}\sin(\beta) \qquad 5.26$$

We assume non conservative forces of the following type:

$$F_{int\,nc}(z(t),\dot{z}(t)) = \Lambda(D + A + A\cos(\omega_d t + \beta))\dot{z}(t)$$
$$= -\omega_d \Lambda(D + A + A\cos(\omega_d t + \beta))A\sin(\omega_d t + \beta) \qquad 5.27$$

Posing $\iota = \omega_d t + \beta$ and $d\iota = \omega_d dt$ we obtain:

$$2\int_0^\pi \Lambda(D + A + A\cos\iota)\sin^2\iota \, d\iota = \frac{k}{2Q_0 f_0} + \frac{kA_0}{Q_0 2Af_d}\sin(\beta) \qquad 5.28$$

Let's introduce the variable change $u = \cos\iota$, $du = -d\iota(1-u^2)^{1/2}$

$$\int_{-1}^1 \Lambda(D + A + Au)\sqrt{1-u^2}\,du = \frac{k}{4Q_0 f_0} + \frac{kA_0}{4Q_0 Af_d}\sin(\beta) \qquad 5.29$$

We use the function $B(x)$:

$$B(x) = 2\int_x^\infty \Lambda(x)dx \qquad 5.30$$

$$\frac{dB(x)}{dx} = 2\Lambda(x) \qquad 5.31$$

Inserting Equation 5.31 into the left term of Equation 5.29:

$$\int_{-1}^1 \Lambda(D + A + Au)\sqrt{1-u^2}\,du = \frac{1}{2}\int_{-1}^1 \frac{dB(D + A + Au)}{d(D + A + Au)}\sqrt{1-u^2}\,du \qquad 5.32$$

Using Equation 5.31:

$$\frac{dB(D + A + Au)}{d(D + A + Au)} = \frac{dB(D + A + Au)}{du}\frac{du}{d(D + A + Au)} = \frac{1}{A}\frac{dB(D + A + Au)}{du} \qquad 5.33$$

$$\int_{-1}^1 \Lambda(D + A + Au)\sqrt{1-u^2}\,du = \frac{1}{2A}\int_{-1}^1 \frac{dB(D + A + Au)}{du}\sqrt{1-u^2}\,du \qquad 5.34$$

Integrating by parts:

$$\int_{-1}^{1} \Lambda(D+A+Au)\sqrt{1-u^2}\,du = \frac{1}{2A}\left[\left[B(D+A+Au)\sqrt{1-u^2}\right]_{-1}^{1} + \int_{-1}^{1}\frac{B(D+A+Au)u}{\sqrt{1-u^2}}\,du\right] \quad 5.35$$

$$\int_{-1}^{1} \Lambda(D+A+Au)\sqrt{1-u^2}\,du = \frac{1}{2A}\int_{-1}^{1}\frac{B(D+A+Au)u}{\sqrt{1-u^2}}\,du \quad 5.36$$

We use the Laplace transform of $M(\lambda)$:

$$B(D) = \int_{0}^{\infty} M(\lambda)e^{-\lambda D}\,d\lambda \quad 5.37$$

So the right term of Equation 5.36 becomes:

$$\frac{1}{2A}\int_{-1}^{1}\frac{B(D+A+Au)u}{\sqrt{1-u^2}}\,du = \frac{1}{2A}\int_{-1}^{1}\int_{0}^{\infty} M(\lambda)e^{-\lambda D-\lambda A-\lambda Au}\,d\lambda\,\frac{u}{\sqrt{1-u^2}}\,du$$
$$= \frac{1}{2A}\int_{0}^{\infty}\int_{-1}^{1} e^{-(A+Au)\lambda}\frac{u}{\sqrt{1-u^2}}\,du\,M(\lambda)e^{-\lambda D}\,d\lambda \quad 5.38$$

Employing Equation (5.7), Equation (5.29) becomes:

$$\int_{0}^{\infty} T(\lambda A)M(\lambda)e^{-D\lambda}\,d\lambda = \frac{Ak}{2Q_0 f_0} + \frac{kA_0}{2Q_0 f_d}\sin(\beta) \quad 5.39$$

Again using the asymptotic limits of $T(\lambda A)$ and a Padé approximant we get:

$$T(\lambda A) \approx -\pi\frac{\lambda A}{2}\left(1 + \alpha\sqrt{\lambda A} + \sqrt{\frac{\pi}{2}}(\lambda A)^{3/2}\right)^{-1} \quad 5.40$$

We choose $\alpha = 1/8$ as the coefficient for the second term. Equation 5.40 into Equation 5.39 gives:

$$\int_{0}^{\infty}\left[\frac{\lambda}{2}\left(1+\frac{1}{8}\sqrt{\lambda A}+\sqrt{\frac{\pi}{2}}(\lambda A)^{3/2}\right)^{-1}\right]M(\lambda)e^{-D\lambda}\,d\lambda = -\frac{k}{Q_0\omega_0} - \frac{kA_0}{AQ_0\omega_d}\sin(\beta) \quad 5.41$$
$$= 2kY$$

If we assume that $\beta$ is negative:

$$Y = \frac{A_0}{2AQ_0\omega_d}\sin(\beta) - \frac{1}{2Q_0\omega_0} \quad 5.42$$

Following the same development as for the conservative forces (i.e. starting from Equation 5.12):

$$B(D) = 2L\left\{\frac{2k}{\lambda}\left(1 + \frac{1}{8}\sqrt{\lambda A} + \sqrt{\frac{\pi}{2}}(\lambda A)^{3/2}\right)L^{-1}\{Y\}\right\} \qquad 5.43$$

$$B(D) = 4k\left[I_-^1\{Y\} + \frac{1}{8}I_-^{1/2}\{\sqrt{A}Y\} + \sqrt{\frac{\pi}{2}}D_-^{1/2}\{A^{3/2}Y\}\right] \qquad 5.44$$

$$B(D) = 4k\left[\int_D^\infty Y(z)dz + \frac{1}{8\sqrt{\pi}}\int_D^\infty \frac{\sqrt{A}Y(z)}{\sqrt{z-D}}dz - \frac{1}{\sqrt{2}}\frac{d}{dD}\int_D^\infty \frac{A^{3/2}Y(z)dz}{\sqrt{z-D}}\right] \qquad 5.45$$

$$\Lambda(D) = 2k\frac{\partial}{\partial D}\left[\int_D^\infty Y(z)dz + \frac{1}{8\sqrt{\pi}}\int_D^\infty \frac{\sqrt{A}Y(z)}{\sqrt{z-D}}dz - \frac{1}{\sqrt{2}}\frac{d}{dD}\int_D^\infty \frac{A^{3/2}Y(z)dz}{\sqrt{z-D}}\right] \qquad 5.46$$

$$\Lambda(D) = 2k\frac{\partial}{\partial D}\int_D^\infty Y(z)dz + \frac{k}{4}\frac{\partial}{\partial D}\int_D^\infty \frac{\sqrt{A}Y(z)}{\sqrt{\pi(z-D)}}dz - 2k\frac{\partial^2}{\partial D^2}\int_D^\infty \frac{A^{3/2}Y(z)dz}{\sqrt{2(z-D)}} \qquad 5.47$$

Equation 5.47 can be employed for amplitude modulation-based force spectroscopy measurements. It can also be used for tuned oscillator atomic force microscopy measurements performed under vacuum conditions.[24]

## 5.2 General formulations without a Padé approximant

A more general derivation can be made without using a Padé approximant.[25]

We start from Equation 5.5 and use $d = D+A$:

$$\int_0^\infty \int_{-1}^1 e^{-Au\lambda} \frac{u}{\sqrt{1-u^2}} du\, G(\lambda) e^{-d\lambda} d\lambda = \left(\frac{\pi A k}{2}\right)\left(\frac{\omega_0^2 - \omega_d^2}{\omega_0^2}\right) - \frac{A_0 \pi k}{2Q_0}\cos(\beta) \qquad 5.48$$

Then with Equation 5.9 we get :

$$-\int_0^\infty \pi I_1(A\lambda) G(\lambda) e^{-d\lambda} d\lambda = \left(\frac{\pi A k}{2}\right)\left(\frac{\omega_0^2 - \omega_d^2}{\omega_0^2}\right) - \frac{A_0 \pi k}{2Q_0}\cos(\beta) \qquad 5.49$$

From Equation 4.42 with $g = 0$:

$$A_0 = a_d Q_0 \qquad 5.50$$

$$F = a_d k = \frac{A_0 k}{Q_0} \qquad 5.51$$

We get:

$$-\int_0^\infty \pi I_1(A\lambda) G(\lambda) e^{-d\lambda} d\lambda = \left(\frac{\pi A}{2}\right)(k - m\omega_d^2) - \frac{\pi F}{2}\cos(\beta) \qquad 5.52$$

Let's shift $\beta$ by $-\pi/2$:

$$\int_0^\infty I_1(A\lambda) G(\lambda) e^{-d\lambda} d\lambda = \frac{F}{2}\sin(\beta) - \left(\frac{A}{2}\right)(k - m\omega_d^2) \qquad 5.53$$

The modified Bessel function of the first kind of order one can be represented by a power series:

$$I_1(A\lambda) = \sum_{l=0}^\infty \frac{(\lambda A)^{2l+1}}{2^{2l+1} l!(l+1)!} \qquad 5.54$$

Inserting Equation 5.54 into the left term of Equation 5.53 gives:

$$\int_0^\infty \sum_{l=0}^\infty \frac{(\lambda A)^{2l+1}}{2^{2l+1} l!(l+1)!} G(\lambda) e^{-d\lambda} d\lambda = \sum_{l=0}^\infty \frac{A^{2l+1}}{2^{2l+1} l!(l+1)!} \int_0^\infty (\lambda)^{2l+1} G(\lambda) e^{-d\lambda} d\lambda \qquad 5.55$$

Since

$$F_{int\,c}(d) = \int_0^\infty G(\lambda) e^{-\lambda d} d\lambda \qquad 5.56$$

$$\frac{dF_{intc}(d)}{dd} = -\int_0^\infty \lambda G(\lambda) e^{-\lambda d} d\lambda \qquad 5.57$$

We obtain:

$$\sum_{l=0}^\infty \frac{A^{2l+1}}{2^{2l+1} l!(l+1)!} \int_0^\infty (\lambda)^{2l+1} G(\lambda) e^{-d\lambda} d\lambda$$
$$= -\sum_{l=0}^\infty \frac{A^{2l+1}}{2^{2l+1} l!(l+1)!} \frac{d^{2l+1} F_{intc}(d)}{dd^{2l+1}} \qquad 5.58$$

Finally, we transform Equation 5.53 using Equation 5.58:

$$\sum_{l=0}^\infty \frac{A^{2l+1}}{2^{2l+1} l!(l+1)!} \frac{d^{2l+1} F_{intc}(d)}{dd^{2l+1}} = \left(\frac{A}{2}\right)(k - m\omega_d^2) - \frac{F}{2}\sin(\beta) \qquad 5.59$$

As $d \to \infty$, $l = 0, 1, 2, \ldots$ boundary conditions must be:

$$\frac{d^l F_{intc}(d)}{dd^l} = 0 \qquad 5.60$$

For a first order term (i.e. $l = 0$), we obtain with Equation 5.59:

$$\frac{A}{2} \frac{dF_{intc}(d)}{dd} = \left(\frac{A}{2}\right)(k - m\omega_d^2) - \frac{F}{2}\sin(\beta) \qquad 5.61$$

Integrating:

$$F_{intc}^{(1)}(d) = \int_d^\infty \left((k - m\omega_d^2) - \frac{F\sin(\beta)}{A}\right) dd \qquad 5.62$$

For non conservative interaction forces, if we assume that they are of the following type:

$$F_{intnc}(z(t),\dot{z}(t)) = -\Lambda(D + A + A\cos(\omega_d t + \beta))\dot{z}(t) \qquad 5.63$$

Using Equation 5.29 with $d = D+A$:

$$-\int_{-1}^1 \Lambda(d + Au)\sqrt{1 - u^2} du = \frac{k}{4Q_0 f_0} + \frac{kA_0}{4Q_0 A f_d}\sin(\beta) \qquad 5.64$$

Again shifting $\beta$ by $-\pi/2$, with Equations 5.50, 5.51 and using:

$$b = \frac{\omega_0 m}{Q_0} \qquad 5.65$$

We get:

$$\int_{-1}^{1} \Lambda(d+Au)\sqrt{1-u^2}\,du = \frac{\pi}{2}\left(\frac{F}{A\omega_d}\cos(\beta)-b\right) \qquad 5.66$$

To further pursue, we use the Laplace transform:

$$\Lambda(d) = \int_0^{\infty} Z(\lambda)e^{-\lambda d}\,d\lambda \qquad 5.67$$

Inserting it in the left term of Equation 5.66:

$$\int_{-1}^{1} \Lambda(d+Au)\sqrt{1-u^2}\,du = \int_{-1}^{1}\int_0^{\infty} Z(\lambda)e^{-\lambda(d+Au)}\,d\lambda\sqrt{1-u^2}\,du \qquad 5.68$$

We use Equation 5.8:

$$\int_{-1}^{1} e^{-\lambda Au}\sqrt{1-u^2}\,du = \left[\frac{-e^{-\lambda Au}}{\lambda A}\sqrt{1-u^2}\right]_{-1}^{1} - \int_{-1}^{1}\frac{e^{-\lambda Au}}{\lambda A}\frac{u\,du}{\sqrt{1-u^2}} = \frac{\pi}{\lambda A}I_1(\lambda A) \qquad 5.69$$

Equation 5.69 becomes:

$$\int_{-1}^{1} \Lambda(d+Au)\sqrt{1-u^2}\,du = \int_0^{\infty}\frac{\pi}{\lambda A}Z(\lambda)I_1(\lambda A)e^{-\lambda d}\,d\lambda \qquad 5.70$$

Then Equation 5.66 becomes:

$$\int_0^{\infty} Z(\lambda)\frac{I_1(\lambda A)}{\lambda A}e^{-\lambda d}\,d\lambda = \frac{1}{2}\left(\frac{F}{A\omega_d}\cos(\phi)-b\right) \qquad 5.71$$

With Equation 5.54, we obtain by inserting it into the left term of Equation 5.71:

$$\int_0^{\infty} Z(\lambda)\sum_{l=0}^{\infty}\frac{(\lambda A)^{2l}}{2^{2l+1}l!(l+1)!}e^{-d\lambda}\,d\lambda = \sum_{k=0}^{\infty}\frac{A^{2k}}{2^{2l+1}l!(l+1)!}\int_0^{\infty}(\lambda)^{2k}Z(\lambda)e^{-d\lambda}\,d\lambda \qquad 5.72$$

Differentiating Equation 5.67:

$$\frac{d\Lambda(d)}{dd} = -\lambda\int_0^{\infty} Z(\lambda)e^{-\lambda d}\,d\lambda \qquad 5.73$$

We finally obtain:

$$\sum_{l=0}^{\infty}\frac{A^{2l}}{2^{2l+1}l!(l+1)!}\frac{d^{2l}\Lambda(d)}{dd^{2l}} = \frac{1}{2}\left(\frac{F}{A\omega_d}\cos(\beta)-b\right) \qquad 5.74$$

As $d\to\infty$, $l = 0, 1, 2, \ldots$ boundary conditions must be:

$$\frac{d^l\Lambda(d)}{dd^l} = 0 \qquad 5.75$$

For a first order term (i.e. $l = 0$), we obtain

$$\Lambda^{(1)}(d) = \frac{F}{A\omega_d}\cos(\beta) - b \qquad 5.76$$

This formulation can replace the small amplitude term with Laplace formulation and enhance the stability of force reconstruction by avoiding constant amplitude approximation of the small amplitude term.[26]

## 5.3 Polynomial-based general formulation

The most general derivation can be made without any assumption on the form of the non-conservative force.[27] Starting from Equation 5.26, the equation for non-conservative forces:

$$\int_0^{1/f_d} F_{int\,nc}(z(t),\dot{z}(t))\sin(\omega_d t+\beta)dt = -\frac{Ak}{2Q_0 f_0} - \frac{kA_0}{Q_0 2f_d}\sin(\beta)$$

If we pose $\theta = \omega_d t + \beta$, $d\theta/\omega_d = dt$, $\Omega = f_d/f_0$, $F = A_0 k/Q_0$, Equation 5.26 is changed to:

$$\int_0^{2\pi} F_{int\,nc}(z(t),\dot{z}(t))\sin\theta d\theta = -\frac{\pi\Omega Ak}{Q_0} - \pi F \sin(\beta) \qquad 5.77$$

We use the variable $\zeta_e$:

$$\zeta_e = \frac{1}{2Q_0} + \frac{1}{2\pi\Omega Ak}\int_0^{2\pi} F_{int\,nc}(z(t),\dot{z}(t))\sin\theta d\theta = -\frac{F\sin(\beta)}{2\Omega Ak} \qquad 5.78$$

So,

$$F\sin(\beta) = -2\Omega Ak\zeta_e \qquad 5.79$$

For conservative forces, we use Equation 5.1:

$$(\pi Ak)\left(\frac{\omega_0^2 - \omega_d^2}{\omega_0^2}\right) = \omega_d \int_0^{1/f_d} F_{int\,c}(d + A\cos(\omega_d t + \beta))\cos(\omega_d t + \beta)dt + \pi k\frac{A_0}{Q_0}\cos(\beta)$$

Then, with Equation 5.51, $\Omega = f_d/f_0$ and $\theta = \omega_d t + \beta$:

$$(\pi Ak)(1-\Omega^2) = \int_0^{2\pi} F_{int\,c}(z(\theta))\cos\theta d\theta + \pi F \cos(\beta) \qquad 5.80$$

$$F\cos(\beta) = Ak\left[1-\Omega^2 - \frac{1}{Ak\pi}\int_0^{2\pi} F_{int\,c}(z(\theta))\cos(\theta)d\theta\right] = Ak\left[\Omega_e^2 - \Omega^2\right] \qquad 5.81$$

Where

$$\Omega_e^2 = 1 - \frac{1}{Ak\pi}\int_0^{2\pi} F_{int\,c}(z(t))\cos(\theta)d\theta \qquad 5.82$$

Far from the sample and with g=0 we find with Equation 2.454.39:

$$A_0 = \frac{a_d}{\sqrt{\left(1-\frac{f_d^2}{f_0^2}\right)^2 + \left(\frac{f_d}{f_0 Q_0}\right)^2}} = \frac{a_d}{\sqrt{(1-\Omega^2)^2 + \left(\frac{\Omega}{Q_0}\right)^2}} \qquad 5.83$$

And

$$F^2 = a_d^2 k^2 = A_0^2 k^2 \left[ \left(1-\Omega^2\right)^2 + \left(\frac{\Omega}{Q_0}\right)^2 \right] \qquad 5.84$$

With $A_{ratio} = A/A_0$ and Equation 5.84, Equation 5.81 becomes:

$$\frac{\cos(\beta)}{A_{ratio}} = \frac{\left[\Omega_e^2 - \Omega^2\right]}{\sqrt{\left(1-\Omega^2\right)^2 + \left(\frac{\Omega}{Q_0}\right)^2}} \qquad 5.85$$

$$\frac{\cos(\beta)}{A_{ratio}} \sqrt{\left(1-\Omega^2\right)^2 + \left(\frac{\Omega}{Q_0}\right)^2} + \Omega^2 - 1 = \Omega_e^2 - 1 \qquad 5.86$$

With $A_{ratio} = A/A_0$ and Equations 5.84, Equation 5.79 becomes:

$$\sin(\beta) = \frac{-2\Omega A k \zeta_e}{A_0 k \sqrt{\left(1-\Omega^2\right)^2 + \left(\frac{\Omega}{Q_0}\right)^2}} \qquad 5.87$$

$$\frac{\sin(\beta)}{A_{ratio}} \sqrt{\left(1-\Omega^2\right)^2 + \left(\frac{\Omega}{Q_0}\right)^2} = -2\Omega \zeta_e \qquad 5.88$$

Then with Equation 5.82:

$$\int_0^{2\pi} F_{intc}(z(t)) \cos(\theta) d\theta = -Ak\pi \left(\Omega_e^2 - 1\right) \qquad 5.89$$

With the Equation 4.44:

$$z(t \gg 0) = d + A\cos(\omega_d t + \beta) = D + A + A\cos\theta$$

We get:

$$\cos\theta = \frac{z - D - A}{A}$$

$$dz = -A\sin\theta d\theta = -\sqrt{A^2 - (z-D-A)^2} d\theta \qquad 5.90$$

Then Equation 5.89 becomes:

$$\int_D^{D+2A} F_{intc}(z) \frac{(z-D-A)dz}{\sqrt{A^2 - (z-D-A)^2}} = \frac{-A^2 k\pi}{2} \left(\Omega_e^2 - 1\right) \qquad 5.91$$

With Equation 5.86, we get:

$$\int_{D}^{D+2A} F_{intc}(z) \frac{(z-D-A)dz}{\sqrt{A^2-(z-D-A)^2}} =$$
$$\frac{-A^2 k\pi}{2}\left(\frac{\cos(\beta)}{A_{ratio}}\sqrt{(1-\Omega^2)^2+\left(\frac{\Omega}{Q_0}\right)^2}+\Omega^2-1\right) \quad 5.92$$

If we drive at resonance then $\Omega = 1$:

$$\int_{D}^{D+2A} F_{intc}(z) \frac{(z-D-A)dz}{\sqrt{A^2-(z-D-A)^2}} = \frac{-A^2 k\pi \cos(\beta)}{2Q_0 A_{ratio}} \quad 5.93$$

We will have one such equation for each set of values $(D, A, \beta)$.

$F_{intc}(z)$ can be expanded using the Chebyshev polynomials of the first kind $R(\hat{z})$. They are defined in the bounded domain [-1,1]. We will use the transformation:

$$F_{intc}(\hat{z}) = F_{intc}\left(\frac{2z-D_1-D_2}{D_1-D_2}\right) = \sum_{n=0}^{\infty} C_n R_n(\hat{z}) \approx \sum_{n=0}^{N} C_n R_n(\hat{z}) \quad 5.94$$

So that $F_{intc}(D_1) = 1$ and $F_{intc}(D_2) = -1$.

Thus, we obtain using Equation 5.93:

$$\sum_{n=0}^{N} C_n \int_{D}^{D_2} R_n(\hat{z}) \frac{(z-D-A)dz}{\sqrt{A^2-(z-D-A)^2}} = \sum_{n=0}^{N} C_n I_n \approx \frac{-A^2 k\pi \cos(\beta)}{2Q_0 A_{ratio}} \quad 5.95$$

With M values of $D$, $A$ and $\beta$, we get an M x N matrix of $I_n$ values that can be inverted to recover the $C_n$ coefficients. Those are then used to reconstruct the force $F_{intc}(z)$.

# References

1. Bures, J. *Cours de mécanique supérieure*. (École Polytechnique de Montréal, 1990).
2. Giessibl, F. J. Forces and frequency shifts in atomic-resolution dynamic-force microscopy. *Physical Review B* **56**, 16010-16015, (1997).
3. Hölscher, H., Schwarz, U. D. & Wiesendanger, R. Calculation of the frequency shift in dynamic force microscopy. *Applied Surface Science* **140**, 344-351, (1999).
4. Dürig, U. Relations between interaction force and frequency shift in large-amplitude dynamic force microscopy. *Applied Physics Letters* **75**, 433-435, (1999).
5. Corben, H. C. & Stehle, P. *Classical mechanics. H.C. Corben, Philip Stehle*. (Wiley, 1960).
6. Goldstein, H., Poole, C. P. & Safko, J. L. *Classical Mechanics*. (Addison Wesley, 2002).
7. Stewart, S. M. *How to Integrate It: A Practical Guide to Finding Elementary Integrals*. (Cambridge University Press, 2017).
8. Bracewell, R. N. *The Fourier transform and its applications*. (McGraw-Hill, 1986).
9. Gotsmann, B., Anczykowski, B., Seidel, C. & Fuchs, H. Determination of tip–sample interaction forces from measured dynamic force spectroscopy curves. *Applied Surface Science* **140**, 314-319, (1999).
10. Hölscher, H., Allers, W., Schwarz, U. D., Schwarz, A. & Wiesendanger, R. Determination of Tip-Sample Interaction Potentials by Dynamic Force Spectroscopy. *Physical Review Letters* **83**, 4780-4783, (1999).
11. Dürig, U. Interaction sensing in dynamic force microscopy. *New Journal of Physics* **2**, 5, (2000).
12. Giessibl, F. J. A direct method to calculate tip–sample forces from frequency shifts in frequency-modulation atomic force microscopy. *Applied Physics Letters* **78**, 123-125, (2001).
13. Sader, J. E. & Jarvis, S. P. Accurate formulas for interaction force and energy in frequency modulation force spectroscopy. *Applied Physics Letters* **84**, 1801-1803, (2004).
14. Sader, J. E. & Jarvis, S. P. Interpretation of frequency modulation atomic force microscopy in terms of fractional calculus. *Physical Review B* **70**, 012303, (2004).
15. Herruzo, E. T. & Garcia, R. Theoretical study of the frequency shift in bimodal FM-AFM by fractional calculus. *Beilstein Journal of Nanotechnology* **3**, 198-206, (2012).
16. Sader, J. E. *et al.* Quantitative force measurements using frequency modulation atomic force microscopy—theoretical foundations. *Nanotechnology* **16**, S94-S101, (2005).
17. Hölscher, H. & Schwarz, U. D. Theory of amplitude modulation atomic force microscopy with and without Q-Control. *International Journal of Non-Linear Mechanics* **42**, 608-625, (2007).
18. Hölscher, H. *et al.* Measurement of conservative and dissipative tip-sample interaction forces with a dynamic force microscope using the frequency modulation technique. *Physical Review B* **64**, 075402, (2001).
19. Kreyszig, E., Kreyszig, H. & Norminton, E. J. *Advanced engineering mathematics*. (John Wiley & Sons, 2011).
20. Rodriguez, T. R. & Garcia, R. Theory of Q control in atomic force microscopy. *Applied Physics Letters* **82**, 4821-4823, (2003).
21. Hölscher, H., Schwarz, A., Allers, W., Schwarz, U. D. & Wiesendanger, R. Quantitative analysis of dynamic-force-spectroscopy data on graphite (0001) in the contact and noncontact regimes. *Physical Review B* **61**, 12678-12681, (2000).

22  Payam, A. F., Martin-Jimenez, D. & Garcia, R. Force reconstruction from tapping mode force microscopy experiments. *Nanotechnology* **26**, 185706, (2015).

23  Katan, A. J., van Es, M. H. & Oosterkamp, T. H. Quantitative force versus distance measurements in amplitude modulation AFM: a novel force inversion technique. *Nanotechnology* **20**, 165703, (2009).

24  Dagdeviren, O. E., Götzen, J., Hölscher, H., Altman, E. I. & Schwarz, U. D. Robust high-resolution imaging and quantitative force measurement with tuned-oscillator atomic force microscopy. *Nanotechnology* **27**, 065703, (2016).

25  Lee, M. & Jhe, W. General Theory of Amplitude-Modulation Atomic Force Microscopy. *Physical Review Letters* **97**, 036104, (2006).

26  Dagdeviren, O. E. Confronting interatomic force measurements. *Review of Scientific Instruments* **92**, 063703 (2021).

27  Hu, S. & Raman, A. Inverting amplitude and phase to reconstruct tip–sample interaction forces in tapping mode atomic force microscopy. *Nanotechnology* **19**, 375704, (2008).